\documentclass[aps,prb,twocolumn]{revtex4}
\usepackage{amssymb,amsbsy,graphicx,times,subfigure,marvosym,color,bm,multirow}
\usepackage[latin1]{inputenc}
\begin{document}

\title{Doping a topological quantum spin liquid: slow holes in the Kitaev honeycomb model}
\author{G\'{a}bor B. Hal\'{a}sz$^1$}
\author{J. T. Chalker$^1$}
\author{R. Moessner$^2$}
\address{$^1$Theoretical Physics, Oxford University, 1 Keble Road, Oxford OX1
3NP, United Kingdom \\
$^2$Max-Planck-Institut f\"{u}r Physik komplexer Systeme,
N\"{o}thnitzer Stra{\ss}e 38, D-01187 Dresden, Germany}


\begin{abstract}

We present a controlled microscopic study of mobile holes in the
spatially anisotropic (Abelian) gapped phase of the Kitaev honeycomb
model. We address the properties of (i) a single hole [its internal
degrees of freedom as well as its hopping properties]; (ii) a pair
of holes [their (relative) particle statistics and interactions];
(iii) the collective state for a finite density of holes. We find
that each hole in the doped model has an eight-dimensional internal
space, characterized by three internal quantum numbers: the first
two ``fractional'' quantum numbers describe the binding to the hole
of the fractional excitations (fluxes and fermions) of the undoped
model, while the third ``spin'' quantum number determines the local
magnetization around the hole. The fractional quantum numbers also
encode fundamentally distinct particle properties, topologically
robust against small local perturbations: some holes are free to hop
in two dimensions, while others are confined to hop in one dimension
only; distinct hole types have different particle statistics, and in
particular, some of them exhibit non-trivial (anyonic) relative
statistics. These particle properties in turn determine the physical
properties of the multi-hole ground state at finite doping, and we
identify two distinct ground states with different hole types that
are stable for different model parameters. The respective hopping
dimensionalities manifest themselves in an electrical conductivity
approximately isotropic in one ground state and extremely
anisotropic in the other one. We also compare our microscopic study
with related mean-field treatments, and discuss the main
discrepancies between the two approaches, which in particular
involve the possibility of binding fractional excitations as well as
the particle statistics of the holes. On a technical level, we
describe the hopping of mobile holes via a quasi-stationary
approach, where effective hopping matrix elements are calculated
between ground states with stationary holes at different positions.
This approach relies on the fact that the model remains exactly
solvable in the presence of stationary holes, and that the motion of
sufficiently slow holes does not generate bulk excitations in a
gapped phase. When the bare hopping amplitude is much smaller than
the energy gap, many of our results, in particular those on the
hopping properties and the particle statistics, are exact.

\end{abstract}


\maketitle


\section{Introduction} \label{sec-int}

The behavior of a Mott insulator upon doping remains one of the
constitutive open questions in the physics of strongly-correlated
electrons.\cite{Dagotto, Lee} Historically, this is in large part
due to the identification of this issue\cite{Lee, Anderson-0} as
being central to the understanding of high-temperature
superconductors.\cite{HTS} Indeed, it has been recognized that Mott
insulators can enter a broad range of spin states, some of which are
considerably more exotic than the familiar antiferromagnetic
N\'{e}el state. \cite{Misguich} In particular, Anderson
suggested\cite{Anderson-1} that the parent state of high-temperature
superconductors is a resonating-valence-bond (RVB) liquid
state\cite{Anderson-2} with no conventional order. This suggestion
in turn provided motivation for the study of such unconventional
spin states,\cite{Lee, Moessner} and it has been
established\cite{Kivelson, QDM-1} that the RVB liquid state belongs
to the class of fractional\cite{Rajaraman} topological\cite{Wen-1}
states. The effective low-energy excitations above these
non-symmetry-breaking topological states are fractional in the sense
that they carry only a fraction of the spin and charge quantum
numbers that characterize a single electron.\cite{Rajaraman} The
simplest example of such low-energy fractionalization is spin-charge
separation in the case of the RVB liquid, where the elementary
excitations are neutral spinful fermions (spinons) and charged
spinless bosons (holons).\cite{Kivelson} For a doped topological
state, it is then natural to ask how the hopping of an extra
electron or a missing electron (hole) translates into the dynamics
of these fractional excitations.

In this work, we provide a controlled and microscopic analysis of
mobile holes hopping in a topological quantum spin liquid containing
such fractional excitations. We are primarily interested in the
internal degrees of freedom possessed by these holes, their
manifestations in the single-particle behavior such as hopping
properties and particle statistics, and their consequences for the
multi-particle ground state that determines the observable physical
properties. Our approach is complementary to previous
phenomenological works on doped topological states as we study the
exactly solvable Kitaev honeycomb model.\cite{Kitaev} This
two-dimensional quantum spin model has a topological spin-liquid
ground state with fractional excitations,\cite{Kitaev} and it also
remains exactly solvable in the presence of vacancies.\cite{Willans}
Since our approach is applicable only in the regime of slow hopping
when the hopping amplitude is much smaller than the energy gap of
the elementary excitations, we restrict our attention to the
spatially anisotropic (Abelian) gapped phase of the model. For a
recent numerical work on the spatially isotropic gapless phase, see
the exact-diagonalization study by Trousselet \emph{et
al.}\cite{Trousselet}

There is an additional methodological interest in this work as the
Kitaev honeycomb model lies at the intersection of an exact
microscopic solution and a standard phenomenological treatment in
terms of RVB trial wave functions\cite{Baskaran-1} that is
applicable to doped Mott insulators in general. The trial wave
function can optimize the magnetic interaction energy via
(anti)ferromagnetic pairing, while a subsequent mean-field
decomposition naturally leads to a BCS-type Hamiltonian. In the
absence of doping, the constraint of single occupancy is enforced by
an appropriate (numerical or approximate) projection
procedure,\cite{Baskaran-1, GP} while in the presence of doping,
this projection procedure requires softening.

In this framework, low-energy fractionalization in unconventional
spin states is typically captured by a slave-particle (parton)
construction, in which electrons are represented by combinations of
fractional degrees of freedom such as spinons and
holons.\cite{Wen-2} Depending on the precise forms of the
slave-particle construction and the subsequent mean-field
decomposition, several distinct slave-particle mean-field theories
can be constructed for the same Hamiltonian. The possible mean-field
saddle points are most efficiently classified in the framework of
projective symmetry groups,\cite{Wen-2} while the fluctuations
around these mean-field saddle points generally give rise to gauge
theories.\cite{Wen-2, Baskaran-2} Importantly, there are an
extremely large number of distinct saddle points,\cite{Essin} and it
is hard to decide which of these saddle points are
stable.\cite{Hermele} Given a Hamiltonian, it is not clear how to
choose the most relevant saddle point, and therefore the
construction of a slave-particle mean-field theory is not a fully
controlled procedure.

The doped Kitaev honeycomb model has been studied extensively in the
framework of slave-particle mean-field theories,\cite{You, MFT} and
in particular, the mean-field construction by You \emph{et al.}
recovers the exact ground-state correlations in the limit of the
undoped model.\cite{You} Since the exact microscopic solution and
the phenomenological mean-field construction coincide at this
natural starting point of the investigation, the setting of the
Kitaev honeycomb model provides a controlled way of clarifying the
relation between the microscopic and the phenomenological
approaches.

\begin{table*}[t]
\begin{tabular*}{1.00\textwidth}{@{\extracolsep{\fill}} c | c | c | c | c | c | c}
\hline \hline
\multicolumn{2}{c |}{Hole type}                          & Interpretation               & \, Superselection sector \,  & \, Hopping dimensionality \,  & \, Absolute statistics \,  & \, Relative statistics \,  \\
\hline
\multirow{2}{*}{\,\, $h = 0$ \,\,}  & \,\, $q = 0$ \,\,  & Bare hole                    & Trivial ($1$)                & 2D (free \& isotropic)        & Fermion                    & Trivial                    \\
                                    & \,\, $q = 1$ \,\,  & Hole + fermion               & Combined ($e \times m$)      & 2D (free \& anisotropic)      & Boson                      & Non-trivial                \\
\hline
\multirow{2}{*}{\,\, $h = 1$ \,\,}  & \,\, $q = 0$ \,\,  & Hole + flux                  & Electric ($e$)               & 1D (confined)                 & Fermion                    & Non-trivial                \\
                                    & \,\, $q = 1$ \,\,  & \, Hole + flux + fermion \,  & Magnetic ($m$)               & 1D (confined)                 & Fermion                    & Non-trivial                \\
\hline  \hline
\end{tabular*}
\caption{Summary of the most important hole properties for different
combinations of the flux quantum number $h = \{ 0,1 \}$ and the
fermion quantum number $q = \{ 0,1 \}$: interpretations in terms of
elementary excitations bound, superselection sectors of equivalent
excitation clusters, generic hopping properties (see details in
Fig.~\ref{fig-8}), absolute particle statistics, and relative
particle statistics (see details in Table \ref{table-7}).
\label{table-1}}
\end{table*}

Our most important results about the properties of single mobile
holes are summarized in Table \ref{table-1}. In particular, we find
that the holes in the doped model possess internal degrees of
freedom because they can bind the fractional excitations of the
undoped model. The holes therefore carry fractional quantum numbers,
and these quantum numbers are robust against small local
perturbations as they are associated with the superselection sectors
of the model. Crucially, the distinct hole types with different
quantum numbers have fundamentally different single-particle
properties. Depending on their quantum numbers, holes can be either
bosons or fermions, while holes with distinct quantum numbers can
have non-trivial (anyonic) relative statistics. Furthermore, the
various hole types have strikingly different hopping properties.
Specifically, the hopping dimensionality is a function of the hole
type: certain holes are free to hop in two dimensions, while others
are confined to hop in one dimension only.

The internal degrees of freedom have a crucial effect on the
physical properties of the doped model, and the fractional quantum
numbers in the multi-particle ground state depend on the model
parameters. This means that bare holes can be induced to bind
fractional excitations in the ground state\cite{QDM-2} and that the
presence of the resulting composite particles is observable in the
physical properties. Importantly, our results can also be juxtaposed
to those obtained from related slave-particle mean-field theories.
The most closely related mean-field treatment in
Ref.~\onlinecite{You} studies the isotropic gapless phase of the
same model, and two significant observations arise from a careful
comparison between the two approaches. First, the mean-field
treatment unsurprisingly fails to capture the formation of composite
particles consisting of bare holes and fractional excitations.
Second, the particle statistics of bare holes are different in the
two approaches: we find that they are fermions, while they are taken
to be bosons by the slave-particle construction of
Ref.~\onlinecite{You}.

The structure of the paper is as follows. In Sec.~\ref{sec-sum}, we
provide an extended summary of our most important results. In
Secs.~\ref{sec-kit} and \ref{sec-gen}, we review the general
properties of the undoped Kitaev honeycomb model and its spatially
anisotropic gapped phase, respectively. In Sec.~\ref{sec-stat}, we
introduce stationary holes into the model and specify their internal
degrees of freedom. In Sec.~\ref{sec-mob}, we discuss the
single-particle behavior of slow mobile holes, including their
hopping properties and particle statistics. In Sec.~\ref{sec-gas},
we describe the multi-particle ground state and the resulting
physical properties of the doped model. In Sec.~\ref{sec-fast}, we
qualitatively consider mobile holes beyond the regime of slow
hopping. In Sec.~\ref{sec-comp}, we compare our exact microscopic
results with the corresponding mean-field results in
Ref.~\onlinecite{You}. Finally, in Sec.~\ref{sec-out}, we conclude
the paper with suggestions for future research.

\section{Extended summary} \label{sec-sum}

We now provide an extended summary of our most important results.
The next two sections review the undoped Kitaev honeycomb model as
background for the new results in the remaining sections. In
Sec.~\ref{sec-kit}, we introduce the model and describe its exact
solution. It is recalled that the ground state of the model has a
topological degeneracy and that the elementary excitations above the
ground state are fractional as they can only be created in pairs.
There are two kinds of elementary excitations: fluxes, which always
have a gapped energy spectrum, and fermions, which have a gapped or
a gapless energy spectrum, depending on the model parameters. From
Sec.~\ref{sec-gen}, we restrict our attention to the gapped phase of
the model, which is characterized by a gapped energy spectrum for
both fluxes and fermions. We refer to a simple limiting point in
this phase, the isolated dimer limit, where the model consists of
infinitesimally coupled spin dimers. Furthermore, we explain the
notion of superselection sectors to quantify the fractional nature
of isolated excitation clusters.

In Sec.~\ref{sec-stat}, we introduce the formalism for describing
holes, and discuss how the elementary degrees of freedom (modes) are
affected by the presence of $n$ holes. The main result of this
section is that each hole in the model has three localized internal
modes at much smaller energies than the remaining bulk modes (fluxes
and fermions). Excitations in these three internal modes are
characterized by three internal quantum numbers: the flux quantum
number $h = \{ 0,1 \}$, the fermion quantum number $q = \{ 0,1 \}$,
and the plaquette quantum number $p = \{ 0,1 \}$. The quantum
numbers $h$ and $q$ specify the kinds of fractional excitations
(fluxes and fermions) bound to the hole. They therefore determine
its superselection sector via an equivalent excitation cluster (see
Table \ref{table-1}). The quantum number $p$ is related to the
discrete spin-rotation symmetry $\sigma^{x,z} \rightarrow
-\sigma^{x,z}$. It therefore acts as a spin quantum number and
determines the local magnetization around the hole. Since $h$ and
$q$ quantify the fractional nature of the hole, they are robust
against arbitrary local perturbations of sufficiently small
strength. This robustness does not extend to $p$ in general, but it
does so in the important special case of a Heisenberg perturbation.
We also consider interactions between holes and find an attractive
two-hole interaction that is diagonal in $h$ and $p$ but not in $q$.
To ensure that holes do not undergo pair formation or phase
separation, we implicitly assume the presence of a sufficiently
strong Coulomb repulsion as well.

In Sec.~\ref{sec-mob}, we introduce the formalism for describing
hole hopping, and discuss the hopping properties of isolated holes
in the model. Our approach is restricted to the regime of slow
hopping, where the bulk modes are not excited as the hopping
amplitude is much smaller than their energy gap. This section has
two main results. First, the internal quantum numbers $h$, $q$, and
$p$ are all conserved by the hopping. The various hole types with
different quantum numbers can therefore be treated as distinct
particles. Second, the hopping properties of a hole are unaffected
by its quantum number $p$ but are strikingly affected by its quantum
numbers $h$ and $q$. Since the model is spatially anisotropic in the
gapped phase, the two perpendicular dimensions of the lattice are
not equivalent. At a generic point of the gapped phase, $h = 0$
holes are free to hop in two dimensions, while $h = 1$ holes are
confined to hop in one dimension only (see Table \ref{table-1}).
Restricting our attention to $h = 0$ holes, the two-dimensional
hopping problem of $q = 0$ holes is approximately isotropic, while
that of $q = 1$ holes is strongly anisotropic. This difference is
amplified in the isolated dimer limit, where $q = 0$ holes remain
free to hop in two dimensions, while $q = 1$ holes become confined
to hop in one dimension only. We also determine the absolute and the
relative particle statistics of the various hole types (see Table
\ref{table-1}), and provide an intuitive explanation for our results
by referring to the fermionic nature of the bare holes and the
anyonic nature of the fractional excitations bound to them.

In Sec.~\ref{sec-gas}, we describe the multi-hole state representing
a finite density of mobile holes, and determine the ground-state
hole quantum numbers $h$, $q$, and $p$ that minimize the energy of
such a multi-hole state. In the absence of hole interactions, there
are two complementary regimes distinguished by the model parameters.
In the first regime, all holes in the ground state are fermions with
quantum numbers $h = 0$ and $q = 0$. They therefore fill two
identical Fermi seas with different quantum numbers $p = \{ 0,1 \}$.
Since these holes are free to hop in two dimensions, the electrical
conductivity is approximately isotropic. In the second regime, all
holes in the ground state are fermions with quantum numbers $h = 1$.
They therefore fill four identical Fermi seas with different quantum
numbers $q = \{ 0,1 \}$ and $p = \{ 0,1 \}$. Since these holes are
confined to hop in one dimension only, the electrical conductivity
is extremely anisotropic. The two complementary regimes remain
applicable in the presence of hole interactions as both the
attractive interaction and the Coulomb repulsion are diagonal in the
quantum number $h$. In the first regime, a mean-field treatment
restricted to $h = 0$ holes reveals that there is a critical hole
density above which $q = 1$ holes appear. Since these holes are
bosons, their coherent condensation leads to charged superfluid
behavior and a spontaneous net magnetization. In the second regime,
a mean-field treatment restricted to $h = 1$ holes reveals that
scattering between coexisting $q = 0$ holes and $q = 1$ holes
facilitates hopping in both dimensions of the lattice. This implies
that the conductivity anisotropy becomes weaker as the hole density
is increased.

In Sec.~\ref{sec-fast}, we qualitatively discuss hole hopping beyond
the regime of slow hopping, where the bulk modes are excited as the
hopping amplitude is larger than their energy gap. Each hole is then
surrounded by a cloud of fluctuating excitations (fluxes and
fermions), but the internal quantum numbers $h$, $q$, and $p$ are
applicable as long as the hole density is sufficiently small so that
the excitation clouds around different holes do not merge. However,
any hole with quantum numbers other than $h = 0$ and $q = 0$ is
unstable against a spontaneous decay into a lower-energy hole with
$h = 0$ and $q = 0$.

In Sec.~\ref{sec-comp}, we compare our results from the exact
description with those in Ref.~\onlinecite{You} that are obtained
from a mean-field treatment. By contrasting the respective ground
states, we find two main discrepancies between the two approaches.
First, the quantum numbers $h$ and $q$ that specify the kinds of
fractional excitations bound to the hole are captured in the exact
description but ignored in the mean-field treatment. Second, the two
approaches predict different particle statistics for holes with $h =
0$ and $q = 0$: they are fermions in the exact description but
bosons in the mean-field treatment.

\section{Kitaev honeycomb model} \label{sec-kit}

\subsection{Introduction of the model} \label{sec-kit-int}

\begin{figure}[t!]
\centering
\includegraphics[width=6.2cm]{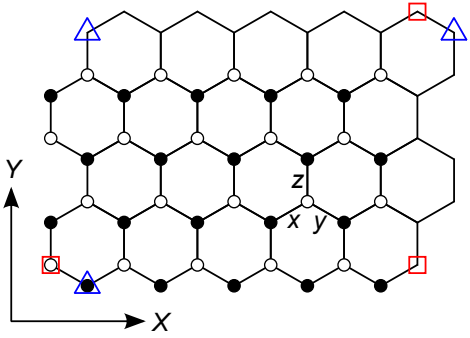}
\caption{(Color online) Illustration of the honeycomb lattice with
dimensions $N_X = 5$ and $N_Y = 4$. Due to the periodic boundary
conditions in the $X$ and $Y$ directions, several sites are
identified with each other, such as the three sites marked by red
rectangles and the three sites marked by blue triangles.
Inequivalent sites in the sublattice $A$ ($B$) are marked by white
(black) dots. Examples of the three bond types ($x$, $y$, $z$) are
also indicated. \label{fig-1}}
\end{figure}

The Kitaev honeycomb model is an exactly solvable two-dimensional
quantum spin model.\cite{Kitaev} Each site of the underlying
honeycomb lattice supports a spin one-half degree of freedom
(particle), and each spin is coupled to its three neighbors by Ising
interactions involving the three different spin components. The
sites of the bipartite lattice can be divided into two sublattices
$A$ and $B$, while the bonds can be divided into three classes $x$,
$y$, and $z$ based on their orientations (see Fig.~\ref{fig-1}). If
$\alpha_{l,l'} = \{x,y,z\}$ gives the type of the bond connecting
two neighboring sites $l$ and $l'$, each site $l$ has three
neighbors $\tilde{\alpha}(l)$ with $\tilde{\alpha} = \{ x,y,z \}$
such that $\alpha_{l,\tilde{\alpha}(l)} = \tilde{\alpha}$. Using
this notation, the Hamiltonian of the model reads as
\begin{equation}
H_{\sigma} = - \sum_{l \in A} \, \sum_{\alpha = x,y,z} J_{\alpha}
\sigma_l^{\alpha} \sigma_{\alpha(l)}^{\alpha}, \label{eq-kit-int-H}
\end{equation}
where $\sigma_l^{\alpha}$ are the physical (Pauli) spin operators,
and $J_{x,y,z}$ are the Ising coupling strengths on the $x$, $y$,
and $z$ bonds, respectively. In the following, we assume without
loss of generality that $0 \leq J_x \leq J_y \leq J_z = 1$.

We consider a lattice with periodic boundary conditions in both the
horizontal ($X$) and the vertical ($Y$) directions. The $N_X \times
N_Y$ lattice has $N \equiv N_X N_Y$ plaquettes, $2N$ sites, and $3N$
bonds (see Fig.~\ref{fig-1}). Based on their relative displacements
in the $X$ direction, the horizontal plaquette stripes of the
lattice can be divided into two classes, even and odd, such that an
even (odd) stripe is neighbored only by odd (even) stripes. We
assume that $N_Y$ is even so that periodic boundary conditions are
applicable in the $Y$ direction without a stripe mismatch between
the top and the bottom of the lattice. Note though that these
boundary conditions are specified only for the purpose of
completeness and that our main results are in fact independent of
the boundary conditions.

\subsection{Flux degrees of freedom} \label{sec-kit-flux}

The Hamiltonian in Eq.~(\ref{eq-kit-int-H}) can be solved exactly by
means of a standard procedure.\cite{Kitaev} The first step is to
notice that there is a commuting non-dynamic observable $W_C$ for
each closed loop $C$ of the lattice. For a loop $C$ containing $L$
sites labeled $\{1, 2, \ldots, L\}$, this non-dynamic observable is
\begin{equation}
W_C = \sigma_1^{\alpha_{1,2}} \sigma_2^{\alpha_{1,2}}
\sigma_2^{\alpha_{2,3}} \sigma_3^{\alpha_{2,3}} \ldots
\sigma_L^{\alpha_{L,1}} \sigma_1^{\alpha_{L,1}}.
\label{eq-kit-flux-W-C}
\end{equation}
Since the lattice is bipartite, the length $L$ of the loop is always
even. We also assume in the following that sites labeled with odd
(even) numbers belong to the sublattice $A$ ($B$).

\begin{figure}[t!]
\centering
\includegraphics[width=5.3cm]{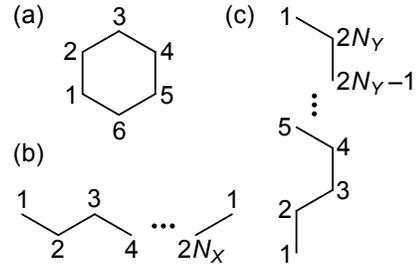}
\caption{Site labeling convention for the generators of the loop
operator group: the plaquette operators $W_P$ (a) and the
topological operators $W_X$ (b) and $W_Y$ (c). \label{fig-2}}
\end{figure}

The loop operators $W_C$ are commuting non-dynamic observables
because they commute with each other as well as the Hamiltonian
$H_{\sigma}$. This means that the different flux sectors
characterized by distinct eigenvalues ($\pm 1$) of the loop
operators can be considered independently. Furthermore, the group
spanned by all loop operators is generated by a finite number of
$\mathbb{Z}_2$ loop operators: those corresponding to the plaquettes
$P$ and the topological strings $X$ and $Y$ going around the lattice
in the $X$ and $Y$ directions. Using the site labeling convention in
Fig.~\ref{fig-2}, these generating loop operators take the forms
\begin{eqnarray}
W_P &=& \sigma_1^{x} \sigma_2^{y} \sigma_3^{z} \sigma_4^{x}
\sigma_5^{y} \sigma_6^{z},
\nonumber \\
W_X &=& -\sigma_1^{z} \sigma_2^{z} \sigma_3^{z} \ldots
\sigma_{2N_X}^{z},
\label{eq-kit-flux-W-PXY} \\
W_Y &=& -\sigma_1^{x} \sigma_2^{y} \sigma_3^{y} \sigma_4^{x}
\sigma_5^{x} \sigma_6^{y} \sigma_7^{y} \sigma_8^{x} \ldots
\sigma_{2N_Y-1}^{y} \sigma_{2N_Y}^{x}. \nonumber
\end{eqnarray}
Importantly, there are only $N-1$ independent plaquette operators
due to the global constraint $\prod_P W_P = 1$. This means that only
$N+1$ flux degrees of freedom are found for the original $2N$ spin
degrees of freedom and that the remaining $N-1$ degrees of freedom
still need to be identified. Note also that the excitation energies
corresponding to the flux degrees of freedom are discussed in
Secs.~\ref{sec-kit-gr} and \ref{sec-gen-lim}.

\subsection{Fermion degrees of freedom} \label{sec-kit-ferm}

To solve the model exactly in each flux sector $\{ W_C = \pm 1 \}$,
four Majorana fermions are introduced at each site $l$ of the
lattice: $c_l$ and $b_l^{\alpha}$ with $\alpha =
x,y,z$.\cite{Kitaev} The corresponding operators satisfy the
standard anticommutation relations
\begin{eqnarray}
\big\{ b_l^{\alpha}, b_{l'}^{\alpha'} \big\} &=& 2 \delta_{ll'}
\delta_{\alpha \alpha'}, \quad \big( b_l^{\alpha} \big)^2 = 1,
\nonumber \\
\big\{ c_l, c_{l'} \big\} &=& 2 \delta_{ll'}, \qquad \quad c_l^2 = 1,
\label{eq-kit-ferm-maj} \\
\big\{ b_l^{\alpha}, c_{l'} \big\} &=& 0.
\nonumber
\end{eqnarray}
The physical spin operators are then expressed in terms of the
Majorana fermions as $\sigma_l^{\alpha} = i b_l^{\alpha} c_l$. From
this expression and the relations in Eq.~(\ref{eq-kit-ferm-maj}),
certain properties of the spin operators can be immediately
recovered: $[\sigma_l^{\alpha}, \sigma_{l'}^{\alpha'}] = 0$ for $l
\neq l'$, $\{ \sigma_l^{\alpha}, \sigma_{l}^{\alpha'} \} = 0$ for
$\alpha \neq \alpha'$, and $(\sigma_l^{\alpha})^2 = 1$.

Since complex fermions are more straightforward to understand than
Majorana fermions, it is useful to construct complex fermions by
pairing up the Majorana fermions in an appropriate manner. Each
Majorana fermion $b_l^{\alpha}$ belongs to an end of a bond, and the
standard choice is to pair up the ones that belong to the two ends
of the same bond. For each site $l \in A$, three complex bond
fermions are then obtained as
\begin{equation}
\chi_l^{\alpha} = \frac{1}{2} \left[ b_l^{\alpha} - i
b_{\alpha(l)}^{\alpha} \right], \quad \left( \chi_l^{\alpha}
\right)^{\dag} = \frac{1}{2} \left[ b_l^{\alpha} + i
b_{\alpha(l)}^{\alpha} \right]. \label{eq-kit-ferm-bond}
\end{equation}
Each Majorana fermion $c_l$ belongs to a site, and the standard
choice is to pair up the ones that belong to any two sites connected
by a $z$ bond. In terms of $c_{l,A} \equiv c_l$ and $c_{l,B} \equiv
c_{z(l)}$ that are defined for each site $l \in A$, one complex
matter fermion is then obtained for each pair of sites as
\begin{equation}
f_l = \frac{1}{2} \left( c_{l,A} + i c_{l,B} \right), \quad
f_l^{\dag} = \frac{1}{2} \left( c_{l,A} - i c_{l,B} \right).
\label{eq-kit-ferm-matt}
\end{equation}
The state of the bond fermion $\chi_l^{\alpha}$ can be measured with
the bond fermion operator $i b_l^{\alpha} b_{\alpha(l)}^{\alpha} = 1
- 2(\chi_l^{\alpha})^{\dag} \chi_l^{\alpha}$, while the state of the
matter fermion $f_l$ can be measured with the matter fermion
operator $-i c_{l,A} c_{l,B} = 1 - 2f_l^{\dag} f_l$. We say that a
bond (matter) fermion is excited if its bond (matter) fermion
operator takes an eigenvalue $-1$ rather than $+1$.

When expressed in terms of the Majorana fermions, the Hamiltonian in
Eq.~(\ref{eq-kit-int-H}) takes the form
\begin{equation}
H_{\hat{u}} = i \sum_{l \in A} \, \sum_{\alpha = x,y,z} J_{\alpha}
\hat{u}_{l,\alpha(l)} c_l c_{\alpha(l)}, \label{eq-kit-ferm-H-1}
\end{equation}
where the $3N$ bond fermion operators $\hat{u}_{l,\alpha(l)} \equiv
i b_l^{\alpha} b_{\alpha(l)}^{\alpha}$ are commuting non-dynamic
observables because they commute with each other as well as the
Hamiltonian $H_{\hat{u}}$. This means that the different bond
fermion sectors characterized by distinct eigenvalues ($\pm 1$) of
the bond fermion operators can be considered independently. On the
other hand, the Hamiltonian $H_{\hat{u}}$ is quadratic and hence
exactly solvable in each bond fermion sector $\{ u_{l,\alpha(l)}
\equiv \langle \hat{u}_{l,\alpha(l)} \rangle = \pm 1 \}$. If the
Majorana fermions $c_l$ corresponding to the two sublattices are
incorporated into two vectors $c_{A,B}$ with elements $(c_A)_l =
c_{l,A}$ and $(c_B)_l = c_{l,B}$, the Hamiltonian in
Eq.~(\ref{eq-kit-ferm-H-1}) becomes
\begin{equation}
H_u = i c_A^T \cdot M \cdot c_B, \qquad M_{ll'} = u_{l,z(l')}
J_{\alpha_{l,z(l')}}, \label{eq-kit-ferm-H-2}
\end{equation}
where $J_{\alpha_{l,z(l')}} = 0$ if $l$ and $z(l')$ are not
neighbors. The matrix $M$ has a singular value decomposition $M = U
\cdot S \cdot V^T$, where $S$ is a positive-semidefinite diagonal
matrix, while $U$ and $V$ are real orthogonal matrices. We assume in
the following that the singular values $S_k \equiv S_{kk}$ are in an
increasing order such that $0 \leq S_1 \leq S_2 \leq \ldots \leq
S_N$. The orthogonal matrices $U$ and $V$ give a new set of Majorana
fermions as
\begin{equation}
\gamma_{k,A} = \sum_{l \in A} U_{lk} c_{l,A}, \quad \gamma_{k,B} =
\sum_{l \in A} V_{lk} c_{l,B}, \label{eq-kit-ferm-maj-new}
\end{equation}
and the corresponding complex matter fermions become
\begin{equation}
\phi_k = \frac{1}{2} \left( \gamma_{k,A} + i \gamma_{k,B} \right),
\,\, \phi_k^{\dag} = \frac{1}{2} \left( \gamma_{k,A} - i
\gamma_{k,B} \right). \label{eq-kit-ferm-matt-new}
\end{equation}
In terms of these new matter fermions $\phi_k$, the Hamiltonian in
Eq.~(\ref{eq-kit-ferm-H-2}) takes the free-fermion form
\begin{equation}
H_u = \sum_{k=1}^N S_k \left( 2 \phi_k^{\dag} \phi_k - 1 \right).
\label{eq-kit-ferm-H-3}
\end{equation}
The ground-state energy in the given bond fermion sector is then
$-\sum_k S_k$, and the elementary excitations are the free matter
fermions $\phi_k$ with excitation energies $2S_k$.

It is important to understand the relation between the commuting
non-dynamic observables in the physical spin picture and the
Majorana fermion picture: the loop operators and the bond fermion
operators, or equivalently, the flux sectors and the bond fermion
sectors. When expressed in terms of the Majorana fermions, the loop
operators $W_C$ take the form
\begin{eqnarray}
W_C &=& b_1^{\alpha_{1,2}} b_2^{\alpha_{1,2}} b_2^{\alpha_{2,3}}
b_3^{\alpha_{2,3}} \ldots b_L^{\alpha_{L,1}} b_1^{\alpha_{L,1}}
\label{eq-kit-ferm-W-C} \\
&=& \hat{u}_{1,2} \hat{u}_{3,2} \hat{u}_{3,4} \hat{u}_{5,4} \ldots
\hat{u}_{L-1,L} \hat{u}_{1,L}, \nonumber
\end{eqnarray}
and in particular, the plaquette operators $W_P$ become
\begin{equation}
W_P = \hat{u}_{1,2} \hat{u}_{3,2} \hat{u}_{3,4} \hat{u}_{5,4}
\hat{u}_{5,6} \hat{u}_{1,6}. \label{eq-kit-ferm-W-P}
\end{equation}
These expressions show that the non-dynamic observables in the
physical spin picture are uniquely determined by those in the
Majorana fermion picture. However, the converse can not be true
because there are $3N$ bond fermion operators in the Majorana
fermion picture for only $N+1$ loop operators in the physical spin
picture. In fact, there is a gauge transformation $D_l \equiv b_l^x
b_l^y b_l^z c_l$ for each site $l$ that flips three bond fermions
but does not flip any loops. This means that the bond fermion
sectors before and after the gauge transformation correspond to the
same flux sector. Since $D \equiv \prod_l D_l$ does not flip any
bond fermions, there are $2N-1$ independent gauge transformations
$D_l$, and the discrepancy between the numbers of non-dynamic
observables is thus explained.

The gauge redundancy in the Majorana fermion picture follows from an
enlarged Hilbert space with respect to the physical spin picture. In
particular, the Hilbert space of a single site is $4$ dimensional in
the Majorana fermion picture and only $2$ dimensional in the
physical spin picture. This discrepancy is consistent with the fact
that the spin identity $-i \sigma_l^x \sigma_l^y \sigma_l^z \equiv
1$ in the physical spin picture translates into the gauge constraint
$D_l = +1$ in the Majorana fermion picture. In fact, all the states
in the Majorana fermion picture that are related to each other by
gauge transformations $D_l$ are equivalent descriptions of the same
state in the physical spin picture. This physical state can be
obtained from any of the gauge-related states by a projection onto
the subspace with $D_l = +1$ for all $l$. The corresponding
projection operator takes the form
\begin{equation}
\mathcal{P} = \prod_l \left( \frac{1 + D_l} {2} \right) =
\mathcal{P}' \left( 1 + D \right), \label{eq-kit-ferm-P}
\end{equation}
where $\mathcal{P}'$ contains all terms in $\mathcal{P}$ that flip
bond fermions in inequivalent ways.\cite{Yao} Since $D =
(-1)^{N_{\chi} + N_f}$ when expressed in terms of the bond fermion
number $N_{\chi} \equiv \sum_{\alpha} \sum_{l \in A}
(\chi_l^{\alpha})^{\dag} \chi_l^{\alpha}$ and the matter fermion
number $N_f \equiv \sum_{l \in A} f_l^{\dag} f_l$, any states with
odd total fermion number are projected to zero. There is a resulting
global constraint for physical states: the total fermion number
$N_{\chi} + N_f$ must be even. In each bond fermion sector with an
even (odd) number of excited bond fermions, the number of excited
matter fermions also must be even (odd). This means that only $N-1$
matter fermions can be excited independently from each other. The
original $2N$ spin degrees of freedom are then fully recovered via
the identification of the $2N$ natural degrees of freedom in the
model: the $N+1$ fluxes and the $N-1$ fermions.

\subsection{Ground state and excitations} \label{sec-kit-gr}

The exact solution of the model provides a simple procedure for
identifying its ground state.\cite{Kitaev} Each flux sector can be
considered individually and represented with one of its
corresponding bond fermion sectors. The ground state in the flux
sector is then projected from that in the bond fermion sector (see
Sec.~\ref{sec-kit-ferm}), and the overall ground state is the lowest
lying of all these individual ground states. Furthermore, it can be
shown using translational invariance that the ground state is in the
trivial flux sector: the one in which $W_P = +1$ for all
plaquettes.\cite{Lieb} The ground-state energy $\Gamma_0$ is then
$-\sum_k S_k$ as obtained from the matrix $M$ in
Eq.~(\ref{eq-kit-ferm-H-2}) using the trivial bond fermion sector:
the one in which $u_{l,\alpha(l)} = +1$ for all bonds. Note that
there are in principle four trivial flux sectors corresponding to
the topological eigenvalues $W_{X,Y} = \pm 1$ and that this leads to
the existence of four degenerate ground states. However, the
topological degrees of freedom are impossible to excite locally. We
therefore neglect them in the following by considering only the
trivial topological sector with $W_X = W_Y = +1$. This means that
the effective number of degrees of freedom is reduced to $2N-2$.

It is also revealed by the exact solution that the elementary
excitations above the ground state are plaquettes (fluxes) and
fermions.\cite{Kitaev} We say that a plaquette $P$ is excited
(carries a flux) if its plaquette operator $W_P$ takes an eigenvalue
$-1$ rather than $+1$. In the presence of flux excitations, the flux
sector can no longer be represented with the trivial bond fermion
sector, and the energy $-\sum_k S_k$ is larger than $\Gamma_0$. This
difference translates into a finite flux excitation energy. Note
that fluxes can only be excited pairwise due to the global
constraint $\prod_P W_P = 1$. The matter fermion excitations
$\phi_k$ have excitation energies $E_k \equiv 2S_k$, and by
considering the distribution of these energies, two distinct phases
of the model can be identified. In the gapless phase with $J_z < J_x
+ J_y$, the smallest excitation energies $E_k$ vanish in the
thermodynamic limit. In the gapped phase with $J_z > J_x + J_y$, the
excitation energies $E_k$ are all finite in the thermodynamic limit.
Note that fermions can only be excited pairwise due to the global
constraint that $N_{\chi} + N_f$ must be even.

\section{Gapped phase of the model} \label{sec-gen}

In the following, we restrict our attention to the gapped phase of
the Kitaev honeycomb model, where the coupling strengths satisfy
$J_x + J_y < J_z$. Since all fluxes and fermions have finite
excitation energies, the ground state in this phase is separated
from the excited states by a finite energy gap. We measure all
energies in units of the largest coupling strength $J_z = 1$ and
choose the two smaller coupling strengths $J_{x,y}$ to be equal. The
model is then parameterized by the dimensionless coupling strength
$J \equiv J_x = J_y < 1/2$.

\subsection{Isolated dimer limit} \label{sec-gen-lim}

When considering the gapped phase, it is useful to start any
discussion in the isolated dimer limit of $J = 0$. In this limit,
the model separates into $N$ isolated (non-interacting) spin dimers
along $z$ bonds.\cite{Dusuel} Since the two spins in any dimer are
coupled by a ferromagnetic Ising term $-\sigma_l^z \sigma_{z(l)}^z$,
they must be either both up or both down in the ground state.
However, there is still an exponentially large ground-state
degeneracy as each dimer can choose from two configurations. This
degeneracy can then be lifted by applying a perturbation theory in
the dimensionless coupling strength $J \ll 1$.\cite{Willans} At
fourth order in $J$, the projection of the Hamiltonian in
Eq.~(\ref{eq-kit-int-H}) onto the degenerate ground-state subspace
is
\begin{equation}
\tilde{H}_{\sigma} = -N - \tilde{C} (N, J) - \frac{J^4} {16} \sum_P
W_P. \label{eq-gen-lim-H}
\end{equation}
The first term is the ground-state energy at $J = 0$ and the
remaining terms are the perturbative corrections: the constant term
$\tilde{C} (N, J)$ shifts the energy of the entire subspace, while
the last term lifts the ground-state degeneracy by specifying the
flux sector. In accordance with Sec.~\ref{sec-kit-gr}, the actual
ground state has $W_P = +1$ for all plaquettes.

It is instructive to write this ground state $| \Omega \rangle$ in
terms of both the physical spins and the Majorana fermions. In the
physical spin picture, it can be obtained by a projection from any
state with $\sigma_l^z \sigma_{z(l)}^z = +1$ for all dimers onto the
subspace with $W_P = +1$ for all plaquettes. For example, by
projecting from the all-spins-up state $| \Uparrow \rangle$, the
ground state becomes
\begin{equation}
| \Omega \rangle = \prod_P \left( \frac{1 + W_P} {2} \right) |
\Uparrow \rangle. \label{eq-gen-lim-gs-1}
\end{equation}
In the Majorana fermion picture, the trivial flux sector is
represented with the trivial bond fermion sector, and the matrix $M$
in Eq.~(\ref{eq-kit-ferm-H-2}) is the unit matrix. Since the free
matter fermions $\phi_k$ in Eq.~(\ref{eq-kit-ferm-matt-new}) are
then identical to the original matter fermions $f_l$ in
Eq.~(\ref{eq-kit-ferm-matt}), the ground state is the vacuum of the
bond fermions and the original matter fermions. Formally, this
vacuum state $| 0 \rangle$ is defined by $\chi_l^{\alpha} | 0
\rangle = 0$ and $f_l | 0 \rangle = 0$ for all $l$ and $\alpha$. The
physical ground state in Eq.~(\ref{eq-gen-lim-gs-1}) is then $|
\Omega \rangle = \mathcal{P} | 0 \rangle$.

The excitations above the ground state can be discussed in a similar
manner. The flux excitations are obtained by projecting onto a
subspace with excited plaquettes $W_P = -1$ in the physical spin
picture and by exciting an appropriate set of bond fermions in the
Majorana fermion picture. Due to the presence of the gauge
transformations $D_l$, it is possible to represent any flux sector
with a bond fermion sector in which only $x$ and $y$ bond fermions
are excited. Since the matrix $M$ in Eq.~(\ref{eq-kit-ferm-H-2})
does not depend on these bond fermions for $J = 0$, we recover the
result in Eq.~(\ref{eq-gen-lim-H}) that the flux excitation energies
$E_P \sim J^4$ vanish when $J \rightarrow 0$. The fermion
excitations are obtained by projecting from a state with broken
dimers $\sigma_l^z \sigma_{z(l)}^z = -1$ in the physical spin
picture and by exciting the corresponding matter fermions in the
Majorana fermion picture. Since $u_{l,z(l)} = +1$ for all dimers
when only $x$ and $y$ bond fermions are excited, the relation
$\sigma_l^z \sigma_{z(l)}^z = \hat{u}_{l,z(l)} (1 - 2f_l^{\dag}
f_l)$ shows that excited matter fermions indeed correspond to broken
dimers. Furthermore, it follows from both pictures that these
fermion excitations all have exactly the same energy $E_f = 2$.

It is a conceptual problem that we require $J > 0$ for a finite
plaquette excitation energy but $J = 0$ for the presence of the
isolated dimers. In fact, since the localized matter fermions at $J
= 0$ all have the same excitation energy, even an infinitesimally
small perturbation $J \ll 1$ is enough to delocalize them across the
entire lattice and have them form a band of a small width $\Delta
E_f \sim J$. This implies that the free (delocalized) matter
fermions $\phi_k$ and the original (localized) matter fermions $f_l$
are entirely different for any $J > 0$. To obtain the ground state
at $J > 0$, the vacuum state $| 0 \rangle$ is then projected onto
the subspace where no free matter fermions $\phi_k$ are excited.
Using this method, the physical ground state takes the form
\begin{equation}
| \Omega \rangle = \mathcal{P} \prod_{k=1}^N \left( \phi_k
\phi_k^{\dag} \right) | 0 \rangle. \label{eq-gen-lim-gs-2}
\end{equation}
Although the perturbation mixes the various creation operators, and
consequently, the various annihilation operators together, it does
not significantly mix the creation operators with the annihilation
operators. This implies that the $J > 0$ ground state in Eq.
(\ref{eq-gen-lim-gs-2}) is close to the $J = 0$ ground state
$\mathcal{P} | 0 \rangle$ and can be described faithfully in terms
of the localized matter fermions. In the following, we therefore
often simultaneously assume a finite plaquette excitation energy and
localized matter fermions, always mentioning when the perturbative
interactions between the matter fermions are important.

\subsection{Global constraints and superselection sectors} \label{sec-gen-sup}

The numbers of independent flux and fermion excitations are limited
by two essential global constraints. In the physical spin picture,
these two constraints can be obtained by noticing that the product
of all plaquette operators $W_P$ corresponding to plaquettes in even
($\eta$) stripes, or alternatively, plaquettes in odd ($\mu$)
stripes is equivalent to the product of all dimer operators
$\lambda_l \equiv \sigma_l^z \sigma_{z(l)}^z$. Mathematically, these
two relations are
\begin{equation}
\prod_{P \in \eta} W_P = \prod_{P \in \mu} W_P = \prod_{l \in A}
\lambda_l. \label{eq-gen-sup-cons}
\end{equation}
Since the $W_P$ and the $\lambda_l$ are all $\mathbb{Z}_2$
variables, the first equality recovers the global constraint
$\prod_P W_P = 1$, while the second equality becomes $\prod_{l \in
A} \lambda_l \prod_{P \in \mu} W_P = 1$. In the Majorana fermion
picture, the first equality is automatically satisfied because
$\hat{u}_{l,\alpha(l)}^2 = 1$ for all bonds. The second equality can
be understood by noticing that an excited $z$ bond fermion
corresponds to two excited plaquettes that are either both in an
even stripe or both in an odd stripe while an excited $x$ or $y$
bond fermion corresponds to one excited plaquette in an even stripe
and one excited plaquette in an odd stripe. Since this property
translates into $\prod_{P \in \mu} W_P = (-1)^{N_{\chi^x} +
N_{\chi^y}}$ and the relation $\lambda_l = \hat{u}_{l,z(l)} (1 -
2f_l^{\dag} f_l)$ implies $\prod_{l \in A} \lambda_l =
(-1)^{N_{\chi^z} + N_f}$, the second equality recovers the global
constraint that $N_{\chi} + N_f$ must be even for physical states.

There is an alternative formulation of the global constraints given
in Eq.~(\ref{eq-gen-sup-cons}) where one electric (magnetic) charge
$e$ ($m$) is assigned to each excited plaquette in an even (odd)
stripe and one from both charges $e$ and $m$ is assigned to each
broken dimer. The global constraints in this formulation are that
the total numbers of electric charges ($N_e$) and magnetic charges
($N_m$) both must be even.\cite{Kitaev} In particular, if there are
isolated clusters of excitations in the lattice, each of them can be
classified into four superselection sectors based on the types of
unpaired charges it contains: trivial ($1$), electric ($e$),
magnetic ($m$), and combined ($\varepsilon \equiv e \times m$). When
different clusters are combined, the superselection sector of the
combined cluster is given by the fusion rules in Table
\ref{table-2}. Using this language, the global constraints mean that
the combination of all clusters belongs to the trivial
superselection sector $1$.

\begin{table}[h]
\begin{tabular*}{0.24\textwidth}{@{\extracolsep{\fill}} c | c c c c }
\hline \hline
                         & \, $1$ \,            & \, $e$ \,            & \, $m$ \,            & \, $\varepsilon$ \,  \\
\hline
\,\, $1$ \,\,            & \, $1$ \,            & \, $e$ \,            & \, $m$ \,            & \, $\varepsilon$ \,  \\
\,\, $e$ \,\,            & \, $e$ \,            & \, $1$ \,            & \, $\varepsilon$ \,  & \, $m$ \,            \\
\,\, $m$ \,\,            & \, $m$ \,            & \, $\varepsilon$ \,  & \, $1$ \,            & \, $e$ \,            \\
\,\, $\varepsilon$ \,\,  & \, $\varepsilon$ \,  & \, $m$ \,            & \, $e$ \,            & \, $1$ \,            \\
\hline  \hline
\end{tabular*}
\caption{Fusion rules governing the combination of superselection
sectors when different excitation clusters are combined.
\label{table-2}}
\end{table}

The most important property of the superselection sectors is that
they are robust against arbitrary local perturbations. Since a local
perturbation acts only within one excitation cluster, it could only
change the superselection sector of the cluster by also violating at
least one global constraint. The superselection sector of an
excitation cluster can then only be changed by a non-local
perturbation that also changes the superselection sector of a
different cluster or creates an additional cluster with a
non-trivial superselection sector.

\section{Stationary holes} \label{sec-stat}

\subsection{Description of holes} \label{sec-stat-int}

We introduce $n$ holes into the Kitaev honeycomb model by removing
the spin one-half particles from $n$ sites of the honeycomb lattice.
For the model with $n > 0$ holes, the exact solution in
Sec.~\ref{sec-kit} is still applicable, but it needs to be performed
in a different way because there are no Majorana fermions at the
hole sites.\cite{Willans} It is then not clear how to construct
complex fermions from the remaining Majorana fermions, and the bond
fermion operators, or equivalently, the plaquette operators acting
on the hole sites become ill-defined.

To fix this problem, we use an alternative description: the spin
one-half particles are not actually removed from the hole sites, but
only their Ising interactions with their neighbors are switched off.
This way, we obtain $2^n$ copies of the original model that
correspond to the different configurations of the $n$
non-interacting hole spins. Since there are still Majorana fermions
at all sites, the bond fermion operators and the plaquette operators
remain well-defined. This means that the exact solution can be
performed in exactly the same way as in Sec.~\ref{sec-kit}. However,
there is an additional $2^n$-fold degeneracy due to the presence of
the non-interacting hole spins, which is unphysical and hence must
be discarded.

Formally, we can demand all hole spins to be in the spin-up state:
$\sigma_l^z = +1$ for all sites $l \in \Delta$, where $\Delta$ is
the set of hole sites. To obtain a physical state, we then need to
use the appropriate projection operator, which takes the form
\begin{equation}
\mathcal{Q}_{\Delta} = \prod_{l \in \Delta} \left( \frac{1 +
\sigma_l^z} {2} \right) = \prod_{l \in \Delta} \left( \frac{1 + i
b_l^z c_l} {2} \right). \label{eq-stat-int-Q}
\end{equation}
Note that the treatment of the unphysical hole spins is completely
analogous to the treatment of the unphysical Majorana fermions. In
the Majorana fermion picture, different states corresponding to the
same state in the physical spin picture are related by gauge
transformations $D_l$. We can work in different gauges and then use
the projector $\mathcal{P}$ to enforce the constraint $D_l = +1$ at
all sites. In the hole spin picture, different states corresponding
to the same state in the actual hole picture are related by gauge
transformations $\sigma_{l \in \Delta}^z$. We can work in different
gauges and then use the projector $\mathcal{Q}_{\Delta}$ to enforce
the constraint $\sigma_l^z = +1$ at all hole sites.

\subsection{Internal degrees of freedom} \label{sec-stat-deg}

We now investigate how the excitations above the ground state as
discussed in Sec.~\ref{sec-kit-gr} are affected by the introduction
of $n > 0$ holes into the model. Since each hole corresponds to one
fewer spin degree of freedom and the topological degrees of freedom
are neglected, the total number of $\mathbb{Z}_2$ degrees of freedom
(modes) is $2N-n-2$. We restrict our attention to the thermodynamic
limit of $N_{X,Y} \rightarrow \infty$ and assume that the holes in
the model are isolated such that the smallest distance between any
two holes is $R \gg 1$.

In the presence of $n > 0$ holes, we distinguish two types of
plaquettes: hole plaquettes that contain one hole site each and bulk
plaquettes that contain no hole sites. Each hole site $l \in \Delta$
is contained by three plaquettes $P_l^{x,y,z}$ whose corresponding
plaquette operators $W_{P_l^{x,y,z}}$ act on the hole site with
$\sigma_l^{x,y,z}$, respectively. The number of hole plaquettes is
therefore $3n$ and the number of bulk plaquettes is $N-3n$. From a
perturbation theory in $J \ll 1$, there is a finite excitation
energy $E_P \sim +J^4$ for bulk plaquettes and no excitation energy
for hole plaquettes. However, at each hole site $l \in \Delta$,
there is a finite excitation energy $E_Q \sim -J^8$ for the hole
loop $Q_l$ surrounding all three hole plaquettes. The negative
excitation energy $E_Q < 0$ means that the hole loop operators
$W_{Q_l}$ preferentially take eigenvalues $-1$ in the ground
state.\cite{Willans} More precisely, since the global constraint
$\prod_P W_P = 1$ translates into $\prod_{l \in \Delta} W_{Q_l} = 1$
when no bulk plaquettes are excited, the hole loop operator
$W_{Q_l}$ is $-1$ for all hole sites when $n$ is even and for all
but one hole sites when $n$ is odd.

\begin{figure}[t!]
\centering
\includegraphics[width=5.5cm]{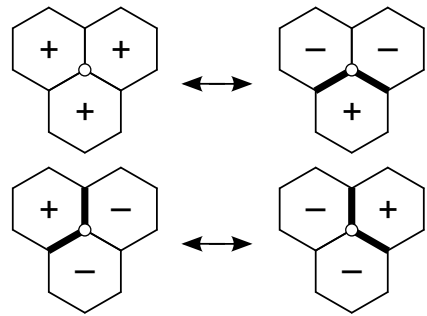}
\caption{Simultaneous gauge transformations $D_l$ and $\sigma_l^z$
relating the bond fermion sectors around a hole site $l \in \Delta$
(white dot) when there is no flux bound to the hole. Each bond
fermion sector is labeled with the excited bond fermions (thick
lines) and the corresponding plaquette operator eigenvalues ($\pm
1$). Our convention is to consider only the two bond fermion sectors
on the left. \label{fig-3}}
\end{figure}

Since the hole loop operator is $W_{Q_l} = W_{P_l^x} W_{P_l^y}
W_{P_l^z}$ in terms of the individual hole plaquette operators, we
say that the hole at site $l$ has a flux bound to it if its hole
loop operator $W_{Q_l}$ takes an eigenvalue $-1$ rather than $+1$.
This relation also suggests that each hole has a hole flux mode
$Q_l$ with a finite excitation energy and two independent hole
plaquette modes $P_l^{x,z}$ with zero excitation energies. In fact,
there is one fewer hole plaquette mode due to the presence of the
unphysical hole spin: the four plaquette sectors corresponding to
$W_{P_l^{x,z}} = \pm 1$ in the hole spin picture are pairwise
related by the gauge transformation $\sigma_l^z$, and the
corresponding bond fermion sectors in the Majorana fermion picture
are pairwise related by the gauge transformations $D_l$ and
$\sigma_l^z$. These gauge transformations are illustrated in
Fig.~\ref{fig-3}. In the following, we use the convention in which
the two remaining bond fermion sectors are related to each other by
the operator $i b_l^x b_l^z$ flipping the $x$ and the $z$ bond
fermions around the hole site $l$. When there is no flux bound to
the hole, this means that the two remaining plaquette sectors with
$W_{P_l^y} = +1$ are distinguished by $W_{P_l^x} = W_{P_l^z} = \pm
1$. In conclusion, if the model contains $n > 0$ holes, there are
$N-3n$ bulk plaquette (flux) modes with excitation energies $E_P
\sim J^4$, there are $n-1$ hole flux modes with excitation energies
$E_Q \sim J^8$, and there are $n$ hole plaquette modes with zero
excitation energies. Note that the number of independent hole flux
modes is reduced by $1$ due to the global constraint $\prod_P W_P =
1$.

In the presence of $n > 0$ holes, we distinguish two types of
fermions: hole fermions and bulk fermions. When $J = 0$, hole
fermions are localized at dimers that contain one hole site each,
while bulk fermions are localized at dimers that contain no hole
sites. Since the bulk dimers have Ising interactions $-\sigma_l^z
\sigma_{z(l)}^z$, there is a finite excitation energy $E_f = 2$ for
the bulk fermions. However, since the Ising interactions of the hole
dimers are switched off, there is no excitation energy for the hole
fermions. When $J > 0$, the bulk fermions delocalize across the
entire lattice (see Sec.~\ref{sec-gen-lim}), but the hole fermions
remain localized at their holes. More precisely, each hole fermion
wave function forms a wedge of opening angle $\pi/3$ around its hole
and its amplitude decays exponentially with distance.\cite{Willans}
Since there is one hole fermion for each hole, there are $N-n$ bulk
fermion modes with excitation energies $E_f \sim 1$, and there are
$n-1$ hole fermion modes with zero excitation energies. Note that
the number of independent hole fermion modes is reduced by $1$ due
to the global constraint that $N_{\chi} + N_f$ must be even.

The independent $\mathbb{Z}_2$ modes of the model with $n > 0$ holes
are summarized in Table \ref{table-3}. We distinguish two classes of
modes depending on their excitation energies and the scaling of
their numbers with $N$ and $n$. The bulk fluxes and the bulk
fermions are external (bulk) modes: they have large excitation
energies $E \gtrsim J^4$ and their numbers scale with the system
size $N$. These modes are extremely hard to treat in the
thermodynamic limit. Conversely, the hole fluxes, the hole fermions,
and the hole plaquettes are internal modes: they have small
excitation energies $E \lesssim J^8$ and their numbers scale with
the hole number $n$. Since these modes are associated with
individual holes, it is straightforward to treat them in the limit
when the holes are isolated. Due to the different energy scales of
the two classes of modes, we can self-consistently neglect the
excitations in the high-energy bulk modes, and concentrate only on
the low-energy internal modes.

\begin{table}[h]
\begin{tabular*}{0.485\textwidth}{@{\extracolsep{\fill}} c | c | c | c | c }
\hline \hline
\multirow{2}{*}{Mode type}  & Excitation  & Number    & Quantum          & Global                        \\
                            & energy      & of modes  & number           & constraint                    \\
\hline
Bulk fermion                & $\sim 1$    & $N-n$     &                  &                               \\
Bulk flux                   & $\sim J^4$  & $N-3n$    &                  &                               \\
Hole flux                   & $\sim J^8$  & $n-1$     & $h = \{ 0,1 \}$  & $\sum_j h_j = \textrm{even}$  \\
Hole fermion                & $0$         & $n-1$     & $q = \{ 0,1 \}$  & $\sum_j q_j = \textrm{even}$  \\
Hole plaquette              & $0$         & $n$       & $p = \{ 0,1 \}$  &                               \\
\hline  \hline
\end{tabular*}
\caption{Energy hierarchy of independent $\mathbb{Z}_2$ modes in the
model with $n > 0$ holes. For the internal modes, the corresponding
quantum numbers are also specified along with any global constraints
on them. The total number of modes is $2N-n-2$ as expected.
\label{table-3}}
\end{table}

\begin{figure}[t!]
\centering
\includegraphics[width=8.2cm]{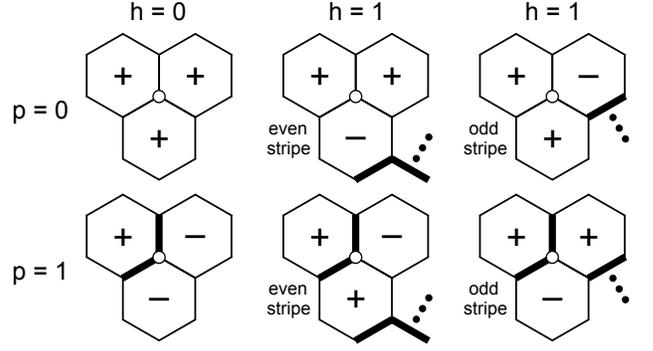}
\caption{Bond fermion sectors around a hole site $l \in \Delta$
(white dot) for different combinations of the flux quantum number $h
= \{ 0,1 \}$ and the plaquette quantum number $p = \{ 0,1 \}$. Each
bond fermion sector is labeled with the excited bond fermions (thick
lines) and the corresponding plaquette operator eigenvalues ($\pm
1$). For $h = 1$, there are two cases depending on whether $P_l^z$
is in an even stripe or in an odd stripe. The triple dots indicate a
string of excited bond fermions connecting two holes with $h = 1$.
\label{fig-4}}
\end{figure}

Each hole in the model has three internal modes, and we characterize
these three internal modes with three $\mathbb{Z}_2$ quantum numbers
$h$, $q$, and $p$. The flux quantum number is $h = 1$ if the hole
has a flux bound to it and $h = 0$ otherwise. The fermion quantum
number is $q = 1$ if the corresponding hole fermion is excited and
$q = 0$ otherwise. The meaning of the plaquette quantum number $p$
depends on the flux quantum number: if $h = 0$, then $p = 0$ means
no hole plaquette excitations and $p = 1$ means two hole plaquette
excitations in two neighboring stripes, while if $h = 1$, then $p =
0$ means one hole plaquette excitation in an even stripe and $p = 1$
means one hole plaquette excitation in an odd stripe. The
corresponding bond fermion sectors are shown in Fig.~\ref{fig-4}.
Importantly, the distinction between even and odd stripes ensures
that $N_{\chi}$ is always even. In the case of $n$ holes labeled $j
= \{ 1, 2, \ldots, n \}$, there are $3n$ internal modes
characterized by $3n$ quantum numbers $h_j$, $q_j$, and $p_j$. Since
fluxes and fermions can only be excited pairwise, the quantum
numbers $h_j$ and $q_j$ are not fully independent from each other.
In particular, the global constraint $\prod_P W_P = 1$ translates
into $\sum_j h_j = \textrm{even}$, while the global constraint that
$N_{\chi} + N_f$ is even, or equivalently, that $N_f$ is even
translates into $\sum_j q_j = \textrm{even}$. The various
formulations of these two global constraints are presented in Table
\ref{table-4}.

\begin{table}[h]
\begin{tabular*}{0.49\textwidth}{@{\extracolsep{\fill}} c | c | c }
\hline \hline
                   & Flux constraint               & Fermion constraint                                     \\
\hline
Physical spins     & $\prod_P W_P = 1$             & $\prod_{l \in A} \lambda_l \prod_{P \in \mu} W_P = 1$  \\
Majorana fermions  & Automatically satisfied       & $N_{\chi} + N_f = \textrm{even}$                       \\
E / M charges      & $N_e + N_m = \textrm{even}$   & $N_m = \textrm{even}$                                  \\
Quantum numbers    & $\sum_j h_j = \textrm{even}$  & $\sum_j q_j = \textrm{even}$                           \\
\hline  \hline
\end{tabular*}
\caption{Formulations of the two essential global constraints in
terms of the physical spins, the Majorana fermions, the electric /
magnetic charges, and the internal quantum numbers. \label{table-4}}
\end{table}

We are now ready to write down the ground states $| \Omega_{h, q,
p}^{\, \Delta} \rangle$ that correspond to the different values of
the internal quantum quantum numbers. Using the method of
Sec.~\ref{sec-gen-lim}, each ground state is obtained from the
vacuum state $| 0 \rangle$ by a projection onto an appropriate
subspace. Formally, the physical ground state for $n > 0$ holes at
sites $\Delta = \{ l_j \}$ with quantum numbers $h \equiv \{ h_j
\}$, $q \equiv \{ q_j \}$, and $p \equiv \{ p_j \}$ reads as
\begin{equation}
\big| \Omega_{h, q, p}^{\, \Delta} \big{\rangle} =
\mathcal{Q}_{\Delta} \mathcal{P} \mathcal{F}_{q; h} \mathcal{B}_p
\mathcal{X}_h | 0 \rangle. \label{eq-stat-deg-gs}
\end{equation}
Before enforcing the gauge constraints with the projection operators
$\mathcal{P}$ and $\mathcal{Q}_{\Delta}$, the vacuum state $| 0
\rangle$ is acted upon by several operators setting the bond fermion
and the matter fermion sectors. The first operator $\mathcal{X}_h$
is responsible for binding fluxes to all holes with $h_j = 1$.
Mathematically, $\mathcal{X}_h$ is an appropriate product of
$(\chi_{l}^\alpha)^{\dag}$ operators along a set of strings
connecting the holes with $h_j = 1$ pairwise. Note that the global
constraint $\sum_j h_j = \textrm{even}$ ensures that the holes with
$h_j = 1$ can always be paired up. Importantly, we choose
$\mathcal{X}_h$ such that it does not excite any $z$ bond fermions
and creates the excited plaquette in an even stripe for each hole
(see Fig.~\ref{fig-4}). In this case, $\mathcal{X}_h$ is a product
of an even number of $(\chi_{l}^\alpha)^{\dag}$ operators, and
therefore it excites an even number of bond fermions. The remaining
two operators in Eq.~(\ref{eq-stat-deg-gs}) are given by
\begin{eqnarray}
\mathcal{B}_p &=& \prod_{j=1}^n \left( i b_{l_j}^x b_{l_j}^z
\right)^{p_j}, \label{eq-stat-deg-op} \\
\mathcal{F}_{q; h} &=& \prod_{k=n+1}^N \left( \phi_k \phi_k^{\dag}
\right) \prod_{j=1}^n \left( \phi_j^{1-q_j} \phi_j^{\dag}
\phi_j^{q_j} \right) \prod_{j=1}^n \left[ f_{\tilde{z} (l_j)}^{\dag}
\right]^{q_j}, \nonumber
\end{eqnarray}
where $\tilde{z}(l) = l$ if $l \in A$ and $\tilde{z}(l) = z(l)$ if
$l \in B$. The operator $\mathcal{B}_p$ sets the bond fermion sector
by flipping an even number of bond fermions around the hole sites,
while the operator $\mathcal{F}_{q; h}$ projects onto one of the
ground states in the given bond fermion sector. The original matter
fermions $f_{\tilde{z} (l_j)}^{\dag}$ are required only to ensure
that $\mathcal{F}_{q; h}$ does not project to zero in the isolated
dimer limit. The free matter fermions $\phi_k$ are obtained from the
matrix $M$ in Eq.~(\ref{eq-kit-ferm-H-2}): there are $n$ hole
fermions $\phi_k$ with $1 \leq k \leq n$ that have zero energies and
$N-n$ bulk fermions $\phi_k$ with $n+1 \leq k \leq N$ that have
finite energies $E_f \sim 1$. We label the hole fermions
consistently such that the hole fermion $\phi_j$ is localized around
the hole site $l_j$. Note that the matrix $M$ is in general a
function of the bond fermions excited by $\mathcal{X}_h$, and
therefore $\mathcal{F}_{q; h}$ depends on the flux quantum numbers
$h_j$ via the free matter fermions $\phi_k$. On the other hand, the
bond fermions flipped by $\mathcal{B}_p$ correspond to bonds with
switched-off interactions, and therefore $\mathcal{F}_{q; h}$ does
not depend on the plaquette quantum numbers $p_j$.

\begin{table}[h]
\begin{tabular*}{0.31\textwidth}{@{\extracolsep{\fill}} c | c | c }
\hline \hline
\multicolumn{2}{c |}{Hole type}                          & \, Superselection sector \,  \\
\hline
\multirow{2}{*}{\,\, $h = 0$ \,\,}  & \,\, $q = 0$ \,\,  & Trivial ($1$)                \\
                                    & \,\, $q = 1$ \,\,  & Combined ($e \times m$)      \\
\hline
\multirow{2}{*}{\,\, $h = 1$ \,\,}  & \,\, $q = 0$ \,\,  & Electric ($e$)               \\
                                    & \,\, $q = 1$ \,\,  & Magnetic ($m$)               \\
\hline  \hline
\end{tabular*}
\caption{Superselection sectors of holes with flux quantum numbers
$h = \{ 0,1 \}$ and fermion quantum numbers $q = \{ 0,1 \}$.
\label{table-5}}
\end{table}

It is useful to interpret the internal quantum numbers in the
isolated dimer limit. In this limit, the free matter fermions
$\phi_k$ are identical to the original matter fermions $f_l$, and
therefore the second operator in Eq.~(\ref{eq-stat-deg-op}) takes
the simplified form $\mathcal{F}_q \equiv \mathcal{F}_{q; h} =
\prod_{j=1}^n [f_{\tilde{z} (l_j)}^{\dag}]^{q_j}$. Note that the
matrix $M$ is no longer a function of the $x$ and $y$ bond fermions
excited by $\mathcal{X}_h$, and therefore $\mathcal{F}_q$ becomes
independent of the flux quantum numbers $h_j$. For a single isolated
hole at site $l$ with quantum numbers $h$, $q$, and $p$, the hole
dimer operator is then $\lambda_l = \sigma_l^z \sigma_{z(l)}^z =
(-1)^{q + p}$. Since the product of the hole plaquette operators is
$\prod_{P_l \in \eta} W_P = (-1)^{h + p}$ in even stripes and
$\prod_{P_l \in \mu} W_P = (-1)^p$ in odd stripes, we conclude that
the different combinations of the quantum numbers $h$ and $q$ are in
one-to-one correspondence with the different superselection sectors
that the hole can belong to. This correspondence is presented in
Table \ref{table-5}. Note that if the bulk modes are not excited,
isolated holes can indeed be thought of as isolated excitation
clusters with well-defined superselection sectors. Furthermore,
since the projection operator $\mathcal{Q}_{\Delta}$ enforces
$\sigma_l^z = +1$ at the hole site $l$, there is a finite local
magnetization $\sigma_{z(l)}^z = (-1)^{q + p}$ at the neighboring
site $z(l)$.\cite{Willans} This magnetization can be reversed by
applying the transformation $\sigma^{x,z} \rightarrow -\sigma^{x,z}$
to all spins except the hole spin. On the other hand, such a
discrete spin rotation is also a symmetry of the model: it flips the
hole plaquettes $P_l^{x,z}$ and changes the sign of the hole dimer
operator $\lambda_l$. It therefore corresponds to a switch in the
plaquette quantum number $p$ only. To summarize, the flux and the
fermion quantum numbers determine the superselection sector, while
the plaquette quantum number determines the local magnetization
around the hole. Importantly, these results are also valid in the
case of $J > 0$ when $(-1)^{q + p}$ is equal to the product of dimer
operators taken over a sufficiently large region around the hole
site $l$.

\subsection{Interactions and bound states} \label{sec-stat-bound}

We now discuss the interactions between two holes at a finite
distance $R$ away from each other. In general, the ground-state
energy is given by $\Gamma_0 = -\sum_k S_k$, where the $N$ singular
values $S_k$ are obtained from the matrix $M$ in
Eq.~(\ref{eq-kit-ferm-H-2}). In the limit of $R \rightarrow \infty$,
there are two vanishing singular values $S_1 = S_2 = 0$
corresponding to the two hole fermions, and the ground-state energy
$\Gamma_0 (\infty)$ is determined by the sum of the remaining $N-2$
non-vanishing singular values. When $R$ is finite, the interaction
energy between the two holes is defined as the change in the
ground-state energy with respect to that in the $R \rightarrow
\infty$ limit: $\Delta \Gamma_0 \equiv \Gamma_0 (R) - \Gamma_0
(\infty)$.

\begin{figure}[t!]
\centering
\includegraphics[width=6.2cm]{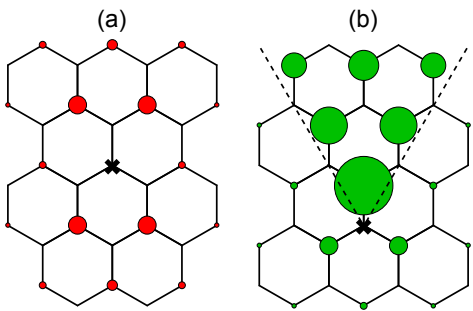}
\caption{(Color online) Absolute interaction energy $| \Delta
\Gamma_0 |$ between two holes as a function of their relative
position when $J_z = 1.0$ and $J \equiv J_x = J_y = 0.2$. One hole
is fixed (black cross) and the other one is moved around (red and
green dots). Each dot has an area proportional to $\sqrt{| \Delta
\Gamma_0 |}$. The interaction is either repulsive with $\Delta
\Gamma_0 > 0$ (a) or attractive with $\Delta \Gamma_0 < 0$ (b). The
wedge of opening angle $\pi/3$ is marked by a dashed line.
\label{fig-5}}
\end{figure}

The interaction energy $\Delta \Gamma_0$ has two contributions
arising from two distinct interaction mechanisms. First, the sum of
the $N-2$ non-vanishing singular values is changed by perturbative
terms similar to those in Eq.~(\ref{eq-gen-lim-H}). Second, the
singular value $S_2$ also becomes non-vanishing due to a
hybridization between the two hole fermions.\cite{Willans} The first
contribution $\Delta \Gamma_0^{(1)}$ is non-zero for both
sublattices and all directions, while the second contribution
$\Delta \Gamma_0^{(2)}$ is non-zero only if the two holes are in
opposite sublattices and their relative direction lies in the wedge
of opening angle $\pi/3$ such that each hole fermion wave function
has a finite amplitude at the hole site of the other hole (see
Fig.~\ref{fig-5}). Importantly, the wedges for the two holes in the
opposite sublattices point in opposite directions, and therefore
this condition for the relative direction is identical from the
point of view of both holes.

Since the two contributions decay as $\Delta \Gamma_0^{(1)} \sim
J^{2R}$ and $\Delta \Gamma_0^{(2)} \sim J^R$ with the distance $R$,
the second contribution is the dominant one at large distances. From
the lowest-order perturbation theory in $J \ll 1$ around the
isolated dimer limit, this contribution takes the general form
\begin{equation}
\Delta \Gamma_0^{(2)} = - \frac{R!} {R_x! R_y!} \, J^R,
\label{eq-stat-bound-int}
\end{equation}
where the string of shortest length $R \equiv R_x + R_y$ connecting
the two holes contains $R_x$ bonds of $x$ type and $R_y$ bonds of
$y$ type.\cite{Willans} The first contribution can also be
calculated from a perturbation theory in $J \ll 1$, but its general
form is more complicated. In particular, $\Delta \Gamma_0^{(1)}$ can
take both signs: the largest negative result $\Delta \Gamma_0^{(1)}
= -J^2 / 4$ is found when the two holes are at nearest-neighbor
sites connected by an $x$ or a $y$ bond, while the largest positive
result $\Delta \Gamma_0^{(1)} = J^2 / 4$ is found when the two holes
are at next-nearest-neighbor sites connected by a $z$ bond and an
$x$ or a $y$ bond. The interaction energy $\Delta \Gamma_0 = \Delta
\Gamma_0^{(1)} + \Delta \Gamma_0^{(2)}$ is always positive when the
two holes are in the same sublattice and always negative when the
two holes are in opposite sublattices. The absolute values of the
interaction energies are plotted in Fig.~\ref{fig-5}.

Importantly, the first interaction mechanism corresponding to
$\Delta \Gamma_0^{(1)}$ is diagonal in the quantum numbers $h$, $q$,
and $p$, while the second interaction mechanism corresponding to
$\Delta \Gamma_0^{(2)}$ is diagonal only in $h$ and $p$ but not in
$q$. In particular, if we set $h_{1,2} = p_{1,2} = 0$ for simplicity
and label the remaining four ground states $| \Omega_{h, q, p}^{\,
\Delta} \rangle$ with the fermion quantum numbers as $| q_1, q_2
\rangle \equiv | \Omega_{q_1, q_2} \rangle$, the second interaction
has identical matrix elements $\sim J^R$ between the states $| 0, 0
\rangle$ and $| 1, 1 \rangle$, and between the states $| 0, 1
\rangle$ and $| 1, 0 \rangle$. This implies that the eigenstates are
in fact $(| 0, 0 \rangle \pm | 1, 1 \rangle) / \sqrt{2}$ and $(| 0,
1 \rangle \pm | 1, 0 \rangle) / \sqrt{2}$. In the strict sense, the
fermion quantum numbers $q$ are then no longer valid quantum numbers
in the presence of hole interactions. However, since the interaction
is exponentially small when the holes are far apart, they are still
practically valid quantum numbers as they are conserved within an
exponentially large timescale $\sim J^{-R}$.

Since the attractive interaction between holes in opposite
sublattices is stronger than the repulsive interaction between holes
in the same sublattice, the overall interaction between two holes is
attractive. The most negative interaction energy $\Delta \Gamma_0 =
-1$ is found when the two holes are at nearest-neighbor sites
connected by a $z$ bond. In the absence of other interactions, this
attraction leads to pair formation, where the holes in the model
form bound pairs along $z$ bonds. It is then useful to investigate
how these hole pairs interact with each other. The interaction
energy $\Delta \Gamma_0' \equiv \Gamma_0' (R) - \Gamma_0' (\infty)$
between two hole pairs is completely analogous to that between two
single holes. In this case, there are two vanishing singular values
for all distances $R$, and the only contribution to the interaction
energy comes from the change in the remaining $N-2$ non-vanishing
singular values. From a perturbation theory around the isolated
dimer limit, we obtain that the interaction energy between two hole
pairs is always negative and decays as $\Delta \Gamma_0' \sim
J^{2R}$ with the distance $R$. The most negative interaction energy
$\Delta \Gamma_0' = -J^2 / 4$ is found when two holes from the
respective hole pairs are at nearest-neighbor sites connected by an
$x$ or a $y$ bond.

In the absence of other interactions, the attraction between hole
pairs leads to phase separation, where the holes are all bound
together to form a large cluster. However, both single holes and
hole pairs are positively charged, and therefore they are also
subject to a Coulomb repulsion. Since the attraction between single
holes is stronger than that between hole pairs, we can distinguish
three complementary regimes in the behavior of the model. If the
Coulomb repulsion is weaker than the attraction between hole pairs,
the model phase separates. If the Coulomb repulsion is stronger than
the attraction between hole pairs but weaker than that between
single holes, the elementary particles of the model are hole pairs.
If the Coulomb repulsion is stronger than the attraction between
single holes, the elementary particles of the model are single
holes. In the following, we restrict our attention to single holes
and implicitly assume a sufficiently strong Coulomb repulsion such
that the model is in the appropriate regime.

\subsection{Robustness against local perturbations} \label{sec-stat-pert}

It is useful to discuss the applicability of the internal quantum
numbers when a local perturbation is applied to the model with $n >
0$ holes. We first notice that two arguments are apparently in
conflict with each other. On one hand, the quantum numbers $q$ and
$h$ are expected to be robust against local perturbations as they
are related to the superselection sectors of the model (see
Secs.~\ref{sec-gen-sup} and \ref{sec-stat-deg}). On the other hand,
the quantum numbers $q$ are not strictly conserved in the presence
of hole interactions (see Sec.~\ref{sec-stat-bound}). Note that the
dimensionless coupling strength $J \ll 1$ is a local perturbation in
the language of the isolated dimer limit.

The resolution of this apparent conflict is that local perturbations
assemble into non-local strings at higher orders of perturbation
theory. If two holes (excitation clusters) are connected by such a
string, only the combined superselection sector is conserved, while
the individual superselection sectors can change. However, when the
two holes are at a distance $R$ away from each other, such a string
can be assembled only at $R$-th order of perturbation theory. For a
local perturbation of strength $\delta E$ that creates excitations
with energies $E_0$, the perturbative term responsible for changing
the superselection sector is then $\sim (\delta E / E_0)^R$. This
means that the superselection sector is conserved within a timescale
$\sim (E_0 / \delta E)^R$ that is exponentially large when $\delta E
\ll E_0$ and $R \gg 1$. Note that the interaction term $\sim J^R$ in
Sec.~\ref{sec-stat-bound} is recovered as a special case with
$\delta E \sim J$ and $E_0 \sim 1$. Since local perturbations excite
bulk fluxes with energies $E_P \sim J^4$ and bulk fermions with
energies $E_f \sim 1$ in general, we conclude that the quantum
numbers $h$ and $q$ are robust against arbitrary local perturbations
of strength $\delta E \ll J^4$ as long as the holes are sufficiently
far away from each other.

\begin{figure}[t!]
\centering
\includegraphics[width=7.2cm]{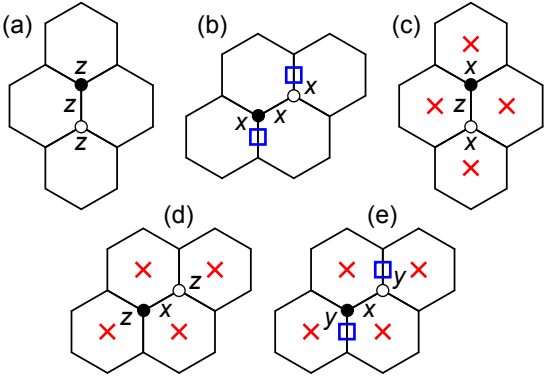}
\caption{(Color online) Effects of the Heisenberg terms $\sigma_l^z
\sigma_{l' = z(l)}^z$ (a), $\sigma_l^x \sigma_{l' = x(l)}^x$ (b),
$\sigma_l^x \sigma_{l' = z(l)}^x$ (c), $\sigma_l^z \sigma_{l' =
x(l)}^z$ (d), and $\sigma_l^y \sigma_{l' = x(l)}^y$ (e) on the
plaquettes and the dimers (fermions) around two neighboring sites
$l$ (white dot) and $l'$ (black dot). Flipped plaquettes are marked
by red crosses and flipped dimers are marked by blue rectangles.
\label{fig-6}}
\end{figure}

It is instructive to examine an explicit example for the
conservation of the internal quantum numbers in the presence of a
local perturbation. To this end, we perturb the Kitaev honeycomb
model with Heisenberg interactions. The contribution of this
perturbation to the Hamiltonian reads as
\begin{equation}
\delta H = \delta E \sum_{\langle l,l' \rangle} \left( \sigma_l^x
\sigma_{l'}^x + \sigma_l^y \sigma_{l'}^y + \sigma_l^z \sigma_{l'}^z
\right), \label{eq-stat-pert-H}
\end{equation}
where $\langle l,l' \rangle$ indicates a summation over bonds, or
equivalently, over pairs of neighboring sites. Based on the type of
the bond and the spin components coupled, there are nine types of
terms in $\delta H$, and these types can be divided into four
distinct classes in the isolated dimer limit. The terms $\sigma_l^z
\sigma_{z(l)}^z$ only renormalize the coupling strength $J_z$ on the
$z$ bonds, and therefore do not flip any plaquettes or dimers
(fermions). The terms $\sigma_l^x \sigma_{x(l)}^x$ and $\sigma_l^y
\sigma_{y(l)}^y$ correspond to the usual couplings with strengths
$J_{x,y}$ on the $x$ and $y$ bonds, and therefore flip no plaquettes
but two dimers each. The terms $\sigma_l^x \sigma_{z(l)}^x$,
$\sigma_l^y \sigma_{z(l)}^y$, $\sigma_l^z \sigma_{x(l)}^z$, and
$\sigma_l^z \sigma_{y(l)}^z$ flip no dimers and four plaquettes
each, while the terms $\sigma_l^y \sigma_{x(l)}^y$ and $\sigma_l^x
\sigma_{y(l)}^x$ flip two dimers and four plaquettes each. The
effects of these types of terms are illustrated in Fig.~\ref{fig-6}.
Since the perturbative terms flip either zero or two dimers, the
number of broken dimers has a conserved parity, and therefore the
parity of $q + p$ does not change either (see
Sec.~\ref{sec-stat-deg}). Since they either flip zero plaquettes or
they flip two plaquettes in even stripes and two plaquettes in odd
stripes, the numbers of excited plaquettes in even and in odd
stripes both have conserved parities, and therefore the parities of
$h + p$ and $p$ do not change either. We conclude that the quantum
numbers $h$, $q$, and $p$ are all conserved in the presence of a
Heisenberg perturbation if its strength satisfies $\delta E \ll J^4$
and the holes are sufficiently far apart.

\section{Isolated mobile holes} \label{sec-mob}

\subsection{Hopping formalism} \label{sec-mob-form}

We consider a hole hopping model in which the holes introduced into
the Kitaev honeycomb model can propagate via nearest-neighbor
hopping. Formally, a spin one-half particle at a site $l'$
neighboring an empty hole site $l$ can hop from $l'$ to $l$ with an
amplitude $-t$. We assume that the spin state of the particle is not
affected by the hopping. In the hole spin picture, the hopping then
exchanges the hole spin at $l$ with the actual spin at $l'$, and
this process can be represented by an exchange operator that takes
the form
\begin{eqnarray}
\mathcal{E}_{l,l'} &=& \frac{1}{2} \left( 1 + \sigma_l^x
\sigma_{l'}^x + \sigma_l^y \sigma_{l'}^y + \sigma_l^z \sigma_{l'}^z
\right)
\label{eq-mob-form-exc} \\
&=& \frac{1}{2} \left( 1 + b_l^y b_{l'}^y b_l^z b_{l'}^z + b_l^z
b_{l'}^z b_l^x b_{l'}^x + b_l^x b_{l'}^x b_l^y b_{l'}^y \right).
\nonumber
\end{eqnarray}
In the following, we restrict our attention to the regime of slow
hopping, where the hopping amplitude is much smaller than the
excitation energies of the bulk modes. We can then neglect the
excitations in the bulk modes and consider only the ground states $|
\Omega_{h, q, p}^{\, \Delta} \rangle$ corresponding to the internal
modes. Since the bulk modes are bulk fluxes with energies $E_P \sim
J^4$ and bulk fermions with energies $E_f \sim 1$ in general, the
condition of slow hopping becomes $t \ll J^4$.

For simplicity, we restrict our attention to only $n = 2$ isolated
holes at sites $l_{1,2}$. However, more holes are assumed to be
present in the background so that the quantum numbers $h_{1,2}$ and
$q_{1,2}$ can be chosen independently without violating the global
constraints. We consider the hopping process in which the hole at
site $l_1$ hops to a neighboring site $l_1'$. The set of hole sites
is $\Delta = \{ l_1, l_2 \}$ before the hopping and $\Delta' = \{
l_1', l_2' \equiv l_2 \}$ after the hopping. The ground states
corresponding to the hole positions $\Delta$ and $\Delta'$ take the
forms
\begin{eqnarray}
\big| \Omega_{h, q, p} \big{\rangle} &\equiv& \big| \Omega_{h, q,
p}^{\, \Delta} \big{\rangle} = \mathcal{Q}_{\Delta} \mathcal{P}
\mathcal{F}_{q; h} \mathcal{B}_p \mathcal{X}_h | 0
\rangle, \label{eq-mob-form-gs} \\
\big| \bar{\Omega}_{h', q', p'} \big{\rangle} &\equiv& \big|
\Omega_{h', q', p'}^{\, \Delta'} \big{\rangle} =
\mathcal{Q}_{\Delta'} \mathcal{P} \bar{\mathcal{F}}_{q'; h'}
\bar{\mathcal{B}}_{p'} \bar{\mathcal{X}}_{h'} | 0 \rangle, \nonumber
\end{eqnarray}
where the operators $\bar{\mathcal{X}}_{h'}$,
$\bar{\mathcal{B}}_{p'}$, and $\bar{\mathcal{F}}_{q'; h'}$ are
completely analogous to $\mathcal{X}_h$, $\mathcal{B}_p$, and
$\mathcal{F}_{q; h}$ as defined in Sec.~\ref{sec-stat-deg}. Since
different bonds have switched-off interactions before and after the
hopping, the operators $\mathcal{F}_{q; h}$ and
$\bar{\mathcal{F}}_{q'; h'}$ contain different free matter fermions
$\phi_k$ and $\bar{\phi}_k$. By considering the hopping between the
respective ground states $| \Omega_{h, q, p} \rangle$ and $|
\bar{\Omega}_{h', q', p'} \rangle$, the effective hopping amplitude
becomes a finite-dimensional matrix. The elements of this matrix are
given by
\begin{equation}
T_{h, q, p}^{h', q', p'} = - \frac{t \, \big{\langle}
\bar{\Omega}_{h', q', p'} \big| \mathcal{E}_{l_1, l_1'} \big|
\Omega_{h, q, p} \big{\rangle}} {\sqrt{\big{\langle}
\bar{\Omega}_{h', q', p'} \big| \bar{\Omega}_{h', q', p'}
\big{\rangle} \big{\langle} \Omega_{h, q, p} \big| \Omega_{h, q, p}
\big{\rangle}}} \, , \label{eq-mob-form-T}
\end{equation}
where the ground-state norms in the denominator are required because
the ground states $| \Omega_{h, q, p} \rangle$ and $|
\bar{\Omega}_{h', q', p'} \rangle$ are not properly normalized in
general.

\subsection{General hopping properties} \label{sec-mob-gen}

We notice that the non-trivial terms in the exchange operator
$\mathcal{E}_{l_1, l_1'}$ are the Heisenberg terms in
Eq.~(\ref{eq-stat-pert-H}). It is therefore directly implied by the
results in Sec.~\ref{sec-stat-pert} that the quantum numbers $h$,
$q$, and $p$ are all conserved by the hopping. Note that the
exchange operator $\mathcal{E}_{l_1, l_1'}$ can in principle change
the plaquette (matter fermion) sector in at most two inequivalent
ways (see Fig.~\ref{fig-6}). The plaquette (matter fermion) sector
after the hopping is then uniquely determined by that before the
hopping via the ground-state constraint that no excited plaquettes
(matter fermions) are allowed to be left behind. Mathematically, the
conservation of the quantum numbers means that the effective hopping
matrix is diagonal.

Furthermore, the diagonal hopping matrix elements that differ only
in their plaquette quantum numbers $p$ are all identical to each
other. Physically, this property follows from the discrete
spin-rotation symmetry discussed in Sec.~\ref{sec-stat-deg} and the
fact that the corresponding transformation switches $p$. However, it
can also be shown explicitly by noticing that $\mathcal{B}_p^{\dag}
\mathcal{B}_p = 1$ and $\bar{\mathcal{B}}_p^{\dag} \mathcal{E}_{l_1,
l_1'} \mathcal{B}_p = \mathcal{E}_{l_1, l_1'}$ for all $p$. We
therefore conclude that the effective hopping matrix elements are
independent of the plaquette quantum numbers $p$ and take the
general form
\begin{equation}
T_{h, q, p}^{h', q', p'} = \delta_{h', h} \delta_{q', q} \delta_{p',
p} \, T_{h, q}, \label{eq-mob-gen-T-1}
\end{equation}
\begin{equation}
T_{h, q} = - \frac{t \, \big{\langle} \bar{\Omega}_{h, q} \big|
\mathcal{E}_{l_1, l_1'} \big| \Omega_{h, q} \big{\rangle}}
{\sqrt{\big{\langle} \bar{\Omega}_{h, q} \big| \bar{\Omega}_{h, q}
\big{\rangle} \big{\langle} \Omega_{h, q} \big| \Omega_{h, q}
\big{\rangle}}} \, , \label{eq-mob-gen-T-2}
\end{equation}
where $| \Omega_{h, q} \rangle \equiv | \Omega_{h, q, p = 0}
\rangle$ and $| \bar{\Omega}_{h, q} \rangle \equiv |
\bar{\Omega}_{h, q, p = 0} \rangle$. In the following, we simplify
our calculations by considering only these ground states with $p_1 =
p_2 = 0$.

Now we derive a formula for the effective hopping matrix element
$\tilde{T}_{h, q}$ in the important case when the bond fermion
sector (plaquette sector) is conserved by the hopping. This
condition is equivalent to $\mathcal{X}_h = \bar{\mathcal{X}}_h$,
and it is always satisfied in the case of trivial flux quantum
numbers $h_1 = h_2 = 0$ when $\mathcal{X}_{h = 0} =
\bar{\mathcal{X}}_{h = 0} = 1$. Since $\langle 0 |
\mathcal{X}_h^{\dag} \mathcal{X}_h | 0 \rangle = 1$ in general, the
ground-state norms in the denominator of Eq.~(\ref{eq-mob-gen-T-2})
become
\begin{widetext}
\begin{eqnarray}
\big{\langle} \Omega_{h, q} \big| \Omega_{h, q} \big{\rangle} &=&
\langle 0 | \mathcal{X}_h^{\dag} \mathcal{F}_{q; h}^{\dag}
\mathcal{P} \mathcal{Q}_{\Delta} \mathcal{F}_{q; h} \mathcal{X}_h |
0 \rangle = \frac{1} {2^{2N+2}} \, \langle 0 | \mathcal{F}_{q;
h}^{\dag} \mathcal{F}_{q; h} | 0 \rangle, \label{eq-mob-gen-T-den} \\
\big{\langle} \bar{\Omega}_{h, q} \big| \bar{\Omega}_{h, q}
\big{\rangle} &=& \langle 0 | \mathcal{X}_h^{\dag}
\bar{\mathcal{F}}_{q; h}^{\dag} \mathcal{P} \mathcal{Q}_{\Delta'}
\bar{\mathcal{F}}_{q; h} \mathcal{X}_h | 0 \rangle = \frac{1}
{2^{2N+2}} \, \langle 0 | \bar{\mathcal{F}}_{q; h}^{\dag}
\bar{\mathcal{F}}_{q; h} | 0 \rangle. \nonumber \\
\nonumber
\end{eqnarray}
By assuming $l_1 \in A$ without loss of generality, using the
property $\sigma_{l_1'}^z \mathcal{E}_{l_1, l_1'} \sigma_{l_1}^z =
\mathcal{E}_{l_1, l_1'}$, and keeping only the terms in
$\mathcal{E}_{l_1, l_1'}$ that do not change the plaquette sector
when $\alpha \equiv \alpha_{l_1, l_1'} = \{ x,y,z \}$, the
ground-state overlap in the numerator of Eq.~(\ref{eq-mob-gen-T-2})
becomes
\begin{eqnarray}
\nonumber \\
\big{\langle} \bar{\Omega}_{h, q} \big| \mathcal{E}_{l_1, l_1'}
\big| \Omega_{h, q} \big{\rangle} &=& \langle 0 |
\mathcal{X}_h^{\dag} \bar{\mathcal{F}}_{q; h}^{\dag} \mathcal{P}
\mathcal{Q}_{\Delta'} \mathcal{E}_{l_1, l_1'} \mathcal{Q}_{\Delta}
\mathcal{F}_{q; h} \mathcal{X}_h | 0 \rangle = \frac{1}{8} \,
\langle 0 | \mathcal{X}_h^{\dag} \bar{\mathcal{F}}_{q; h}^{\dag}
\mathcal{P} \left( 1 + b_{l_1}^{\alpha} b_{l_1'}^{\alpha} c_{l_1}
c_{l_1'} \right) \mathcal{F}_{q; h} \mathcal{X}_h | 0 \rangle
\label{eq-mob-gen-T-num} \\
&=& \frac{1}{2^{2N+3}} \, \langle 0 | \mathcal{X}_h^{\dag}
\bar{\mathcal{F}}_{q; h}^{\dag} \left( 1 + b_{l_1}^{\alpha}
b_{l_1'}^{\alpha} c_{l_1} c_{l_1'} \right) \mathcal{F}_{q; h}
\mathcal{X}_h | 0 \rangle = \frac{1}{2^{2N+3}} \, \langle 0 |
\bar{\mathcal{F}}_{q; h}^{\dag} \left( 1 - i u_{l_1, l_1'} c_{l_1}
c_{l_1'} \right) \mathcal{F}_{q; h} | 0 \rangle. \nonumber \\
\nonumber
\end{eqnarray}
Note that $u_{l_1, l_1'} \equiv \langle 0 | \mathcal{X}_h^{\dag}
\hat{u}_{l_1, l_1'} \mathcal{X}_h | 0 \rangle$ is determined by the
bond fermion sector. Finally, the effective hopping matrix element
in the case of a conserved bond fermion sector (plaquette sector)
takes the form
\begin{equation}
\tilde{T}_{h, q} = - \frac{t \, \langle 0 | \bar{\mathcal{F}}_{q;
h}^{\dag} \left( 1 - i u_{l_1, l_1'} c_{l_1} c_{l_1'} \right)
\mathcal{F}_{q; h} | 0 \rangle} {2 \, \sqrt{\langle 0 |
\bar{\mathcal{F}}_{q; h}^{\dag} \bar{\mathcal{F}}_{q; h} | 0 \rangle
\langle 0 | \mathcal{F}_{q; h}^{\dag} \mathcal{F}_{q; h} | 0
\rangle}} \, . \label{eq-mob-gen-T-fin}
\end{equation}
Since the operators $\mathcal{F}_{q; h}^{(\dag)}$ and
$\bar{\mathcal{F}}_{q; h}^{(\dag)}$ are all simple products of
matter fermion operators, the vacuum expectation values in
Eq.~(\ref{eq-mob-gen-T-fin}) can be evaluated using Wick's theorem.
The state $| 0 \rangle$ is the vacuum of the original matter
fermions $f_l$, and the orthogonal matrices $U$ and $V$ are
therefore used to express the free matter fermions $\phi_k$ and
$\bar{\phi}_k$ in terms of $f_l$.

\subsection{Hopping in the isolated dimer limit} \label{sec-mob-lim}

We now consider the isolated dimer limit ($J = 0$) and evaluate the
effective hopping matrix elements explicitly. In this limit, the
operators $\mathcal{F}_q \equiv \mathcal{F}_{q; h}$ and
$\bar{\mathcal{F}}_q \equiv \bar{\mathcal{F}}_{q; h}$ no longer
depend on $h$ and take the simplified forms $\mathcal{F}_q =
[f_{\tilde{z} (l_1)}^{\dag}]^{q_1} [f_{\tilde{z}
(l_2)}^{\dag}]^{q_2}$ and $\bar{\mathcal{F}}_q = [f_{\tilde{z}
(l_1')}^{\dag}]^{q_1} [f_{\tilde{z} (l_2)}^{\dag}]^{q_2}$ (see
Sec.~\ref{sec-stat-deg}). The vacuum expectation values in
Eq.~(\ref{eq-mob-gen-T-fin}) thus become
\begin{eqnarray}
\nonumber \\
\langle 0 | \mathcal{F}_q^{\dag} \mathcal{F}_q | 0 \rangle &=&
\langle 0 | \bar{\mathcal{F}}_q^{\dag} \bar{\mathcal{F}}_q | 0
\rangle = 1, \nonumber \\ \nonumber \\
\langle 0 | \bar{\mathcal{F}}_{q_1 = 0}^{\dag} \left( 1 - i u_{l_1,
l_1'} c_{l_1} c_{l_1'} \right) \mathcal{F}_{q_1 = 0} | 0 \rangle &=&
\langle 0 | \Big[ 1 - u_{l_1, l_1'} \big[ f_{l_1} + f_{l_1}^{\dag}
\big] \big[ f_{z(l_1')} - f_{z(l_1')}^{\dag} \big] \Big] | 0 \rangle
= \Big\{ \begin{array}{c} 2 \quad \, \, (\alpha_{l_1, l_1'} = z)
\quad \, \, \\ 1 \quad \, \, (\alpha_{l_1, l_1'} = x,y),
\end{array}
\label{eq-mob-lim-T-den-num} \\
\langle 0 | \bar{\mathcal{F}}_{q_1 = 1}^{\dag} \left( 1 - i u_{l_1,
l_1'} c_{l_1} c_{l_1'} \right) \mathcal{F}_{q_1 = 1} | 0 \rangle &=&
\langle 0 | f_{z(l_1')} \Big[ 1 - u_{l_1, l_1'} \big[ f_{l_1} +
f_{l_1}^{\dag} \big] \big[ f_{z(l_1')} - f_{z(l_1')}^{\dag} \big]
\Big] f_{l_1}^{\dag} | 0 \rangle = \Big\{ \begin{array}{c} 0 \qquad
\quad \, \, (\alpha_{l_1, l_1'} = z) \quad \, \\ -u_{l_1, l_1'}
\quad (\alpha_{l_1, l_1'} = x,y), \end{array} \nonumber
\end{eqnarray}
\end{widetext}
and the corresponding hopping matrix elements take the form
\begin{eqnarray}
\tilde{T}_{h, q_1 = 0} &=& \Big\{ \begin{array}{c} -t \qquad (\alpha_{l_1, l_1'} = z) \quad \, \, \\
-t/2 \quad (\alpha_{l_1, l_1'} = x,y), \end{array}
\label{eq-mob-lim-T-fin} \\ \nonumber \\
\tilde{T}_{h, q_1 = 1} &=& \Big\{ \begin{array}{c} 0 \qquad \qquad \, (\alpha_{l_1, l_1'} = z) \quad \, \, \\
u_{l_1, l_1'} \, t/2 \quad \, (\alpha_{l_1, l_1'} = x,y).
\end{array} \nonumber
\end{eqnarray}
Note that $u_{l_1, l_1'} = +1$ for $\alpha_{l_1, l_1'} = z$ because
$\mathcal{X}_h$ excites only $x$ and $y$ bond fermions. Furthermore,
the matrix elements in Eq.~(\ref{eq-mob-lim-T-fin}) are independent
of the quantum number $q_2$. Since the two holes are isolated, the
hole hopping between $l_1$ and $l_1'$ is not affected by the other
hole at $l_2$.

\begin{figure*}[t!]
\centering
\includegraphics[width=16.4cm]{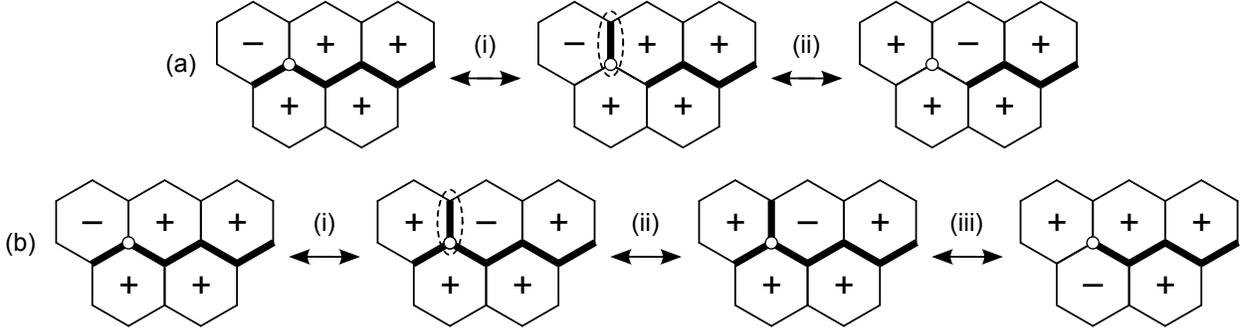}
\caption{Different types of transformations relating bond fermion
and matter fermion sectors around a hole site $l \in \Delta$ (white
dot) when there is a flux bound to the hole. Each bond fermion
sector is labeled with the excited bond fermions (thick lines) and
the corresponding plaquette operator eigenvalues ($\pm 1$), while
each matter fermion sector is labeled with the excited matter
fermions (dashed ellipses). (a) Gauge transformations $D_l$ (i) and
$\sigma_l^z$ (ii) for shifting the excited plaquette within a
stripe. (b) Transformations $\sigma_l^z$ (i) [gauge], $c_{\tilde{z}
(l)}$ (ii) [$q$-switch], and $i b_l^x b_l^z$ (iii) [$p$-switch] for
shifting the excited plaquette between neighboring stripes.
\label{fig-7}}
\end{figure*}

\begin{figure*}[t!]
\centering
\includegraphics[width=11.9cm]{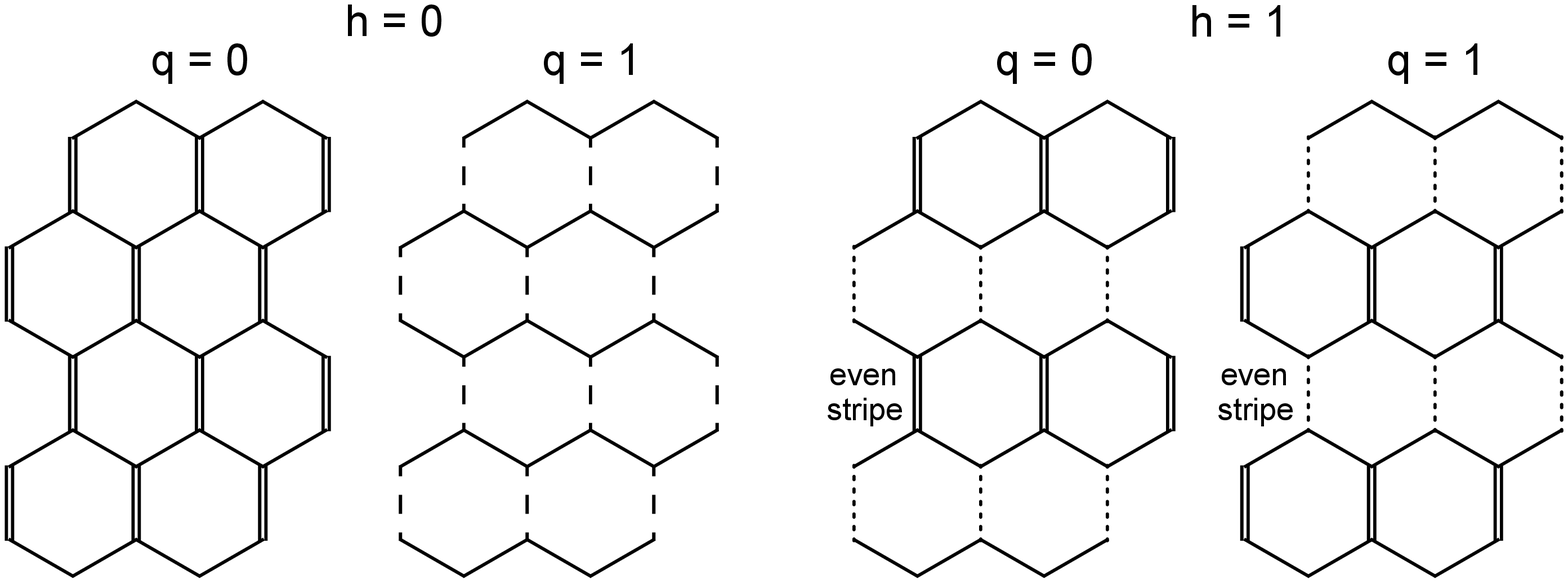}
\caption{Hole hopping problems in the isolated dimer limit for
different combinations of the flux quantum number $h = \{ 0,1 \}$
and the fermion quantum number $q = \{ 0,1 \}$. Each bond is labeled
according to the effective hopping amplitude $T$ along it: double
solid lines indicate $T = -t$, single solid lines indicate $T =
-t/2$, while dashed and dotted lines indicate $T = 0$. For dashed
lines, the effective hopping amplitude vanishes only in the isolated
dimer limit, while for dotted lines, it vanishes in the entire
gapped phase. \label{fig-8}}
\end{figure*}

It is crucial to emphasize that the matrix elements in
Eq.~(\ref{eq-mob-lim-T-fin}) are valid only if the bond fermion
sector is the same before and after the hopping. However, we
demonstrate in the following that the hopping problem for a single
isolated hole with quantum numbers $h = \{ 0,1 \}$ and $q = \{ 0,1
\}$ can be constructed by referring to these matrix elements only.
The most important steps of the construction are illustrated in
Fig.~\ref{fig-7}, while the resulting hopping problems for the
different quantum numbers are summarized in Fig.~\ref{fig-8}.

For a hole with no flux bound to it ($h = 0$), the bond fermion
sector is always trivial, and the matrix elements in
Eq.~(\ref{eq-mob-lim-T-fin}) are therefore directly applicable. This
means that hopping along $x$ and $y$ bonds is allowed for both
values of the quantum number $q$, while hopping along $z$ bonds is
allowed for $q = 0$ but not for $q = 1$. In other words, $q = 0$
holes can hop in both the $X$ and the $Y$ directions, while $q = 1$
holes can hop only in the $X$ direction. Since $u_{l,\alpha(l)} =
+1$ for all bonds around an $h = 0$ hole, the hopping problem in the
$X$ direction is in fact the same for $q = 0$ and $q = 1$. Note that
the opposite sign in the matrix element $\tilde{T}_{h, q_1 = 1}$ is
irrelevant because the honeycomb lattice is bipartite.

For a hole with a flux bound to it ($h = 1$), the hopping problem is
more complicated because the bond fermion sector depends on the hole
position. However, if the hole hops only around the excited
plaquette, the bond fermion sector can be chosen to remain the same,
and the matrix elements in Eq.~(\ref{eq-mob-lim-T-fin}) are
therefore applicable. Remember that the excited plaquette is in an
even stripe for a $p = 0$ hole (see Fig.~\ref{fig-4}). Furthermore,
the excited plaquette can be shifted along its stripe by applying
two simultaneous gauge transformations $D_l$ and $\sigma_l^z$ at the
hole site $l$. After these transformations illustrated in
Fig.~\ref{fig-7}(a), the bond fermion and the matter fermion sectors
around the hole site look the same from the point of view of the new
plaquette as they did before from the point of view of the old
plaquette. This implies that the hopping problem for $h = 1$ is
identical to that for $h = 0$ as long as the hole hops only around
the plaquettes of one particular even stripe. On the other hand, the
excited plaquette can only be shifted into a neighboring odd stripe
by applying the gauge transformation $\sigma_l^z$ along with the
transformations $c_{\tilde{z} (l)}$ and $i b_l^x b_l^z$ that switch
the quantum numbers $q$ and $p$. After these transformations
illustrated in Fig.~\ref{fig-7}(b), the bond fermion and the matter
fermion sectors around the hole site $l$ look the same from the
point of view of the new stripe as they did before from the point of
view of the old stripe. Since the hopping is independent of $p$,
this implies that the hopping problem for $q = 1$ around the
plaquettes of odd (even) stripes is identical to that for $q = 0$
around the plaquettes of even (odd) stripes. Unlike in the case of
$h = 0$, holes with different values of $q$ do not have
fundamentally different hopping problems in the case of $h = 1$:
they can both hop along $x$ and $y$ bonds in the $X$ direction,
while hopping along $z$ bonds in the $Y$ direction is allowed for $q
= 0$ in even stripes and for $q = 1$ in odd stripes.

\subsection{Hopping in the gapped phase} \label{sec-mob-phase}

We are now ready to discuss the hopping problem for a single
isolated hole at a generic point of the gapped phase away from the
isolated dimer limit ($J > 0$). From a perturbation theory in $J \ll
1$, there are possible corrections to the matrix elements in
Eq.~(\ref{eq-mob-lim-T-fin}), and the importance of these
corrections depends on whether the original matrix element is zero
or non-zero. If there is a finite matrix element at $J = 0$, the
perturbative corrections can be neglected as they only renormalize
the matrix element. However, if the matrix element vanishes at $J =
0$, these corrections are extremely important as they determine the
matrix element in the lowest order.

\begin{figure}[t!]
\centering
\includegraphics[width=3.5cm]{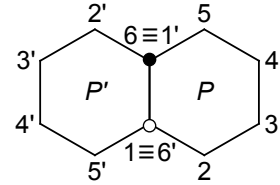}
\caption{Site and plaquette labeling conventions around the sites
$l_1$ (white dot) and $l_1'$ (black dot) when considering the
hopping along the $z$ bond between $l_1$ and $l_1'$. \label{fig-9}}
\end{figure}

According to Eq.~(\ref{eq-mob-lim-T-fin}), the only vanishing matrix
elements in the isolated dimer limit are $\tilde{T}_{h, q_1 = 1}$
along $z$ bonds. Any such matrix element is zero because the matter
fermion corresponding to the two sites $l_1$ and $l_1'$ connected by
the $z$ bond is excited: $-i c_{l_1} c_{l_1'} = -1$. To obtain a
non-zero correction for the matrix element, we need to find
corrections with a non-zero overlap for the ground states before and
after the hopping such that $-i c_{l_1} c_{l_1'} = +1$ for both
corrections. In general, these two corrections belong to two
complementary sections of an open string connecting the sites $l_1$
and $l_1'$. For example, if we use the site labeling convention in
Fig.~\ref{fig-9} around the sites $l_1$ and $l_1'$, one such pair of
corrections is
\begin{eqnarray}
\mathcal{F}_{q_1 = 1} | 0 \rangle &=& \frac{J} {6} \left( b_5^x
b_6^x c_5 c_6 \right) \frac{J} {4} \left( b_3^x b_2^x c_3 c_2
\right) f_1^{\dag} | 0 \rangle,
\label{eq-mob-phase-corr-1} \\
\bar{\mathcal{F}}_{q_1 = 1} | 0 \rangle &=& \frac{J} {6} \left(
b_1^y b_2^y c_1 c_2 \right) \frac{J} {4} \left( b_5^y b_4^y c_5 c_4
\right) f_1^{\dag} | 0 \rangle, \nonumber
\end{eqnarray}
and the resulting correction to the ground-state overlap is
\begin{eqnarray}
\langle 0 | \bar{\mathcal{F}}_{q_1 = 1}^{\dag} \left( 1 - i c_1 c_6
\right) \mathcal{F}_{q_1 = 1} | 0 \rangle &=& J^4 u_{1,2}
u_{3,2} u_{5,4} u_{5,6} Z_P \nonumber \\
&=& J^4 W_P Z_P, \label{eq-mob-phase-corr-2}
\end{eqnarray}
\begin{eqnarray}
Z_P &=& \frac{1}{576} \langle 0 | f_1 c_5 c_4 c_1 c_2 \left( 1 - i
c_1 c_6 \right) c_5 c_6 c_3 c_2 f_1^{\dag} | 0 \rangle \nonumber \\
&=& \frac{1}{288} \langle 0 | f_1 c_5 c_4 c_1 c_2 c_5 c_6 c_3 c_2
f_1^{\dag} | 0 \rangle = \frac{1}{288} \, .
\label{eq-mob-phase-corr-3}
\end{eqnarray}
Note that $u_{1,6} = u_{3,4} = +1$ because $\mathcal{X}_h$ is
defined such that it excites only $x$ and $y$ bond fermions.

For a generic open string connecting the sites $1$ and $1'$, we
define a closed loop $C$ consisting of the open string and the $z$
bond between $1$ and $1'$. Any correction to the ground-state
overlap due to the open string is then proportional to $W_C Z_C$,
where $W_C$ is the corresponding loop operator eigenvalue and $Z_C$
is an expectation value similar to that in
Eq.~(\ref{eq-mob-phase-corr-3}). By means of a reflection across the
middle of the $z$ bond, we also define a dual loop $C'$ with a loop
operator eigenvalue $W_{C'}$ and a dual correction with an
expectation value $Z_{C'}$. Note that the dual correction strictly
corresponds to a backward hopping because the reflection exchanges
the sites $1$ and $1'$. On the other hand, $Z_{C'} \in \mathbb{R}$
means that there is an equivalent dual correction for the forward
hopping as well. If we identify the site labels of the loop $C$ with
the dual site labels of the loop $C'$, the explicit forms of the
expectation values $Z_C$ and $Z_{C'}$ are identical, and thus $Z_C =
Z_{C'}$. Since this equality is true for all corrections, we
conclude that the total corrections due to the loops $C$ and $C'$
have equal magnitudes, while their relative signs are determined by
the loop eigenvalues $W_{C,C'}$.

This result has already strong implications for holes with a flux
bound to them ($h = 1$). When the bond fermion sector is conserved
by the hopping, the flux is necessarily bound to either of the
plaquettes $P$ or $P'$. If we then choose any two dual loops $C$ and
$C'$ that do not enclose any other holes, one of them contains one
excited plaquette and the other one contains no excited plaquettes.
This implies $W_C + W_{C'} = 0$, and therefore the corrections due
to all of the paired-up dual loops vanish because $(W_C + W_{C'})
Z_C = 0$. The only non-zero corrections are then due to loops that
are large enough such that they enclose at least one other hole with
a flux bound to it. If the smallest distance between any two holes
is $R$, the length of such a loop is at least $2R$, and therefore
the lowest-order corrections to the matrix element are $\sim
J^{2R}$. Since this quantity is exponentially small in the $R \gg 1$
limit, holes with $h = 1$ can hop only along their respective
stripes as long as we are in the gapped phase with $J < 1/2$.

For holes with no flux bound to them ($h = 0$), there are no excited
plaquettes, and all loops have $W_C = +1$. This means that the
lowest-order corrections to the ground-state overlap are due to the
plaquettes $P$ and $P'$. These two corrections are identical because
$W_P = W_{P'} = +1$. The total expectation value $Z_P$ is obtained
by considering all possible ways of dividing the open string between
the sites $1$ and $1'$ into two complementary sections and all
possible ways of ordering the $x$ and $y$ bonds within the resulting
two sections. Note that the choice of the complementary sections is
limited by the fact that some bonds have switched-off interactions:
the bonds around the site $1$ can only be used after the hopping,
while the bonds around the site $1'$ can only be used before the
hopping. Exploiting symmetry to reduce the number of inequivalent
terms, the total expectation value becomes
\begin{widetext}
\begin{eqnarray}
Z_P &=& \frac{2}{16} \langle 0 | f_1 c_1 c_2 \epsilon_{1,6} c_3 c_2
c_5 c_4 c_5 c_6 f_1^{\dag} | 0 \rangle + \frac{2}{48} \langle 0 |
f_1 c_1 c_2 \epsilon_{1,6} c_5 c_4 c_3 c_2 c_5 c_6 f_1^{\dag} | 0
\rangle + \frac{2}{32} \langle 0 | f_1 c_1 c_2 \epsilon_{1,6} c_3
c_2 c_5 c_6 c_5 c_4 f_1^{\dag} | 0 \rangle \nonumber \\
&& + \, \frac{2}{64} \langle 0 | f_1 c_1 c_2 \epsilon_{1,6} c_5 c_6
c_3 c_2 c_5 c_4 f_1^{\dag} | 0 \rangle + \frac{2}{96} \langle 0 |
f_1 c_1 c_2 \epsilon_{1,6} c_5 c_4 c_5 c_6 c_3 c_2 f_1^{\dag} | 0
\rangle + \frac{2}{64} \langle 0 | f_1 c_1 c_2 \epsilon_{1,6}
c_5 c_6 c_5 c_4 c_3 c_2 f_1^{\dag} | 0 \rangle \nonumber \\
&& + \, \frac{1}{16} \langle 0 | f_1 c_1 c_2 c_3 c_2 \epsilon_{1,6}
c_5 c_4 c_5 c_6 f_1^{\dag} | 0 \rangle + \frac{2}{32} \langle 0 |
f_1 c_1 c_2 c_3 c_2 \epsilon_{1,6} c_5 c_6 c_5 c_4 f_1^{\dag} | 0
\rangle + \frac{1}{64} \langle 0 | f_1 c_3 c_2 c_1 c_2
\epsilon_{1,6} c_5 c_6 c_5 c_4 f_1^{\dag} | 0 \rangle \nonumber \\
&& + \, \frac{1}{144} \langle 0 | f_1 c_1 c_2 c_5 c_4 \epsilon_{1,6}
c_3 c_2 c_5 c_6 f_1^{\dag} | 0 \rangle + \frac{2}{288} \langle 0 |
f_1 c_1 c_2 c_5 c_4 \epsilon_{1,6} c_5 c_6 c_3 c_2 f_1^{\dag} | 0
\rangle + \frac{1}{576} \langle 0 | f_1 c_5 c_4 c_1 c_2
\epsilon_{1,6} c_5 c_6 c_3 c_2 f_1^{\dag} | 0 \rangle \nonumber \\
\nonumber \\
&=& \frac{1}{4} + \frac{1}{12} - \frac{1}{8} - \frac{1}{16} +
\frac{1}{24} - \frac{1}{16} + \frac{1}{8} - \frac{1}{8} +
\frac{1}{32} + \frac{1}{72} + \frac{1}{72} + \frac{1}{288} =
\frac{3}{16}, \label{eq-mob-phase-corr-fin} \\
\nonumber
\end{eqnarray}
\end{widetext}
where $\epsilon_{1,6} \equiv 1 - i c_1 c_6 = 2$ in all the terms
above. Since $W_P = W_{P'} = +1$ and $Z_P = Z_{P'} = 3/16$, the
corresponding lowest-order correction to the ground-state overlap is
$\langle 0 | \bar{\mathcal{F}}_{q_1 = 1}^{\dag} \epsilon_{1,6}
\mathcal{F}_{q_1 = 1} | 0 \rangle = 3 J^4 / 8$. On the other hand,
the ground-state norms $\langle 0 | \mathcal{F}_q^{\dag}
\mathcal{F}_q | 0 \rangle$ and $\langle 0 |
\bar{\mathcal{F}}_q^{\dag} \bar{\mathcal{F}}_q | 0 \rangle$ are
still approximately $1$, and therefore the lowest-order correction
to the hopping matrix element takes the form
\begin{equation}
\tilde{T}_{h_1 = 0, q_1 = 1} = -\frac{3} {16} \, J^4 \, t \qquad
(\alpha_{l_1, l_1'} = z). \label{eq-mob-phase-T}
\end{equation}
This result shows that holes with $h = 0$ and $q = 1$ are only
confined to hop in the $X$ direction in the limit of $J \rightarrow
0$. At a generic point of the gapped phase, holes with $h = 0$ are
free to hop in both the $X$ and the $Y$ directions.

\begin{figure}[t!]
\centering
\includegraphics[width=8.6cm]{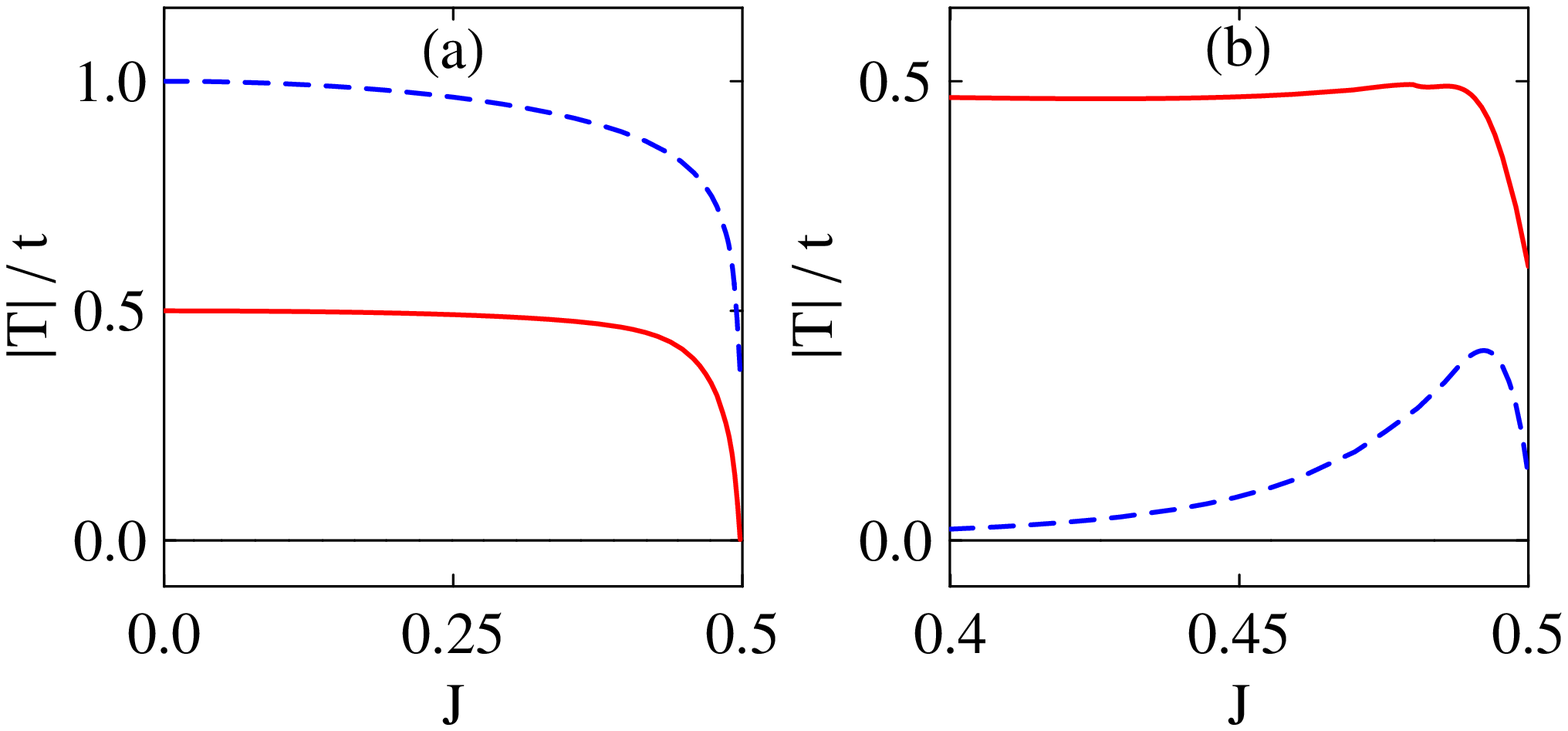}
\caption{(Color online) Effective hopping amplitude as a function of
the $x$ and $y$ bond coupling strengths within the gapped phase for
a hole with $h = 0$ and either $q = 0$ (a) or $q = 1$ (b) along $x$
and $y$ bonds (solid lines) and along $z$ bonds (dashed lines). The
lattice dimensions are $N_X = N_Y = 20$ in all cases.
\label{fig-10}}
\end{figure}

It is instructive to investigate the hole hopping problems across
the entire gapped phase with $0 < J < 1/2$. Since the perturbation
theory in $J \ll 1$ is not applicable in general, we need to
evaluate the hopping matrix elements in Eq.~(\ref{eq-mob-gen-T-fin})
numerically. The resulting hopping matrix elements for quantum
numbers $h = 0$ and $q = \{ 0,1 \}$ are plotted across the gapped
phase in Fig.~\ref{fig-10}. In the limit of $J \rightarrow 0$, when
the perturbation theory is valid, the hopping matrix elements in
Eqs.~(\ref{eq-mob-lim-T-fin}) and (\ref{eq-mob-phase-T}) are
accurately recovered. In the opposite limit of $J \rightarrow 1/2$,
when the phase transition to the gapless phase is close, the hopping
matrix elements for all quantum numbers $h = \{ 0,1 \}$ and $q = \{
0,1 \}$ become strongly dependent on the system size and exhibit a
sudden drop towards zero. These results are both explained by the
vanishing energy gap of the bulk fermion excitations: finite-size
effects become important due to the divergent correlation length,
while the hopping matrix elements vanish due to the hybridization
between the hole fermions and the lowest-energy bulk fermions.

Note that the condition of slow hopping breaks down in the limit of
$J \rightarrow 1/2$ as the lowest-energy bulk fermions no longer
have finite excitation energies. The hopping process in this limit
involves not only the respective ground states as in
Sec.~\ref{sec-mob-form}, but also the excited states in which some
of the lowest-energy bulk fermions are excited. On the other hand,
this means that the hopping matrix elements in
Eq.~(\ref{eq-mob-gen-T-fin}) underestimate the actual hopping
amplitudes, and therefore the vanishing hopping matrix elements at
$J \rightarrow 1/2$ do not imply that the holes become stationary at
the phase transition point.

\subsection{Particle statistics} \label{sec-mob-stat}

Since the quantum numbers $h$, $q$, and $p$ are conserved by the
hopping process, we can treat holes with different quantum numbers
as distinct particles and determine their respective particle
statistics. To this end, we consider an exchange process in which
two isolated identical holes at sites $0$ and $\ell$ are exchanged
along a closed loop $C$ that contains $L$ sites labeled $\{ 1, 2,
\ldots, \ell - 1, \ell, \ell + 1, \ldots, L \equiv 0 \}$. If the
exchange process is adiabatically slow, the final state is identical
to the initial state up to a complex phase factor $\exp (i
\varphi)$. The corresponding phase $\varphi$ has two contributions:
a dynamic phase from the time integral of the governing Hamiltonian
that depends on the details of the exchange process, and a geometric
phase $\theta_2$ that depends only on the loop $C$. To determine the
particle statistics, we first need to obtain the phase $\theta_2$.

\begin{figure}[t!]
\centering
\includegraphics[width=6.7cm]{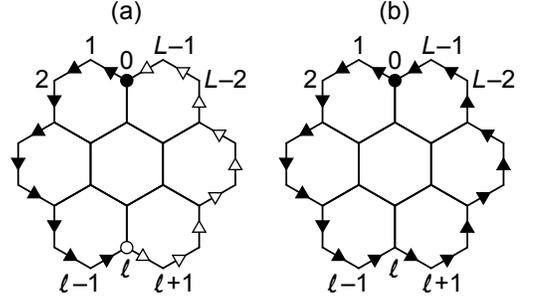}
\caption{Illustrations of the two processes that are used to
evaluate the statistical angle $\vartheta = \theta_2 - \theta_1$.
(a) Exchange process for obtaining $\theta_2$. Two identical holes
with the given quantum numbers at sites $0$ and $\ell = 9$ (black
and white dots) are exchanged along a closed loop of length $L =
18$. The subsequent hopping processes for the respective holes are
marked by black and white arrows. (b) Looping process for obtaining
$\theta_1$. One hole with the same quantum numbers at site $0$
(black dot) is moved around the same closed loop. The subsequent
hopping processes are marked by black arrows. \label{fig-11}}
\end{figure}

The adiabatic exchange process along the loop $C$ starts from the
initial ground state $| \Omega_{h, q, p}^{ \{ 0, \ell \} } \rangle$,
ends at the exchanged ground state $| \Omega_{h, q, p}^{ \{ \ell, 0
\} } \rangle$, and happens via subsequent nearest-neighbor hopping
processes through intermediate ground states $| \Omega_{h, q, p}^{
\{ l, l' \} } \rangle$, where $0 \leq l \leq \ell$ and $\ell \leq l'
\leq L$. These hopping processes are illustrated in
Fig.~\ref{fig-11}(a). The geometric phase $\theta_2$ arises from the
geometric connections between the intermediate ground states $|
\Omega_{h, q, p}^{ \{ l, l' \} } \rangle$. On the other hand, it can
be argued theoretically and verified numerically that these
geometric connections are given by the hopping matrix elements in
Eq.~(\ref{eq-mob-gen-T-2}). Since there is exactly one intermediate
hopping process for each section of the loop, this suggests that the
phase $\theta_2$ is the phase of the product of all the hopping
matrix elements around the loop $C$. In fact, we need to consider
two additional phase factors due to the two holes being exchanged.
First, the exchanged ground state $| \Omega_{h, q, p}^{ \{ \ell, 0
\} } \rangle$ can contain a non-trivial phase factor with respect to
the initial ground state $| \Omega_{h, q, p}^{ \{ 0, \ell \} }
\rangle$. Second, our hopping formalism in the hole spin picture
ignores the inherent fermionic nature of the holes. Since the two
hole spins are removed from both ground states $| \Omega_{h, q, p}^{
\{ 0, \ell \} } \rangle$ and $| \Omega_{h, q, p}^{ \{ \ell, 0 \} }
\rangle$ by fermionic annihilation operators, the exchange between
the two holes corresponds to a non-trivial phase factor $-1$ in the
actual hole picture. The geometric phase $\theta_2$ thus takes the
form
\begin{equation}
\theta_2 = \mathrm{arg} \left[ - \big{\langle} \Omega_{h, q, p}^{ \{
0, \ell \} } \big| \Omega_{h, q, p}^{ \{ \ell, 0 \} } \big{\rangle}
\prod_{l=0}^{L-1} \mathrm{T}_{l, l+1}^{(2)} \right],
\label{eq-mob-stat-theta-2-1}
\end{equation}
\begin{equation}
\mathrm{T}_{l, l+1}^{(2)} = - \frac{t \, \big{\langle} \Omega_{h, q,
p}^{ \{ l+1, l' \} } \big| \mathcal{E}_{l, l+1} \big| \Omega_{h, q,
p}^{ \{ l, l' \} } \big{\rangle}} {\sqrt{\big{\langle} \Omega_{h, q,
p}^{ \{ l+1, l' \} } \big| \Omega_{h, q, p}^{ \{ l+1, l' \} }
\big{\rangle} \big{\langle} \Omega_{h, q, p}^{ \{ l, l' \} } \big|
\Omega_{h, q, p}^{ \{ l, l' \} } \big{\rangle}}} \, .
\label{eq-mob-stat-theta-2-2}
\end{equation}
Note that the matrix element $\mathrm{T}_{l, l+1}^{(2)}$ does not
depend on the site $l'$ of the other hole as the two holes are
assumed to be isolated at each step of the exchange process.

The geometric phase of the exchange process can be written as a sum
of two terms: $\theta_2 = \vartheta + \theta_1$. The first term
$\vartheta$ is the actual statistical phase that specifies the
particle statistics, while the second term $\theta_1$ is the
geometric (Berry) phase of a looping process in which a single hole
at site $0$ with the same quantum numbers $h$, $q$, and $p$ is moved
adiabatically slowly around the same closed loop $C$. Since the
statistical angle is given by $\vartheta = \theta_2 - \theta_1$ in
terms of the two geometric phases, we also need to obtain the second
phase $\theta_1$.

The adiabatic looping process around the loop $C$ starts from the
initial ground state $| \Omega_{h, q, p}^{ \{ 0 \} } \rangle$, ends
at the final ground state $| \Omega_{h, q, p}^{ \{ L \} } \rangle
\equiv | \Omega_{h, q, p}^{ \{ 0 \} } \rangle$, and happens via
subsequent nearest-neighbor hopping processes through intermediate
ground states $| \Omega_{h, q, p}^{ \{ l \} } \rangle$, where $0 < l
< L$. These hopping processes are illustrated in
Fig.~\ref{fig-11}(b). As in the case of the exchange process, the
geometric phase $\theta_1$ of the looping process arises from the
geometric connections between the intermediate ground states, and is
therefore related to the product of the hopping matrix elements
around the loop $C$. However, the two additional phase factors are
absent because no holes are being exchanged. The geometric phase
$\theta_1$ thus takes the form
\begin{equation}
\theta_1 = \mathrm{arg} \left[ \prod_{l=0}^{L-1} \mathrm{T}_{l,
l+1}^{(1)} \right], \label{eq-mob-stat-theta-1-1}
\end{equation}
\begin{equation}
\mathrm{T}_{l, l+1}^{(1)} = - \frac{t \, \big{\langle} \Omega_{h, q,
p}^{ \{ l+1 \} } \big| \mathcal{E}_{l, l+1} \big| \Omega_{h, q, p}^{
\{ l \} } \big{\rangle}} {\sqrt{\big{\langle} \Omega_{h, q, p}^{ \{
l+1 \} } \big| \Omega_{h, q, p}^{ \{ l+1 \} } \big{\rangle}
\big{\langle} \Omega_{h, q, p}^{ \{ l \} } \big| \Omega_{h, q, p}^{
\{ l \} } \big{\rangle}}} \, . \label{eq-mob-stat-theta-1-2}
\end{equation}
Importantly, the matrix element $\mathrm{T}_{l, l+1}^{(1)}$ is in
most cases identical to the matrix element $\mathrm{T}_{l,
l+1}^{(2)}$ because the presence of the other isolated hole is
irrelevant. The only exception is the case of $h = 1$ and $q = 1$
when there is a string of excited bond fermions connected to the
hole and the hopping is sensitive to excited bond fermions [see
Eq.~(\ref{eq-mob-lim-T-fin})]. It is then relevant for at least one
section of the loop $C$ whether the other end of the string is at
the other hole moving around the same loop or at a stationary hole
in the background.

We are now ready to determine the particle statistics of the various
hole types. From a direct comparison between
Eqs.~(\ref{eq-mob-stat-theta-2-1}) and
(\ref{eq-mob-stat-theta-1-1}), the statistical phase becomes
\begin{equation}
\vartheta = \mathrm{arg} \left[ - \big{\langle} \Omega_{h, q, p}^{
\{ 0, \ell \} } \big| \Omega_{h, q, p}^{ \{ \ell, 0 \} }
\big{\rangle} \prod_{l=0}^{L-1} \left( \frac{\mathrm{T}_{l,
l+1}^{(2)}} {\mathrm{T}_{l, l+1}^{(1)}} \right) \right].
\label{eq-mob-stat-vartheta-1}
\end{equation}
Furthermore, if the holes have quantum numbers other than $h = 1$
and $q = 1$, the matrix elements $\mathrm{T}_{l, l+1}^{(1)}$ and
$\mathrm{T}_{l, l+1}^{(2)}$ are identical, and therefore
Eq.~(\ref{eq-mob-stat-vartheta-1}) reduces to
\begin{equation}
\vartheta = \mathrm{arg} \left[ - \big{\langle} \Omega_{h, q, p}^{
\{ 0, \ell \} } \big| \Omega_{h, q, p}^{ \{ \ell, 0 \} }
\big{\rangle} \right]. \label{eq-mob-stat-vartheta-2}
\end{equation}
By evaluating $\vartheta$ for all hole types, we can then directly
obtain their particle statistics: $\vartheta = 0$ is indicative of
bosons, while $\vartheta = \pi$ is indicative of fermions.

For holes with $h = 0$ and $q = 0$, the initial ground state $|
\Omega_{0, 0, p}^{ \{ 0, \ell \} } \rangle$ and the final ground
state $| \Omega_{0, 0, p}^{ \{ \ell, 0 \} } \rangle$ are identical
by construction. In the isolated dimer limit, the two ground states
for $p = 0$ holes are $| \Omega_{0, 0, 0}^{ \{ 0, \ell \} } \rangle
= | \Omega_{0, 0, 0}^{ \{ \ell, 0 \} } \rangle = \mathcal{Q}_{ \{ 0,
\ell \} } \mathcal{P} | 0 \rangle$. In the general case, there are
additional operators $B_p \neq 1$ and $\mathcal{F}_{0; 0} \neq 1$
that set the bond fermion and the matter fermion sectors. On the
other hand, these operators are the same for both ground states, and
therefore the relation $| \Omega_{0, 0, p}^{ \{ 0, \ell \} } \rangle
= | \Omega_{0, 0, p}^{ \{ \ell, 0 \} } \rangle$ remains true. Since
applying Eq.~(\ref{eq-mob-stat-vartheta-2}) then gives $\vartheta =
\pi$, we conclude that holes with $h = 0$ and $q = 0$ are fermions.

For holes with $h = 0$ and $q = 1$, the initial ground state $|
\Omega_{0, 1, p}^{ \{ 0, \ell \} } \rangle$ and the final ground
state $| \Omega_{0, 1, p}^{ \{ \ell, 0 \} } \rangle$ are only
identical up to a minus sign as the two ground states have the two
hole fermions at sites $0$ and $\ell$ excited in an opposite order.
In the isolated dimer limit, the two ground states for $p = 0$ holes
are $| \Omega_{0, 1, 0}^{ \{ 0, \ell \} } \rangle = -| \Omega_{0, 1,
0}^{ \{ \ell, 0 \} } \rangle = \mathcal{Q}_{ \{ 0, \ell \} }
\mathcal{P} f_{\tilde{z} (0)}^{\dag} f_{\tilde{z} (\ell)}^{\dag} | 0
\rangle$. In the general case, there are additional operators $B_p
\neq 1$ and $\mathcal{F}_{1; 0} \neq 1$ that set the bond fermion
and the matter fermion sectors. On the other hand, these operators
are the same for both ground states, and therefore the relation $|
\Omega_{0, 1, p}^{ \{ 0, \ell \} } \rangle = -| \Omega_{0, 1, p}^{
\{ \ell, 0 \} } \rangle$ remains true. Since applying
Eq.~(\ref{eq-mob-stat-vartheta-2}) then gives $\vartheta = 0$, we
conclude that holes with $h = 0$ and $q = 1$ are bosons.

It is crucial that holes with $h = 1$ can move only around the
plaquettes of particular stripes: even stripes for $q = 0$ and odd
stripes for $q = 1$. Furthermore, it is shown by Fig.~\ref{fig-7}
that $q = 0$ holes in even stripes are equivalent to $q = 1$ holes
in odd stripes. This implies that these two hole types have the same
particle statistics, and therefore it is enough to consider one of
them. We choose to consider holes with $h = 1$ and $q = 0$ because
Eq.~(\ref{eq-mob-stat-vartheta-2}) is then applicable. For these
holes, the only difference in the ground state with respect to holes
with $h = 0$ and $q = 0$ is the presence of an additional
flux-binding operator $\mathcal{X}_1 \neq 1$. On the other hand,
this operator is the same for the initial and the final ground
states, and therefore the two ground states are identical: $|
\Omega_{1, 0, p}^{ \{ 0, \ell \} } \rangle = | \Omega_{1, 0, p}^{ \{
\ell, 0 \} } \rangle$. Since applying
Eq.~(\ref{eq-mob-stat-vartheta-2}) then gives $\vartheta = \pi$, we
conclude that holes with $h = 1$ and $q = \{ 0,1 \}$ are fermions.

To supplement the above derivations, we also provide an intuitive
explanation for the particle statistics found. The main principle is
that the holes in the model can bind the elementary excitations of
the model: fluxes and fermions. The various hole types with
different quantum numbers $h$ and $q$ are then distinguished only by
the kinds of elementary excitations that are bound to them. In
particular, a hole with a non-trivial flux quantum number $h = 1$
has a bound flux, while a hole with a non-trivial fermion quantum
number $q = 1$ has a bound fermion. Holes with $h = 0$ and $q = 0$
are interpreted as bare holes with no elementary excitations bound
to them. Since bare holes are missing spin one-half fermions, it is
natural that they are fermions themselves. Conversely, the remaining
three types of holes are interpreted as composite holes made out of
bare holes and elementary excitations. Due to the presence of the
bound excitations, their statistics can be different from that of
bare holes. For holes with $h = 0$ and $q = 1$, the binding of a
fermion leads to a statistical transmutation, and therefore these
holes are bosons. For holes with $h = 1$ and $q = 0$, the binding of
a flux has no effect on the statistics, and therefore these holes
are fermions. We might then naively expect that holes with $h = 1$
and $q = 1$ should be bosons because there is a statistical
transmutation due to the binding of a fermion. However, the bound
flux and the bound fermion have semionic relative statistics. Since
this corresponds to an additional transmutation for the composite
hole, these holes are in fact fermions. The particle statistics of
the various hole types along with their interpretations in terms of
the bound excitations are summarized in Table \ref{table-6}.

\begin{table}[h]
\begin{tabular*}{0.42\textwidth}{@{\extracolsep{\fill}} c | c | c | c }
\hline \hline
\multicolumn{2}{c |}{Hole type}                          & \,\, Statistics \,\,  & Interpretation               \\
\hline
\multirow{2}{*}{\,\, $h = 0$ \,\,}  & \,\, $q = 0$ \,\,  & Fermion               & Bare hole                    \\
                                    & \,\, $q = 1$ \,\,  & Boson                 & Hole + fermion               \\
\hline
\multirow{2}{*}{\,\, $h = 1$ \,\,}  & \,\, $q = 0$ \,\,  & Fermion               & Hole + flux                  \\
                                    & \,\, $q = 1$ \,\,  & Fermion               & \, Hole + flux + fermion \,  \\
\hline  \hline
\end{tabular*}
\caption{Absolute statistics of holes with flux quantum numbers $h =
\{ 0,1 \}$ and fermion quantum numbers $q = \{ 0,1 \}$ from a
process when two identical holes are exchanged. Interpretations are
given in terms of elementary excitations bound. \label{table-6}}
\end{table}

It is also useful to investigate the relative statistics between the
various hole types. To this end, we consider two looping processes
in which a hole with quantum numbers $h$ and $q$ is moved around a
closed loop $C$. In the first case, there is no hole enclosed by the
loop, and Eq.~(\ref{eq-mob-stat-theta-1-1}) gives a geometric phase
$\theta_1$. In the second case, there is one stationary hole with
quantum numbers $h'$ and $q'$ enclosed by the loop, and
Eq.~(\ref{eq-mob-stat-theta-1-1}) gives a geometric phase
$\theta_1'$. The relative statistics between holes with quantum
numbers $h$ and $q$ and holes with quantum numbers $h'$ and $q'$ is
then specified by the relative statistical phase $\vartheta' =
\theta_1' - \theta_1$. For $\vartheta' = 0$, the two hole types have
trivial relative statistics, while for $\vartheta' \neq 0$, the two
hole types have anyonic relative statistics. Importantly, the
relative statistical phase $\vartheta'$ is symmetric in the two hole
types: it does not depend on which one is kept stationary and which
one is moved around the loop.

We first notice that the two looping processes giving the phases
$\theta_1$ and $\theta_1'$ are not both possible for all
combinations of the quantum numbers. For mobile holes with $h = 1$,
the exchange process and the looping process with no hole enclosed
are barely possible, but the looping process with a stationary hole
enclosed is impossible. Since these holes can move only around the
plaquettes of particular stripes, there is no space for a stationary
hole inside any loop they can possibly move around. This means that
the mobile hole must have a trivial flux quantum number $h = 0$. On
the other hand, the stationary hole can then only influence the
hopping of the mobile hole if the stationary hole is connected to a
string of excited bond fermions and the hopping of the mobile hole
is sensitive to excited bond fermions. This corresponds to quantum
numbers $h' = 1$ and $q' = \{ 0,1 \}$ for the stationary hole and
quantum numbers $h = 0$ and $q = 1$ for the mobile hole. In these
cases, one hopping matrix element picks up a minus sign at the
intersection point of the loop and the string of excited bond
fermions [see Eq.~(\ref{eq-mob-lim-T-fin})]. This implies that the
relative statistical phase is $\vartheta' = \pi$, and therefore the
two hole types have semionic relative statistics. In all other
possible cases, the hopping of the mobile hole is not influenced by
the stationary hole. This implies that the relative statistical
phase is $\vartheta' = 0$, and therefore the two hole types have
trivial relative statistics. The results for the relative statistics
between the various hole types are summarized in Table
\ref{table-7}.

\begin{table}[h]
\begin{tabular*}{0.44\textwidth}{@{\extracolsep{\fill}} c | c | c c | c c }
\hline \hline
\multicolumn{2}{c |}{\multirow{2}{*}{Hole type}}         & \multicolumn{2}{c |}{$h' = 0$}            & \multicolumn{2}{c}{$h' = 1$}              \\
\cline{3-6}
\multicolumn{2}{c |}{}                                   & \,\, $q' = 0$ \,\,  & \,\, $q' = 1$ \,\,  & \,\, $q' = 0$ \,\,  & \,\, $q' = 1$ \,\,  \\
\hline
\multirow{2}{*}{\,\, $h = 0$ \,\,}  & \,\, $q = 0$ \,\,  & $0$                 & $0$                 & $0$                 & $0$                 \\
                                    & \,\, $q = 1$ \,\,  & $0$                 & $0$                 & $\pi$               & $\pi$               \\
\hline
\multirow{2}{*}{\,\, $h = 1$ \,\,}  & \,\, $q = 0$ \,\,  & $-$                 & $-$                 & $-$                 & $-$                 \\
                                    & \,\, $q = 1$ \,\,  & $-$                 & $-$                 & $-$                 & $-$                 \\
\hline  \hline
\end{tabular*}
\caption{Relative statistics between holes with quantum numbers $h$
and $q$ and holes with quantum numbers $h'$ and $q'$ from a process
when the former hole type is moved around the latter hole type:
$\vartheta' = 0$ indicates trivial statistics, $\vartheta' = \pi$
indicates semionic statistics, while there is no value if the
process is impossible. \label{table-7}}
\end{table}

We can also interpret the relative statistics in terms of the
elementary excitations bound to the holes. First, two bare holes or
two identical elementary excitations have trivial relative
statistics. Second, the relative statistics between a bare hole and
an elementary excitation is trivial, while that between a flux and a
fermion is semionic. As a result of these properties, the relative
statistics between two identical holes and that between a bare hole
and a composite hole is trivial, while that between two distinct
composite holes is semionic. The entries of Table \ref{table-7} can
then be obtained, even the ones that correspond to impossible
processes: the diagonal entries and the entries of the first row or
the first column are $\vartheta' = 0$, while the remaining entries
are $\vartheta' = \pi$. We finally remark that all of our results
for the absolute and the relative particle statistics are consistent
with the correspondence between the various hole types and the
superselection sectors (see Table \ref{table-5}).

\section{Finite density of mobile holes} \label{sec-gas}

\subsection{Non-interacting treatment} \label{sec-gas-free}

We now consider the Kitaev honeycomb model with a finite density of
mobile holes. The hole density $\rho = n / 2N$ gives the fraction of
sites $l$ that are hole sites $l \in \Delta$. For simplicity, we
assume a small hole density $\rho \ll 1$ and neglect any hole
interactions. The ground state of the model is then a multi-hole
state of $n$ non-interacting holes: depending on their particle
statistics, these holes either form a Bose condensate or fill up a
Fermi sea. By evaluating the multi-hole energy for all combinations
of $h = \{ 0,1 \}$ and $q = \{ 0,1 \}$, we can determine the
ground-state quantum numbers.

The most straightforward way to represent the multi-hole state is to
use appropriate single-hole creation and annihilation operators. If
the operator $a_{h, q, p}^{(\dag)} (\mathbf{R}_l)$ annihilates
(creates) a hole at site $l$ with quantum numbers $h$, $q$, and $p$,
the multi-hole state of $n$ stationary holes at sites $\Delta = \{
l_j \}$ with quantum numbers $\{ h_j \}$, $\{ q_j \}$, and $\{ p_j
\}$ reads as
\begin{equation}
\big| \Omega_{h, q, p}^{\, \Delta} \big{\rangle} = \prod_{j=1}^n
a_{h_j, q_j, p_j}^{\dag} (\mathbf{R}_{l_j}) \, | \Omega \rangle,
\label{eq-gas-free-state}
\end{equation}
where $| \Omega \rangle$ is the ground state of the model with no
holes, and the lattice position $\mathbf{R}_l = (X_l, Y_l)$ of the
site $l$ is measured in units of the lattice constant. We now assume
and later verify that holes with distinct flux quantum numbers $h =
\{ 0,1 \}$ are not simultaneously present in this multi-hole state.
Since Table \ref{table-7} shows that no anyonic relative statistics
manifests itself between holes with identical flux quantum numbers,
the single-hole operators $a_{h, q, p}^{(\dag)} (\mathbf{R}_l)$ can
then be treated as standard bosonic and fermionic operators. In
particular, they satisfy bosonic commutation relations in the case
of $h = 0$ and $q = 1$, and fermionic anticommutation relations in
all other cases, except for an overall hard-core constraint that
there can be at most one hole of any type at each site. However, if
the hole density $\rho$ is sufficiently small, this hard-core
constraint is practically irrelevant. We can then write an effective
Hamiltonian for the model with $n$ mobile holes in terms of the
standard bosonic and fermionic operators $a_{h, q, p}^{(\dag)}
(\mathbf{R}_l)$. In the absence of hole interactions, this
Hamiltonian is quadratic: it contains an onsite potential term
corresponding to the flux-binding energy discussed in
Sec.~\ref{sec-stat-deg} and several hopping terms corresponding to
the hopping problems in Fig.~\ref{fig-8}. Taking the isolated dimer
limit and keeping only the lowest-order terms in $J \ll 1$, the
effective Hamiltonian takes the form
\begin{widetext}
\begin{eqnarray}
H_a &=& \Gamma_0 - \frac{9 J^8} {1024} \sum_l \sum_{q, p}
\hat{n}_{1, q, p} (\mathbf{R}_l) - \frac{t}{2} \sum_{l \in A} \,
\sum_{\alpha = x,y} \, \sum_{h, q, p} \left[ a_{h, q, p}^{\dag}
(\mathbf{R}_l) \, a_{h, q, p} (\mathbf{R}_{\alpha(l)}) +
\mathrm{H.c.} \right] \nonumber \\
&& - \, t \sum_{l \in A} \sum_p \left[ a_{0, 0, p}^{\dag}
(\mathbf{R}_l) \, a_{0, 0, p} (\mathbf{R}_{z(l)}) + \mathrm{H.c.}
\right] - \frac{3 J^4 \, t} {16} \sum_{l \in A} \sum_p \left[ a_{0,
1, p}^{\dag} (\mathbf{R}_l) \, a_{0, 1, p} (\mathbf{R}_{z(l)}) +
\mathrm{H.c.}
\right] \label{eq-gas-free-H-1} \\
&& - \, t \sum_{l \in A'} \sum_p \left[ a_{1, 0, p}^{\dag}
(\mathbf{R}_l) \, a_{1, 0, p} (\mathbf{R}_{z(l)}) + \mathrm{H.c.}
\right] - t \sum_{l \in A''} \sum_p \left[ a_{1, 1, p}^{\dag}
(\mathbf{R}_l) \, a_{1, 1, p} (\mathbf{R}_{z(l)}) + \mathrm{H.c.}
\right], \nonumber \\
\nonumber
\end{eqnarray}
where $\Gamma_0$ is the ground-state energy of the model with $n$
stationary $h = 0$ holes, and $\hat{n}_{h, q, p} (\mathbf{R}_l)
\equiv a_{h, q, p}^{\dag} (\mathbf{R}_l) a_{h, q, p} (\mathbf{R}_l)$
is the number operator. The fixed total number of holes is enforced
by the constraint $n = \sum_l \sum_{h, q, p} \langle \hat{n}_{h, q,
p} (\mathbf{R}_l) \rangle$. Note that the coefficients of the
hopping terms in Eq.~(\ref{eq-gas-free-H-1}) are the hopping
amplitudes in Fig.~\ref{fig-8}: those along the bonds marked by
dashed lines are given by Eq.~(\ref{eq-mob-phase-T}), while those
along the bonds marked by dotted lines are exactly zero. Since the
hopping problems for $h = 1$ holes break the translational symmetry
of the lattice, it is necessary to divide each sublattice $A$ and
$B$ into two further sublattices: $A = A' \cup A''$ and $B = B' \cup
B''$, where sites in the sublattices $A'$ and $B'$ are pairwise
connected by $z$ bonds in even stripes, and sites in the sublattices
$A''$ and $B''$ are pairwise connected by $z$ bonds in odd stripes.

The Hamiltonian in Eq.~(\ref{eq-gas-free-H-1}) is quadratic, and
therefore it becomes diagonal after an appropriate transformation of
the single-hole operators $a_{h, q, p}^{(\dag)} (\mathbf{R}_l)$. Due
to the translational symmetry of the hopping problems, the new
single-hole operators $\tilde{a}_{h, q, p}^{(\dag)} (\mathbf{k},
\nu)$ are labeled with the lattice momentum $\mathbf{k} = (k_X,
k_Y)$ conjugate to the lattice position $\mathbf{R} = (X, Y)$. In
terms of the original real-space operators, these new momentum-space
operators are given by
\begin{eqnarray}
\tilde{a}_{0, q, p} (\mathbf{k}, \nu) &=& \frac{1} {\sqrt{N}}
\sum_{l \in A} \Big{[} \beta_{0, q}^A (\mathbf{k}, \nu) \, a_{0, q,
p} (\mathbf{R}_l) \, e^{-i \mathbf{k} \cdot \mathbf{R}_l} +
\beta_{0, q}^B (\mathbf{k}, \nu) \, a_{0, q, p} (\mathbf{R}_{z(l)})
\, e^{-i \mathbf{k} \cdot \mathbf{R}_{z(l)}} \Big{]},
\nonumber \\
\tilde{a}_{1, q, p} (\mathbf{k}, \nu) &=& \sqrt{\frac{2} {N}}
\sum_{l \in A'} \Big{[} \beta_{1, q}^{A'} (\mathbf{k}, \nu) \, a_{1,
q, p} (\mathbf{R}_l) \, e^{-i \mathbf{k} \cdot \mathbf{R}_l} +
\beta_{1, q}^{B'} (\mathbf{k}, \nu) \, a_{1, q, p}
(\mathbf{R}_{z(l)}) \, e^{-i \mathbf{k} \cdot \mathbf{R}_{z(l)}}
\label{eq-gas-free-op} \\
&& + \, \beta_{1, q}^{A''} (\mathbf{k}, \nu) \, a_{1, q, p}
(\mathbf{R}_{x(z(l))}) \, e^{-i \mathbf{k} \cdot
\mathbf{R}_{x(z(l))}} + \beta_{1, q}^{B''} (\mathbf{k}, \nu) \,
a_{1, q, p} (\mathbf{R}_{y(l)}) \, e^{-i \mathbf{k} \cdot
\mathbf{R}_{y(l)}} \Big{]}, \nonumber
\end{eqnarray}
\end{widetext}
where the coefficients $\beta_{h, q} (\mathbf{k}, \nu) \sim 1$ for
the different sublattices distinguish two bands $\nu = \{ 1,2 \}$ in
the case of $h = 0$ and four bands $\nu = \{ 1,2,3,4 \}$ in the case
of $h = 1$. In terms of the momentum-space operators $\tilde{a}_{h,
q, p}^{(\dag)} (\mathbf{k}, \nu)$, the Hamiltonian in
Eq.~(\ref{eq-gas-free-H-1}) takes the free-particle form
\begin{equation}
H_a = \Gamma_0 + \sum_{h, q, p} \sum_{\mathbf{k}, \nu} \Lambda_{h,
q} (\mathbf{k}, \nu) \, \tilde{n}_{h, q, p} (\mathbf{k}, \nu),
\label{eq-gas-free-H-2}
\end{equation}
where $\tilde{n}_{h, q, p} (\mathbf{k}, \nu) \equiv \tilde{a}_{h, q,
p}^{\dag} (\mathbf{k}, \nu) \tilde{a}_{h, q, p} (\mathbf{k}, \nu)$
is the number operator in momentum space. The constraint on the
total number of holes is then $n = \sum_{h, q, p} \sum_{\mathbf{k},
\nu} \langle \tilde{n}_{h, q, p} (\mathbf{k}, \nu) \rangle$.

To evaluate the multi-hole energy as the expectation value of the
Hamiltonian in Eq.~(\ref{eq-gas-free-H-2}), we need to know the
energies $\Lambda_{h, q} (\mathbf{k}, \nu)$ and the occupation
numbers $\langle \tilde{n}_{h, q, p} (\mathbf{k}, \nu) \rangle$ of
the single-hole states. In the ground state of the model, holes
occupy only the lowest-energy single-hole states, and it is
therefore enough to determine the single-hole energy $\Lambda_{h, q}
(\mathbf{k}, \nu)$ around its overall minimum. On the other hand,
the overall minimum of $\Lambda_{h, q} (\mathbf{k}, \nu)$ in the
lowest band $\nu = 1$ is at zero momentum because the hopping
amplitudes in Fig.~\ref{fig-8} are all non-positive. Expanding
$\Lambda_{h, q} (\mathbf{k}, 1)$ up to quadratic order in the
momentum around $\mathbf{k} = \mathbf{0}$, and keeping the
lowest-order terms in $J \ll 1$, the single-hole dispersion
relations for the different quantum numbers are given by
\begin{eqnarray}
\Lambda_{0, 0} (\mathbf{k}, 1) &=& \left[ -2 + \frac{3 k_X^2} {8} +
\frac{9 k_Y^2} {16} \right] t,
\nonumber \\
\Lambda_{0, 1} (\mathbf{k}, 1) &=& \left[ -1 + \frac{3 k_X^2} {8} +
\frac{27 J^4 k_Y^2} {128} \right] t,
\label{eq-gas-free-disp} \\
\Lambda_{1, q} (\mathbf{k}, 1) &=& - \frac{9 J^8} {1024} + \left[ -
\frac{1 + \sqrt{5}} {2} + \frac{3 k_X^2} {4 \sqrt{5}} \right] t.
\nonumber
\end{eqnarray}
Since $h = 1$ holes are not allowed to hop at all between their
stripes, their dispersion relation is independent of the component
$k_Y$ at all orders of the momentum.

\begin{figure}[t!]
\centering
\includegraphics[width=8.2cm]{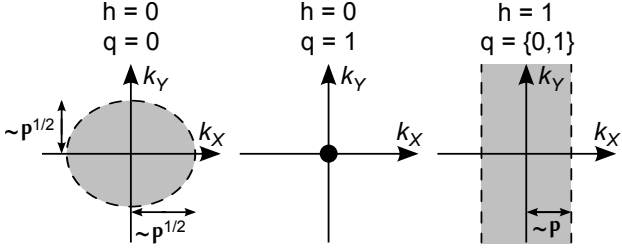}
\caption{Occupations of the single-hole states for different
combinations of the flux quantum number $h = \{ 0,1 \}$ and the
fermion quantum number $q = \{ 0,1 \}$. In the bosonic case, the
lowest-energy state of macroscopic occupation is marked by a black
dot. In the fermionic cases, the Fermi sea states of constant
occupation are marked by gray shading, while the Fermi surface
separating occupied and unoccupied states is marked by a dashed
line. \label{fig-12}}
\end{figure}

When turning our attention to the corresponding occupation numbers
$\langle \tilde{n}_{h, q, p} (\mathbf{k}, 1) \rangle$ around zero
momentum, we assume that all holes in the multi-hole state have
identical quantum numbers $h$ and $q$. It is then crucial to notice
that certain holes are bosons, while others are fermions. If the
holes are bosons, they all occupy the zero-momentum state. For holes
with $h = 0$ and $q = 1$, the average single-hole energy in the
multi-hole state is then $\langle \Lambda_{0, 1} (\mathbf{k}, \nu)
\rangle = -t$. If the holes are fermions, they fill up a Fermi sea
around the zero-momentum state: each state inside the Fermi surface
is occupied by two holes with different quantum numbers $p = \{ 0,1
\}$, while the states outside the Fermi surface are unoccupied. For
holes with $h = 0$ and $q = 0$, the equipotential curves are
ellipses of similar half-axes. The Fermi sea is therefore an ellipse
of half-axes $\Delta k_X \sim \Delta k_Y \sim \sqrt{\rho}$, and the
average single-hole energy is $\langle \Lambda_{0, 0} (\mathbf{k},
\nu) \rangle = -2t + \kappa_1 t \rho$, where $\kappa_1 \sim 1$. For
holes with $h = 1$ and $q = \{ 0,1 \}$, the equipotential curves are
lines parallel to the $k_Y$ direction. The Fermi sea is therefore a
strip of half-width $\Delta k_X \sim \rho$, and the average
single-hole energy is $\langle \Lambda_{1, q} (\mathbf{k}, \nu)
\rangle = -9 J^8 / 1024 -(1 + \sqrt{5}) t / 2 + \kappa_2 t \rho^2$,
where $\kappa_2 \sim 1$. The occupation numbers of the single-hole
states for the different quantum numbers $h$ and $q$ are illustrated
in Fig.~\ref{fig-12}, while the resulting average single-hole
energies in the multi-hole state are summarized in Table
\ref{table-8}.

\begin{table}[h]
\begin{tabular*}{0.43\textwidth}{@{\extracolsep{\fill}} c | c | c }
\hline \hline
\multicolumn{2}{c |}{Hole type}                          & Average single-hole energy                                                         \\
\hline
\multirow{2}{*}{\,\, $h = 0$ \,\,}  & \,\, $q = 0$ \,\,  & $-2t + \kappa_1 t \rho$                                                            \\
                                    & \,\, $q = 1$ \,\,  & $-t$                                                                               \\
\hline
\multirow{2}{*}{\,\, $h = 1$ \,\,}  & \,\, $q = 0$ \,\,  & \, \multirow{2}{*}{$-9 J^8 / 1024 - (1 + \sqrt{5}) t / 2 + \kappa_2 t \rho^2$} \,  \\
                                    & \,\, $q = 1$ \,\,  &                                                                                    \\
\hline  \hline
\end{tabular*}
\caption{Average single-hole energy $\langle \Lambda_{h, q}
(\mathbf{k}, \nu) \rangle$ in the model with a density $\rho \ll 1$
of mobile holes with flux quantum numbers $h = \{ 0,1 \}$ and
fermion quantum numbers $q = \{ 0,1 \}$. \label{table-8}}
\end{table}

We are now ready to identify the ground-state quantum numbers of the
model. Since the total number of holes is fixed, the average
single-hole energies $\langle \Lambda_{h, q} (\mathbf{k}, \nu)
\rangle$ for the different quantum numbers can be compared directly.
Furthermore, the assumption of small hole density $\rho \ll 1$ means
that the energies $\sim t \rho$ and $\sim t \rho^2$ are negligible
compared to the energies $\sim t$. The results in Table
\ref{table-8} then indicate two complementary regimes in the
behavior of the model. In the first regime with $J^8 \ll t \ll J^4$,
holes with $h = 0$ and $q = 0$ have the lowest average energy. This
means that all holes in the ground state have quantum numbers $h =
0$ and $q = 0$. At each momentum $\mathbf{k}$ within the Fermi
ellipse of Fig.~\ref{fig-12}, there are two holes with quantum
numbers $p = \{ 0,1 \}$. In the second regime with $t \ll J^8$,
holes with $h = 1$ and $q = \{ 0,1 \}$ have the lowest average
energy. This means that all holes in the ground state have quantum
numbers $h = 1$. At each momentum $\mathbf{k}$ within the Fermi
strip of Fig.~\ref{fig-12}, there are four holes with quantum
numbers $q = \{ 0,1 \}$ and $p = \{ 0,1 \}$. Note that our original
assumption of no anyonic relative statistics is self-consistent as
all holes in the ground state are $h = 0$ holes in the first regime
and $h = 1$ holes in the second regime.

Due to the distinct ground states, the model also has different
physical properties in the two regimes. We consider the net
magnetization and the electrical conductivities in the $X$ and $Y$
directions. The net magnetization is the sum of the local hole
magnetizations $(-1)^{q + p}$ and is zero in both regimes because
each hole with quantum numbers $h$, $q$, and $p$ has a pair with
quantum numbers $h$, $q$, and $1 - p$. In terms of the partial
densities $\rho_{h, q} = \sum_p \sum_{\mathbf{k}} \langle
\tilde{n}_{h, q, p} (\mathbf{k}, 1) \rangle / 2N$ of the various
hole types, the conductivities in the two directions are
\begin{equation}
\sigma_{X, Y}^{*} = e_{*}^2 \tau \sum_{h, q} \rho_{h, q} \left[
\frac{\partial^2 \Lambda_{h, q} (\mathbf{k}, 1)} {\partial k_{X,
Y}^2} \right]_{\mathbf{k} = \mathbf{0}}, \label{eq-gas-free-cond}
\end{equation}
where $e_{*}$ is the hole charge, and $\tau$ is the elastic
scattering time. In the first regime with $J^8 \ll t \ll J^4$, the
partial hole densities are $\rho_{0, 0} = \rho$ and $\rho_{1, q} =
\rho_{h, 1} = 0$. Since the effective masses $[\partial^2
\Lambda_{0, 0} (\mathbf{k}, 1) /
\partial k_{X, Y}^2]^{-1}$ are similar in the $X$ and $Y$
directions, the conductivity is approximately isotropic:
$\sigma_X^{*} \sim \sigma_Y^{*} \sim t \rho e_{*}^2 \tau$. In the
second regime with $t \ll J^8$, the partial hole densities are
$\rho_{1, q} = \rho / 2$ and $\rho_{0, q} = 0$. Since the effective
masses $[\partial^2 \Lambda_{1, q} (\mathbf{k}, 1) / \partial k_{X,
Y}^2]^{-1}$ are finite in the $X$ direction and infinite in the $Y$
direction, the conductivity is extremely anisotropic: $\sigma_X^{*}
\sim t \rho e_{*}^2 \tau$ and $\sigma_Y^{*} = 0$.

\subsection{Mean-field treatment of interactions} \label{sec-gas-int}

We now consider hole interactions in the model with a small density
$\rho \ll 1$ of mobile holes. To represent an interaction of
strength $\Delta \Gamma_0$ between two holes at a relative lattice
position $\mathbf{R} = \mathbf{R}_2 - \mathbf{R}_1$, we need to add
an appropriate quartic term to the Hamiltonian in
Eq.~(\ref{eq-gas-free-H-1}). Restricting our attention to the
Coulomb repulsion and the two attraction mechanisms discussed in
Sec.~\ref{sec-stat-bound}, this quartic term takes the general form
\begin{eqnarray}
\Delta H_a &=& \Delta \Gamma_0 \sum_{\mathbf{R}_1} a_{h_1, q_1',
p_1}^{\dag} (\mathbf{R}_1) \, a_{h_2, q_2', p_2}^{\dag}
(\mathbf{R}_1 + \mathbf{R}) \nonumber \\
&& \times \, a_{h_2, q_2, p_2} (\mathbf{R}_1 + \mathbf{R}) \,
a_{h_1, q_1, p_1} (\mathbf{R}_1). \label{eq-gas-int-H-1}
\end{eqnarray}
The flux quantum numbers $h_{1,2}$ and the plaquette quantum numbers
$p_{1,2}$ are conserved by such a general hole interaction, while
the fermion quantum numbers $q_{1,2} \neq q_{1,2}'$ satisfy the
relation $q_1' + q_2' = q_1 + q_2$ modulo $2$ so that the global
constraint $\sum_j q_j = \textrm{even}$ is not violated.

Since the flux quantum numbers are conserved, the two complementary
regimes found in Sec.~\ref{sec-gas-free} remain applicable in the
presence of hole interactions: all holes in the ground state have
quantum numbers $h = 0$ in the first regime with $J^8 \ll t \ll
J^4$, while they all have quantum numbers $h = 1$ in the second
regime with $t \ll J^8$. We can then consider the two regimes
independently from each other with only $h \equiv h_{1,2} = 0$ holes
in the first regime and only $h \equiv h_{1,2} = 1$ holes in the
second regime. On the other hand, this means that our assumption of
no anyonic relative statistics remains self-consistent in the
presence of hole interactions. Expressing the real-space operators
$a_{h, q, p}^{(\dag)} (\mathbf{R}_l)$ in terms of the momentum-space
operators $\tilde{a}_{h, q, p}^{(\dag)} (\mathbf{k}, \nu)$, and
considering only the lowest band $\nu = 1$, the quartic term in
Eq.~(\ref{eq-gas-int-H-1}) becomes
\begin{eqnarray}
\Delta H_a &=& \frac{\Delta \Gamma_0} {N} \sum_{\mathbf{k}_1,
\mathbf{k}_2, \mathbf{k}'} \Upsilon_{h, \hat{q}} (\mathbf{k}) \,
e^{i \mathbf{k}' \cdot \mathbf{R}} \, \, \tilde{a}_{h, q_1',
p_1}^{\dag} (\mathbf{k}_1 + \mathbf{k}')
\nonumber \\
&& \times \, \tilde{a}_{h, q_2', p_2}^{\dag} (\mathbf{k}_2 -
\mathbf{k}') \, \tilde{a}_{h, q_2, p_2} (\mathbf{k}_2) \,
\tilde{a}_{h, q_1, p_1} (\mathbf{k}_1), \nonumber \\
\label{eq-gas-int-H-2}
\end{eqnarray}
where $\tilde{a}_{h, q, p}^{(\dag)} (\mathbf{k}) \equiv
\tilde{a}_{h, q, p}^{(\dag)} (\mathbf{k}, 1)$, and the quantity
$\Upsilon_{h, \hat{q}} (\mathbf{k}) \sim 1$ with $\hat{q} \equiv \{
q_1, q_2, q_1', q_2' \}$ depends on the various coefficients
$\beta_{h, q} (\mathbf{k}) \equiv \beta_{h, q} (\mathbf{k}, 1) \sim
1$ for the different sublattices.

The behavior of the model is influenced by hole interactions in
several ways. We aim to specify the extent of applicability of the
results in Sec.~\ref{sec-gas-free} for the ground-state quantum
numbers and the corresponding physical properties. To this end, we
investigate how hole interactions renormalize the average
single-hole energies in Table \ref{table-8} as a function of the
partial hole densities $\rho_{h, q}$. In practice, we apply a
standard mean-field decomposition to the Hamiltonian: each quartic
term in Eq.~(\ref{eq-gas-int-H-2}) is decomposed into two
constituent quadratic terms, and each quadratic term is coupled to
the expectation value of the other one. The single-hole energies
$\Lambda_{h, q} (\mathbf{k}) \equiv \Lambda_{h, q} (\mathbf{k}, 1)$
in Eq.~(\ref{eq-gas-free-H-2}) are then renormalized by the
mean-field decomposition of any quartic term that is a product of
two number operators $\tilde{n}_{h, q, p} (\mathbf{k}) \equiv
\tilde{n}_{h, q, p} (\mathbf{k}, 1)$. In general, if we keep all
such quartic terms in Eq.~(\ref{eq-gas-int-H-2}), and include all
the equivalent quartic terms that differ only in their conserved
plaquette quantum numbers $p_{1,2}$, the resulting mean-field
decomposition takes the approximate form
\begin{eqnarray}
\Delta \tilde{H}_a &\sim& \frac{\Delta \Gamma_0} {N}
\sum_{\mathbf{k}_1, \mathbf{k}_2} \Big{[} \langle \tilde{n}_{h, q_1}
(\mathbf{k}_1) \rangle \,
\tilde{n}_{h, q_2} (\mathbf{k}_2) \nonumber \\
&& + \, \langle \tilde{n}_{h, q_2} (\mathbf{k}_2) \rangle \,
\tilde{n}_{h, q_1} (\mathbf{k}_1) \Big{]}, \label{eq-gas-int-mft}
\end{eqnarray}
where $\tilde{n}_{h, q} (\mathbf{k}) \equiv \sum_p \tilde{n}_{h, q,
p} (\mathbf{k})$. Since $\rho = \sum_{h, q} \rho_{h, q}$ and
$\sum_{\mathbf{k}} \langle \tilde{n}_{h, q} (\mathbf{k}) \rangle =
2N \rho_{h, q}$, the single-hole energies in
Eq.~(\ref{eq-gas-free-H-2}) are renormalized by $\Delta \Lambda_{h,
q} (\mathbf{k}) \sim \Delta \Gamma_0 \, \rho$.

\subsubsection{First regime: $J^8 \ll t \ll J^4$} \label{sec-gas-int-1}

In the first regime with $J^8 \ll t \ll J^4$, all holes in the
ground state have flux quantum numbers $h = 0$, and therefore all
quartic terms in Eq.~(\ref{eq-gas-int-H-1}) have $h \equiv h_{1,2} =
0$. In the region around zero momentum occupied by holes, the
coefficients $\beta_{0, q} (\mathbf{k})$ for the two sublattices $A$
and $B$ are
\begin{eqnarray}
\beta_{0, 0}^A (\mathbf{k}) &=& \frac{1} {\sqrt{2}} \, , \quad
\beta_{0, 0}^B (\mathbf{k}) = \frac{1} {\sqrt{2}} \, e^{i k_Y / 4},
\nonumber \\
\beta_{0, 1}^A (\mathbf{k}) &=& \frac{1} {\sqrt{2}} \, , \quad
\beta_{0, 1}^B (\mathbf{k}) = \frac{1} {\sqrt{2}} \, e^{-i k_Y / 2}.
\label{eq-gas-int-1-beta}
\end{eqnarray}
Furthermore, the total hole density is $\rho = \rho_{0, 0} +
\rho_{0, 1}$ in terms of the partial hole densities $\rho_{h, q}$.

It is instructive to first consider hole interactions that conserve
the fermion quantum numbers $q_{1,2} = q_{1,2}'$ and are also
independent of them. Hole interactions of this type include the
Coulomb repulsion and the first attraction mechanism of
Sec.~\ref{sec-stat-bound}. In this case, each quartic term in
Eq.~(\ref{eq-gas-int-H-2}) with $\mathbf{k}' = \mathbf{0}$ is a
product of two number operators. Keeping only the terms with
$\mathbf{k}' = \mathbf{0}$, using that $\Upsilon_{0, \hat{q}}
(\mathbf{k}) = 1/4$ for all such terms, and summing over the quantum
numbers $q_{1,2}$ and $p_{1,2}$, the mean-field decomposition
becomes
\begin{eqnarray}
\Delta \tilde{H}_a &=& \frac{\Delta \Gamma_0} {4N} \sum_{q_1, q_2}
\sum_{\mathbf{k}_1, \mathbf{k}_2} \Big{[} \langle \tilde{n}_{0, q_1}
(\mathbf{k}_1) \rangle \, \tilde{n}_{0, q_2} (\mathbf{k}_2)
\nonumber \\
&& + \, \langle \tilde{n}_{0, q_2} (\mathbf{k}_2) \rangle \,
\tilde{n}_{0, q_1} (\mathbf{k}_1) \Big{]}.
\label{eq-gas-int-1-mft-1}
\end{eqnarray}
Since $\sum_q \sum_{\mathbf{k}} \langle \tilde{n}_{0, q}
(\mathbf{k}) \rangle = 2N \rho$ for both equivalent terms inside its
square brackets, Eq.~(\ref{eq-gas-int-1-mft-1}) reduces to
\begin{equation}
\Delta \tilde{H}_a = \Delta \Gamma_0 \, \rho \sum_q
\sum_{\mathbf{k}} \tilde{n}_{0, q} (\mathbf{k}).
\label{eq-gas-int-1-mft-2}
\end{equation}
From a comparison between Eqs.~(\ref{eq-gas-free-H-2}) and
(\ref{eq-gas-int-1-mft-2}), we conclude that the single-hole
energies for $h = 0$ and $q = \{ 0,1 \}$ are renormalized by $\Delta
\Lambda_{0, q} (\mathbf{k}) = \Delta \Gamma_0 \, \rho$. Since this
energy depends only on the total hole density $\rho$, it corresponds
to a constant shift for all the single-hole energies. This means
that the results for the ground state in Sec.~\ref{sec-gas-free} are
not affected by hole interactions of this type.

\begin{figure}[t!]
\centering
\includegraphics[width=6.3cm]{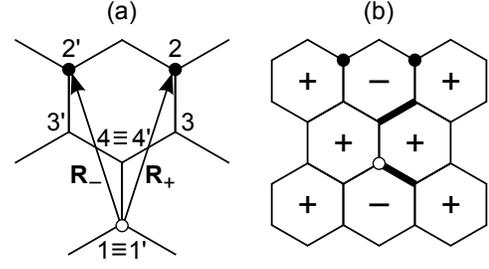}
\caption{(a) Site labeling convention around two interacting holes
(white and black dots) at the relative lattice positions
$\mathbf{R}_{\pm}$. (b) Bond fermion sector around the same two
holes when fluxes are bound to them: it is labeled with the excited
bond fermions (thick lines) and the corresponding plaquette operator
eigenvalues ($\pm 1$). \label{fig-13}}
\end{figure}

Importantly, the second attraction mechanism of
Sec.~\ref{sec-stat-bound} switches the fermion quantum numbers
$q_{1,2} = 1 - q_{1,2}'$. It is therefore represented by quartic
terms in Eq.~(\ref{eq-gas-int-H-2}) where either $q_1 = q_2$ and
$q_1' = q_2'$ or $q_1 = q_2'$ and $q_1' = q_2$. In the first case,
the quartic term is never a product of two number operators, while
in the second case, it is a product of two number operators when
$\mathbf{k}' = \mathbf{k}_2 - \mathbf{k}_1$. According to the
discussion in Sec.~\ref{sec-stat-bound}, this interaction has the
largest strength $|\Delta \Gamma_0| = 1$ when the two holes are at
neighboring sites connected by a $z$ bond, or equivalently, at a
relative lattice position $\mathbf{R}_z = (0, 1)$. However, the two
holes in this case have a mutual hole fermion, and therefore their
fermion quantum numbers become ill-defined. If we require the
fermion quantum numbers to be well-defined, the interaction has the
largest strength $|\Delta \Gamma_0| = J$ when the two holes are at
the relative lattice positions $\mathbf{R}_{\pm} = (\pm \sqrt{3} /
2, 5 / 2)$ shown in Fig.~\ref{fig-13}. Setting $q_1 = q_2' =
\tilde{q}$ and $q_1' = q_2 = 1 - \tilde{q}$, keeping only the terms
with $\mathbf{k}' = \mathbf{k}_2 - \mathbf{k}_1$ for both relative
positions $\mathbf{R}_{\pm}$, and summing over $\tilde{q}$ and
$p_{1,2}$, the mean-field decomposition becomes
\begin{eqnarray}
\Delta \tilde{H}_a &=& \frac{1} {N} \sum_{\tilde{q}}
\sum_{\mathbf{k}_1, \mathbf{k}_2} \sum_{\pm} \Delta \Gamma_0
(\tilde{q}, \pm) \, \Upsilon_{0, \hat{q}} (\mathbf{k}) \, e^{i
(\mathbf{k}_2 - \mathbf{k}_1) \cdot \mathbf{R}_{\pm}}
\nonumber \\
&& \times \Big{[} \langle \tilde{n}_{0, \tilde{q}} (\mathbf{k}_1)
\rangle \, \tilde{n}_{0, 1 - \tilde{q}} (\mathbf{k}_2) + \langle
\tilde{n}_{0, 1 - \tilde{q}} (\mathbf{k}_2) \rangle \, \tilde{n}_{0,
\tilde{q}} (\mathbf{k}_1) \Big{]}, \nonumber
\end{eqnarray}
\begin{equation}
\Upsilon_{0, \hat{q}} (\mathbf{k}) = \Big\{ \begin{array}{c}
\frac{1}{4} e^{i (\mathbf{k}_1 + 2 \mathbf{k}_2) \cdot \mathbf{R}_z
/ 4} \qquad
(\tilde{q} = 0) \quad \,\, \\
\frac{1}{4} e^{-i (2 \mathbf{k}_1 + \mathbf{k}_2) \cdot \mathbf{R}_z
/ 4} \quad \,\,\, (\tilde{q} = 1). \quad \end{array}
\label{eq-gas-int-1-mft-3}
\end{equation}
The four interaction strengths $\Delta \Gamma_0 (\tilde{q}, \pm)$
can be obtained by treating the Ising interactions $-J \sigma_l^x
\sigma_{x(l)}^x$ and $-J \sigma_l^y \sigma_{y(l)}^y$ as
perturbations around the isolated dimer limit. Using the site
labeling convention in Fig.~\ref{fig-13}, the interaction strength
in the case of $\tilde{q} = 0$ for the relative position
$\mathbf{R}_{+}$ is
\begin{eqnarray}
\Delta \Gamma_0 (0, +) &=& +J \langle 0 | f_1 (i b_3^x c_3) (i b_4^x
c_4) f_3^{\dag} | 0 \rangle \label{eq-gas-int-1-str} \\
&=& -i J \langle 0 | \hat{u}_{3,4} f_1 c_3 c_4 f_3^{\dag} | 0
\rangle = - J u_{3,4} = -J, \nonumber
\end{eqnarray}
where an additional minus sign arises because the corresponding
quartic term in Eq.~(\ref{eq-gas-int-H-1}) does not only transfer
the bound matter fermion from $\mathbf{R}_2$ to $\mathbf{R}_1$ but
also exchanges the two bare holes at $\mathbf{R}_{1,2}$. Note that
$u_{l, \alpha(l)} = +1$ for all bonds because no bond fermions are
excited. The interaction strength in the other case $\tilde{q} = 1$
is the Hermitian conjugate of Eq.~(\ref{eq-gas-int-1-str}), while
that for the other relative position $\mathbf{R}_{-}$ is equivalent
to it via site relabeling, and thus $\Delta \Gamma_0 (\tilde{q},
\pm) = -J$ in both cases and for both relative positions. On the
other hand, this implies that the mean-field decomposition in
Eq.~(\ref{eq-gas-int-1-mft-3}) becomes
\begin{eqnarray}
\Delta \tilde{H}_a &=& -\frac{J} {N} \sum_{\mathbf{k}_1,
\mathbf{k}_2} \cos \left[ \textstyle \frac{\sqrt{3}} {2} \tilde{k}_X
\right] \cos \left[ \textstyle \frac{9} {4}
k_{1,Y} - 3 k_{2,Y} \right] \nonumber \\
&& \times \Big{[} \langle \tilde{n}_{0, 0} (\mathbf{k}_1) \rangle \,
\tilde{n}_{0, 1} (\mathbf{k}_2) + \langle \tilde{n}_{0, 1}
(\mathbf{k}_2) \rangle \, \tilde{n}_{0, 0} (\mathbf{k}_1) \Big{]},
\nonumber \\
\label{eq-gas-int-1-mft-4}
\end{eqnarray}
where $\tilde{\mathbf{k}} = (\tilde{k}_X, \tilde{k}_Y) \equiv
\mathbf{k}_1 - \mathbf{k}_2$ is the relative lattice momentum. Since
the original single-hole energies $\Lambda_{0, q} (\mathbf{k})$ in
Eq.~(\ref{eq-gas-free-disp}) and their renormalizations $\Delta
\Lambda_{0, q} (\mathbf{k})$ resulting from
Eq.~(\ref{eq-gas-int-1-mft-4}) are both minimal for
$\mathbf{k}_{1,2} = \mathbf{0}$, the smallest renormalized
single-hole energies $\Lambda_{0, q}' (\mathbf{k}) \equiv
\Lambda_{0, q} (\mathbf{k}) + \Delta \Lambda_{0, q} (\mathbf{k})$
are obtained if holes occupy the single-hole states around zero
momentum. By approximating $\sum_{\mathbf{k}} \psi (\mathbf{k})
\langle \tilde{n}_{0, q} (\mathbf{k}) \rangle$ with $\psi (\langle
\mathbf{k} \rangle) \sum_{\mathbf{k}} \langle \tilde{n}_{0, q}
(\mathbf{k}) \rangle$ for any function $\psi (\mathbf{k})$ in terms
of the respective central momenta $\langle \mathbf{k}_{1,2} \rangle
= \mathbf{0}$, and making use of $\sum_{\mathbf{k}} \langle
\tilde{n}_{0, q} (\mathbf{k}) \rangle = 2N \rho_{0, q}$,
Eq.~(\ref{eq-gas-int-1-mft-4}) reduces to
\begin{eqnarray}
\Delta \tilde{H}_a &=& -2J \rho_{0, 1} \sum_{\mathbf{k}} \cos \left[
\textstyle \frac{\sqrt{3}} {2} k_X \right] \cos \left[ \textstyle
\frac{9} {4} k_Y \right] \tilde{n}_{0, 0} (\mathbf{k}) \quad
\nonumber \\
&& - 2J \rho_{0, 0} \sum_{\mathbf{k}} \cos \left[ \textstyle
\frac{\sqrt{3}} {2} k_X \right] \cos \big{[} 3 k_Y \big{]} \,
\tilde{n}_{0, 1} (\mathbf{k}). \nonumber \\
\label{eq-gas-int-1-mft-5}
\end{eqnarray}
From a comparison between Eqs.~(\ref{eq-gas-free-H-2}) and
(\ref{eq-gas-int-1-mft-5}), we conclude that the single-hole
energies for $h = 0$ and $q = \{ 0,1 \}$ are approximately
renormalized by $\Delta \Lambda_{0, q} (\mathbf{0}) = -2J \rho_{0, 1
- q}$ in the region around zero momentum. If we keep only the
leading-order terms in $\rho \ll 1$, the average single-hole
energies in Table \ref{table-8} are then given by $\langle
\Lambda_{0, 0}' (\mathbf{k}) \rangle = -2t - 2J \rho_{0, 1}$ and
$\langle \Lambda_{0, 1}' (\mathbf{k}) \rangle = -t - 2J \rho_{0,
0}$.

Assuming that the results for the ground state in
Sec.~\ref{sec-gas-free} remain applicable so that $\rho_{0, 0} =
\rho$ and $\rho_{0, 1} = 0$, the two average single-hole energies
$\langle \Lambda_{0, q}' (\mathbf{k}) \rangle$ become equal at the
critical hole density $\rho = \rho_C = t / 2J$. At subcritical
densities $\rho < \rho_C$, we find that $\langle \Lambda_{0, 0}'
(\mathbf{k}) \rangle < \langle \Lambda_{0, 1}' (\mathbf{k}) \rangle$
for all possible values of the partial densities $\rho_{0, q}$. This
means that the ground-state values are $\rho_{0, 0} = \rho$ and
$\rho_{0, 1} = 0$, and that the results in Sec.~\ref{sec-gas-free}
indeed remain applicable. At supercritical densities $\rho >
\rho_C$, there are equilibrium values of the partial densities
$\rho_{0, q}$ at which $\langle \Lambda_{0, 0}' (\mathbf{k}) \rangle
= \langle \Lambda_{0, 1}' (\mathbf{k}) \rangle$. By solving $\langle
\Lambda_{0, 0}' (\mathbf{k}) \rangle = \langle \Lambda_{0, 1}'
(\mathbf{k}) \rangle$ and $\rho = \rho_{0, 0} + \rho_{0, 1}$ for the
two unknowns $\rho_{0, q}$, the ground-state values are
\begin{equation}
\rho_{0, 0} = \frac{1}{2} \left( \rho + \rho_C \right), \quad \,\,
\rho_{0, 1} = \frac{1}{2} \left( \rho - \rho_C \right).
\label{eq-gas-int-1-dens}
\end{equation}
To summarize, only holes with $h = 0$ and $q = 0$ are present in the
low-density limit $\rho \rightarrow 0$, while holes with $h = 0$ and
$q = 1$ appear above the critical density $\rho = \rho_C$. Note that
$\rho_C = t / 2J$ is small due to $t \ll J^4$.

The subcritical and the supercritical regimes are also distinct in
terms of their physical properties. At subcritical densities, the
physical properties of the model are as discussed in
Sec.~\ref{sec-gas-free}, except for a renormalization of the
effective masses and hence the electrical conductivities. At
supercritical densities, the physical properties are changed in an
essential way by the presence of holes with $h = 0$ and $q = 1$.
Since these holes are bosons, they all condense into the
lowest-energy single-hole state at zero momentum. This condensation
then leads to charged superfluid behavior in the presence of the
Coulomb repulsion. Furthermore, due to the coherent condensation of
both $p = 0$ holes and $p = 1$ holes, the model spontaneously
develops a net magnetization.

\subsubsection{Second regime: $t \ll J^8$} \label{sec-gas-int-2}

In the second regime with $t \ll J^8$, all holes in the ground state
have flux quantum numbers $h = 1$, and therefore all quartic terms
in Eq.~(\ref{eq-gas-int-H-1}) have $h \equiv h_{1,2} = 1$. The
coefficients $\beta_{1, q} (\mathbf{k})$ for the four sublattices
$A'$, $B'$, $A''$, and $B''$ are
\begin{eqnarray}
\beta_{1, 0}^{A'} (\mathbf{k}) &=& \beta_{1, 1}^{A''} (\mathbf{k}) =
\xi_1 (k_X), \nonumber \\
\beta_{1, 0}^{B'} (\mathbf{k}) &=& \beta_{1, 1}^{B''} (\mathbf{k}) =
\xi_1 (k_X) \, e^{i k_Y}, \label{eq-gas-int-2-beta} \\
\beta_{1, 0}^{A''} (\mathbf{k}) &=& \beta_{1, 1}^{A'} (\mathbf{k}) =
\xi_2 (k_X) \, e^{3i k_Y / 2}, \nonumber \\
\beta_{1, 0}^{B''} (\mathbf{k}) &=& \beta_{1, 1}^{B'} (\mathbf{k}) =
\xi_2 (k_X) \, e^{-i k_Y / 2}, \nonumber
\end{eqnarray}
where $\xi_{1,2} (k_X) \in \mathbb{R}$, and $\xi_1^2 (k_X) + \xi_2^2
(k_X) = 1/2$. Furthermore, the total hole density is $\rho =
\rho_{1, 0} + \rho_{1, 1}$ in terms of the partial hole densities
$\rho_{h, q}$.

We first notice that the mean-field decomposition of a hole
interaction that is independent of the conserved fermion quantum
numbers $q_{1,2}$ no longer takes the form of
Eq.~(\ref{eq-gas-int-1-mft-2}). Since $\Upsilon_{1, \hat{q}}
(\mathbf{k})$ is not $1/4$ for all quartic terms with $\mathbf{k}' =
\mathbf{0}$ in Eq.~(\ref{eq-gas-int-H-2}), the renormalizations
$\Delta \Lambda_{1, q} (\mathbf{k})$ of the single-hole energies
become dependent on the individual partial hole densities $\rho_{1,
q}$. If we consider the Coulomb repulsion and the first attraction
mechanism of Sec.~\ref{sec-stat-bound} for all relative lattice
positions $\mathbf{R}$, the single-hole energies in
Eq.~(\ref{eq-gas-free-H-2}) are renormalized by $\Delta \Lambda_{1,
q} (\mathbf{k}) = \Delta \Gamma_0' \, \rho_{1, q} + \Delta
\Gamma_0'' \, \rho_{1, 1 - q}$, where the exact values of $\Delta
\Gamma_0'$ and $\Delta \Gamma_0''$ depend on the detailed form of
the Coulomb repulsion. In the case of $\Delta \Gamma_0' > \Delta
\Gamma_0''$, the partial hole densities remain $\rho_{1, 0} =
\rho_{1, 1} = \rho / 2$, while in the case of $\Delta \Gamma_0' <
\Delta \Gamma_0''$, the partial hole densities become either
$\rho_{1, 0} = \rho$ and $\rho_{1, 1} = 0$ or $\rho_{1, 0} = 0$ and
$\rho_{1, 1} = \rho$. We assume the first case in the following so
that there are equal densities of $q = 0$ holes and $q = 1$ holes in
the ground state.

\begin{figure}[t!]
\centering
\includegraphics[width=6.7cm]{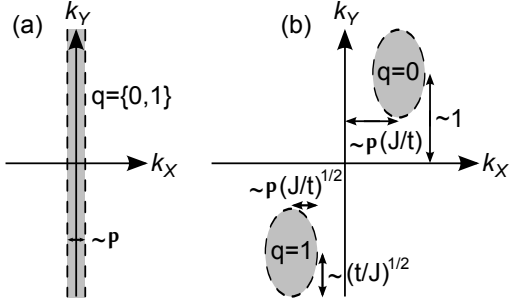}
\caption{Single-hole states occupied by holes with quantum numbers
$h = 1$ and $q = \{ 0,1 \}$ in the non-interacting treatment (a) and
in the interacting treatment (b). Fermi sea states of constant
occupation are marked by gray shading, while the Fermi surface
separating occupied and unoccupied states is marked by a dashed
line. \label{fig-14}}
\end{figure}

For the second attraction mechanism of Sec.~\ref{sec-stat-bound} at
the relative lattice positions $\mathbf{R}_{\pm}$, the mean-field
decomposition takes the form of Eq.~(\ref{eq-gas-int-1-mft-3}) with
$\Upsilon_{1, \hat{q}} (\mathbf{k}) = \frac{1}{4} \Xi (k_X) e^{i
(\mathbf{k}_2 - \mathbf{k}_1) \cdot \mathbf{R}_z / 2}$ and $\Xi
(k_X) = 16 \xi_1 (k_{1,X}) \xi_2 (k_{1,X}) \xi_1 (k_{2,X}) \xi_2
(k_{2,X})$. Since Fig.~\ref{fig-13} shows that $u_{3,4} = -1$ and
$u_{3',4'} = +1$, the interaction strengths are $\Delta \Gamma_0
(\tilde{q}, \pm) = \pm J$. On the other hand, this implies that
mean-field decomposition becomes
\begin{eqnarray}
\Delta \tilde{H}_a &=& -\frac{J} {N} \sum_{\mathbf{k}_1,
\mathbf{k}_2} \Xi (k_X) \, \sin \left[ \textstyle \frac{\sqrt{3}}
{2} \tilde{k}_X \right] \sin \big{[} 3 \tilde{k}_Y \big{]}
\label{eq-gas-int-2-mft-1} \\
&& \times \Big{[} \langle \tilde{n}_{1, 0} (\mathbf{k}_1) \rangle \,
\tilde{n}_{1, 1} (\mathbf{k}_2) + \langle \tilde{n}_{1, 1}
(\mathbf{k}_2) \rangle \, \tilde{n}_{1, 0} (\mathbf{k}_1) \Big{]}.
\nonumber
\end{eqnarray}
Unlike the original single-hole energies $\Lambda_{1, q}
(\mathbf{k})$ in Eq.~(\ref{eq-gas-free-disp}), their
renormalizations $\Delta \Lambda_{1, q} (\mathbf{k})$ resulting from
Eq.~(\ref{eq-gas-int-2-mft-1}) are not minimal for $\mathbf{k}_1 =
\mathbf{k}_2 = \mathbf{0}$. We therefore need to determine the
ground-state occupations of the single-hole states that correspond
to the smallest renormalized single-hole energies $\Lambda_{1, q}'
(\mathbf{k}) \equiv \Lambda_{1, q} (\mathbf{k}) + \Delta \Lambda_{1,
q} (\mathbf{k})$. Exploiting the equivalence between $q = 0$ holes
at momenta $\mathbf{k}_1$ and $q = 1$ holes at momenta
$\mathbf{k}_2$, and noticing that both sine factors in
Eq.~(\ref{eq-gas-int-2-mft-1}) depend only on the relative momentum
$\tilde{\mathbf{k}} \equiv \mathbf{k}_1 - \mathbf{k_2}$, we conclude
that the respective central momenta are related by $\mathbf{K} =
(K_X, K_Y) \equiv \langle \mathbf{k}_1 \rangle = - \langle
\mathbf{k}_2 \rangle$, and minimize the single-hole energies with
respect to $\mathbf{K}$. Since $\Lambda_{1, q} (\mathbf{k})$ does
not depend on the momentum component $k_Y$, the second sine factor
in Eq.~(\ref{eq-gas-int-2-mft-1}) can be maximized independently. In
particular, its maximum $\sin [3 \langle \tilde{k}_Y \rangle] = +1$
corresponds to the ground-state value $K_Y = \pi / 12$. Furthermore,
if we assume $K_X \ll 1$, the first sine factor is approximately
$\sqrt{3} K_X$. Due to $\sum_{\mathbf{k}} \langle \tilde{n}_{1, q}
(\mathbf{k}) \rangle = 2N \rho_{1, q} = N \rho$, the single-hole
energies around the central momenta $\langle \mathbf{k}_{1,2}
\rangle = \pm \mathbf{K}$ are then renormalized by $\Delta
\Lambda_{1, q} (\mathbf{k}) \sim - J \rho \, K_X$, and the average
single-hole energies in Table \ref{table-8} take the form
\begin{equation}
\langle \Lambda_{1, q}' (\mathbf{k}) \rangle = C' (t, J) - \kappa_0
J \rho \, K_X + \frac{3} {4 \sqrt{5}} \, t \, K_X^2,
\label{eq-gas-int-2-lambda}
\end{equation}
where $C' (t, J)$ and $\kappa_0 \sim 1$ are independent of $K_X$.
The minimum of $\langle \Lambda_{1, q}' (\mathbf{k}) \rangle$ with
respect to $K_X$ corresponds to the ground-state value $K_X \sim J
\rho / t$. By approximating $\sum_{\mathbf{k}} \psi (\mathbf{k})
\langle \tilde{n}_{1, q} (\mathbf{k}) \rangle$ with $\psi (\langle
\mathbf{k} \rangle) \sum_{\mathbf{k}} \langle \tilde{n}_{1, q}
(\mathbf{k}) \rangle$ for any function $\psi (\mathbf{k})$ in terms
of the central momenta $\langle \mathbf{k}_{1,2} \rangle = \pm
\mathbf{K}$, and assuming $K_X \sim J \rho / t \ll 1$,
Eq.~(\ref{eq-gas-int-2-mft-1}) reduces to
\begin{eqnarray}
\Delta \tilde{H}_a &=& -\frac{J^2 \rho^2} {t} \sum_q
\sum_{\mathbf{k}} \tilde{\Xi} \big{[} k_X - (-1)^q K_X \big{]}
\nonumber \\
&& \times \, \cos \big{\{} 3 [k_Y - (-1)^q K_Y] \big{\}} \,
\tilde{n}_{1, q} (\mathbf{k}), \label{eq-gas-int-2-mft-2}
\end{eqnarray}
where $\tilde{\Xi} [k_X - (-1)^q K_X] \sim 1$ contains all
dependence on the momentum component $k_X$. Importantly, the
renormalized single-hole energies $\Lambda_{1, q}' (\mathbf{k})$
resulting from Eq.~(\ref{eq-gas-int-2-mft-2}) depend on the momentum
component $k_Y$ as well. In fact, the single-hole dispersion
relations for $q = \{ 0,1 \}$ holes around their respective central
momenta $\pm \mathbf{K}$ are quadratic in both the $k_X$ and the
$k_Y$ directions: the leading-order terms are $\sim t (k_X \mp
K_X)^2$ and $\sim J^2 \rho^2 (k_Y \mp K_Y)^2 / t$. This implies that
the Fermi seas for the two hole types are ellipses of half-axes
$\Delta k_X \sim \rho \sqrt{J / t}$ and $\Delta k_Y \sim \sqrt{t /
J}$ centered at $\pm \mathbf{K}$. Note that $\Delta k_Y \ll 1$ due
to $t \ll J^8$ and that $\Delta k_X \sim K_X \, \Delta k_Y \ll 1$
due to $K_X \ll 1$ and $\Delta k_Y \ll 1$. Since our calculation
resulting in these Fermi ellipses is valid for any hole density
$\rho > 0$, the Fermi strip described in Sec.~\ref{sec-gas-free} is
unstable against an arbitrarily small hole interaction. The Fermi
ellipses of the interacting treatment and the Fermi strip of the
non-interacting treatment are contrasted in Fig.~\ref{fig-14}.

In terms of physical properties, the main difference with respect to
the results in Sec.~\ref{sec-gas-free} is a finite electrical
conductivity in the $Y$ direction. The conductivities in the $X$ and
$Y$ directions are still calculated by Eq.~(\ref{eq-gas-free-cond}),
except that we use the renormalized single-hole energies
$\Lambda_{1, q}' (\mathbf{k})$ and take their second derivatives at
the central momenta $\pm \mathbf{K}$. Since the partial hole
densities are $\rho_{1, q} = \rho / 2$, and the second derivatives
are $\partial^2 \Lambda_{1, q}' (\mathbf{k}) /
\partial k_X^2 \sim t$ and $\partial^2 \Lambda_{1, q}' (\mathbf{k})
/ \partial k_Y^2 \sim J^2 \rho^2 / t$, the conductivities in the two
directions become
\begin{equation}
\sigma_X^{*} \sim t \rho e_{*}^2 \tau, \quad \,\, \sigma_Y^{*} \sim
\frac{J^2 \rho^3 e_{*}^2 \tau} {t} \, . \label{eq-gas-int-2-cond}
\end{equation}
Note in particular that $\sigma_X^{*} \propto \rho$ and
$\sigma_Y^{*} \propto \rho^3$. Since the ratio of the two
conductivities is $\sigma_Y^{*} / \sigma_X^{*} \sim (J \rho / t)^2
\ll 1$, the model has a strong conductivity anisotropy that becomes
weaker as we increase the hole density $\rho$.

\section{Mobile holes beyond slow hopping} \label{sec-fast}

By relaxing the condition of slow hopping, we qualitatively describe
the Kitaev honeycomb model with mobile holes in the regimes of
intermediate hopping ($J^4 \ll t \ll 1$) and fast hopping ($t \gg
1$). We first consider a single isolated hole and investigate the
applicability of the internal quantum numbers $h$, $q$, and $p$.
Since the original definitions of these quantum numbers in
Sec.~\ref{sec-stat-deg} are in terms of the internal modes only,
they are not applicable beyond the limit of slow hopping when the
excitations in the bulk modes can no longer be neglected. In the
regime of intermediate hopping when $t \gg E_P \sim J^4$, the bulk
flux excitations are no longer negligible, and the hole is
surrounded by a cloud of fluctuating fluxes. In the regime of fast
hopping when $t \gg E_f \sim 1$, the bulk fermion excitations are no
longer negligible, and the hole is also surrounded by a cloud of
fluctuating fermions. On the other hand, the hole combined with
these excitation clouds has a well-defined superselection sector
that is conserved by the hopping process due to the locality of the
exchange operator $\mathcal{E}_{l,l'}$. This means that the
definitions of the quantum numbers $h$ and $q$ can be generalized in
terms of their correspondence to these conserved superselection
sectors. Furthermore, the only non-trivial terms in the exchange
operator $\mathcal{E}_{l,l'}$ are the Heisenberg terms in
Eq.~(\ref{eq-stat-pert-H}) that conserve the product of dimer
operators $\lambda_l = \sigma_l^z \sigma_{z(l)}^z$ and the products
of plaquette operators $W_P$ both in even stripes and in odd stripes
(see Sec.~\ref{sec-stat-pert}). This means that the definitions of
all quantum numbers $h$, $q$, and $p$ can be generalized in terms of
these products such that they are conserved by the hopping process:
$(-1)^{q + p} = \prod_{l \in A} \lambda_l$, $(-1)^{h + p} = \prod_{P
\in \eta} W_P$, and $(-1)^p = \prod_{P \in \mu} W_P$, where each
product is taken over a sufficiently large region that contains the
clouds of fluctuating fluxes and fermions. The quantum numbers $h$,
$q$, and $p$ are then valid if the distances between holes exceed
the radii of these excitation clouds.

To provide an upper bound on the radius of each excitation cloud, we
notice that the fluctuating fluxes and fermions increase the
potential energy and decrease the kinetic energy of the hole. This
means that the two radii are determined by a balance between the
potential and the kinetic energies. Since the excitation energy of a
bulk flux is $E_P \sim J^4$ and that of a bulk fermion is $E_f \sim
1$, the increase in the potential energy is on the order of $R_P^2
E_P \sim R_P^2 J^4$ for a flux cloud of radius $R_P$ and on the
order of $R_f^2 E_f \sim R_f^2$ for a fermion cloud of radius $R_f$.
On the other hand, since the decrease in the kinetic energy due to
both excitation clouds is at most $\sim t$, the increase in the
potential energy due to either excitation cloud must be bounded by
$\lesssim t$. We therefore conclude that the upper bound on the flux
cloud radius is $R_P \lesssim \sqrt{t / J^4}$ and that on the
fermion cloud radius is $R_f \lesssim \sqrt{t}$.

We are now ready to investigate the ground-state quantum numbers $h$
and $q$ for a finite density of mobile holes. For simplicity, we
consider the case of $J^4 \ll t \ll 1$ when only the plauqettes are
fluctuating around the holes. To ensure that the quantum numbers $h$
and $q$ are valid, we assume a small hole density $\rho \ll
R_P^{-2}$ and neglect any hole interactions. Since the hopping
matrix elements are independent of the quantum numbers $p$, we also
set $p = q$ for each hole without loss of generality. Due to the
lack of broken dimers in the isolated dimer limit, the plaquette
sector is then conserved by all hopping processes along $z$ bonds.
Since the hopping processes along $x$ and $y$ bonds either flip no
plaquettes or two plaquettes in each of two neighboring stripes (see
Fig.~\ref{fig-6}), this implies that the number of excited
plaquettes has a conserved parity in each stripe. If there are an
odd number of excited plaquettes in any stripe around any hole, the
given hole can hop in the $Y$ direction only if it leaves behind an
excited plaquette in the given stripe. Since the kinetic energy
decreases by $\sim t$ for each hole that can hop in the $Y$
direction, this process happens spontaneously for $t \gg J^4$, and
there remain an even number of excited plaquettes in each stripe
around each hole. Due to the relations $\prod_{P \in \eta} W_P =
(-1)^{h + q}$ and $\prod_{P \in \mu} W_P = (-1)^q$, this means that
any hole with quantum numbers other than $h = 0$ and $q = 0$ is
unstable against a spontaneous decay into a hole with quantum
numbers $h = 0$ and $q = 0$. Note that this result remains valid
away from the isolated dimer limit and in the case of $t \gg 1$ when
the fermions are also fluctuating. Holes with quantum numbers $h =
0$ and $q = 0$ are then energetically favorable because their
hopping is the least constrained in the $Y$ direction.

In conclusion, the quantum numbers $h$, $q$, and $p$ generalize
beyond the regime of slow hopping, but they are valid only for
smaller hole densities due to the clouds of fluctuating excitations
around holes. Furthermore, any hole with quantum numbers other than
$h = 0$ and $q = 0$ is unstable against a spontaneous decay into a
hole with quantum numbers $h = 0$ and $q = 0$. This means that all
holes in the ground state have quantum numbers $h = 0$ and $q = 0$,
and that the ground state is identical to that in the case of $J^8
\ll t \ll J^4$. The only difference is that there are clouds of
fluctuating excitations around each hole. Importantly, when the hole
density becomes $\rho \gtrsim R_{P,f}^{-2}$, the clouds of
fluctuating fluxes (fermions) around different holes merge, and the
holes hop in an entire lattice of fluctuating fluxes (fermions).

\section{Comparison with mean-field results} \label{sec-comp}

\subsection{Holes in the parton description} \label{sec-comp-part}

We now discuss the relation between our exact results for the Kitaev
honeycomb model with mobile holes and the corresponding mean-field
results in Ref.~\onlinecite{You}. In their description, the physical
operators $\mathbf{c}_{l, \uparrow}^{(\dag)}$ and $\mathbf{c}_{l,
\downarrow}^{(\dag)}$ that annihilate (create) a spin-up and a
spin-down particle at site $l$, respectively, are expressed in terms
of the fermionic spinon operators $\mathbf{f}_{l,
\uparrow}^{(\dag)}$ and $\mathbf{f}_{l, \downarrow}^{(\dag)}$ and
the bosonic holon operators $\mathbf{b}_{l, 1}^{(\dag)}$ and
$\mathbf{b}_{l, 2}^{(\dag)}$. The resulting relations between the
physical operators and the parton (holon and spinon) operators can
be summarized in the matrix form $\mathbf{C}_l = \mathbf{F}_l \cdot
\mathbf{B}_l / \sqrt{2}$, where the physical-operator matrix is
\begin{equation}
\mathbf{C}_l = \left( \begin{array}{cc} \mathbf{c}_{l, \uparrow} &
-\mathbf{c}_{l, \downarrow}^{\dag} \\ \mathbf{c}_{l, \downarrow} &
\mathbf{c}_{l, \uparrow}^{\dag}
\end{array} \right), \label{eq-comp-part-C}
\end{equation}
while the spinon-operator and the holon-operator matrices are
\begin{equation}
\mathbf{F}_l = \left( \begin{array}{cc} \mathbf{f}_{l, \uparrow} &
-\mathbf{f}_{l, \downarrow}^{\dag} \\ \mathbf{f}_{l, \downarrow} &
\mathbf{f}_{l, \uparrow}^{\dag}
\end{array} \right), \quad \,\, \mathbf{B}_l = \left( \begin{array}{cc}
\mathbf{b}_{l, 1}^{\dag} & -\mathbf{b}_{l, 2} \\
\mathbf{b}_{l, 2}^{\dag} & \mathbf{b}_{l, 1} \end{array} \right).
\label{eq-comp-part-FB}
\end{equation}
Importantly, the physical-operator matrix $\mathbf{C}_l$ is
invariant under the combined gauge transformation $\mathbf{F}_l
\rightarrow \mathbf{F}_l \cdot \mathbf{G}_l$ and $\mathbf{B}_l
\rightarrow \mathbf{G}_l \cdot \mathbf{B}_l$ for any $\mathbf{G}_l
\in$ SU(2). Since a physical state should also be invariant under
such an SU(2) gauge transformation at any site $l$, it must satisfy
$\mathcal{K}_l^{\alpha} = 0$ for all SU(2) generators
$\mathcal{K}_l^{\alpha}$ with $\alpha = \{ x,y,z \}$. If the spinon
operators are related to the Majorana fermions introduced in
Sec.~\ref{sec-kit-ferm} by $\mathbf{f}_{l, \uparrow} = (c_l + i
b_l^z) / 2$ and $\mathbf{f}_{l, \downarrow} = (i b_l^x - b_l^y) /
2$, these SU(2) generators take the form [see Eq.~(20) in
Ref.~\onlinecite{You}]
\begin{equation}
\mathcal{K}_l^{\alpha} = \frac{i}{4} b_l^{\alpha} c_l - \frac{i}{8}
\sum_{\alpha_1, \alpha_2} \tilde{\epsilon}_{\alpha \alpha_1
\alpha_2} b_l^{\alpha_1} b_l^{\alpha_2} - \frac{1}{2} \sum_{\zeta_1,
\zeta_2} \mathbf{b}_{l, \zeta_1} \tilde{\sigma}_{\zeta_1
\zeta_2}^{\alpha} \mathbf{b}_{l, \zeta_2}^{\dag},
\label{eq-comp-part-K}
\end{equation}
where $\tilde{\sigma}^{\alpha}$ are the Pauli matrices,
$\tilde{\epsilon}$ is the completely antisymmetric tensor, and the
summations are over $\alpha_{1,2} = \{ x,y,z \}$ and $\zeta_{1,2} =
\{ 1,2 \}$. For a single site $l$, there are only three physical
states: the empty hole state $| \times_l \rangle$, the spin-up
particle state $| \uparrow_l \rangle$, and the spin-down particle
state $| \downarrow_l \rangle$. The projection of any state in the
parton description onto the physical subspace with
$\mathcal{K}_l^{\alpha} = 0$ is then a superposition of $| \times_l
\rangle$, $| \uparrow_l \rangle$, and $| \downarrow_l \rangle$. In
terms of the parton operators, these three physical states are given
by [see Eq.~(18) in Ref.~\onlinecite{You}]
\begin{eqnarray}
| \times_l \rangle &=& \frac{1} {\sqrt{2}} \left( \mathbf{b}_{l,
1}^{\dag} + \mathbf{b}_{l, 2}^{\dag} \mathbf{f}_{l, \uparrow}^{\dag}
\mathbf{f}_{l, \downarrow}^{\dag} \right) | \mathbf{0}_l \rangle,
\nonumber \\
| \uparrow_l \rangle &=& \mathbf{c}_{l, \uparrow}^{\dag} | \times_l
\rangle = \mathbf{f}_{l, \uparrow}^{\dag} | \mathbf{0}_l \rangle,
\label{eq-comp-part-state} \\
| \downarrow_l \rangle &=& \mathbf{c}_{l, \downarrow}^{\dag} |
\times_l \rangle = \mathbf{f}_{l, \downarrow}^{\dag} | \mathbf{0}_l
\rangle, \nonumber
\end{eqnarray}
where $| \mathbf{0}_l \rangle$ is the vacuum of the parton operators
that is defined by $\mathbf{f}_{l, \uparrow} | \mathbf{0}_l \rangle
= \mathbf{f}_{l, \downarrow} | \mathbf{0}_l \rangle = 0$ and
$\mathbf{b}_{l, 1} | \mathbf{0}_l \rangle = \mathbf{b}_{l, 2} |
\mathbf{0}_l \rangle = 0$. Note that these three states are indeed
physical because they satisfy $\mathcal{K}_l^{\alpha} | \times_l
\rangle = \mathcal{K}_l^{\alpha} | \uparrow_l \rangle =
\mathcal{K}_l^{\alpha} | \downarrow_l \rangle = 0$ for all $\alpha =
\{ x,y,z \}$.

Before investigating the mean-field treatment, we consider a single
stationary hole and aim to make a connection between its parton
description and its internal quantum numbers $h$, $q$, and $p$. In
the isolated dimer limit, there is an effective two-site system
around the hole consisting of the hole site $l$ and the neighboring
site $l' = z(l)$. We assume $l \in A$ without loss of generality.
Since there is a hole at site $l$ and there is no hole at site $l'$,
one holon is excited at site $l$ and no holon is excited at site
$l'$. Due to the four spinons at sites $l$ and $l'$ that are either
excited or not and the two holons at site $l$ from which exactly one
is excited, the Hilbert space of the two-site system in the parton
description is then $32$ dimensional. However, the physical Hilbert
space of the two-site system is only $2$ dimensional because it is
spanned by the two physical states $| \times_l \rangle \otimes |
\uparrow_{l'} \rangle$ and $| \times_l \rangle \otimes |
\downarrow_{l'} \rangle$. This means that the projection of any
state in the parton description onto the subspace with
$\mathcal{K}_l^{\alpha} = \mathcal{K}_{l'}^{\alpha} = 0$ is a
superposition of these two physical states. Since the effective
Hamiltonian of the two-site system is $H = -b_l^z b_{l'}^z c_l
c_{l'}$, its ground state has expectation values $\langle i b_l^z
b_{l'}^z \rangle = \langle -i c_l c_{l'} \rangle = \pm 1$. In fact,
there are $16$ such ground states in the parton description that
take the form
\begin{eqnarray}
\big| \Psi_{\zeta, \pm}^{r_1, r_2} \big{\rangle} &=& \left( 1 \mp i
\mathbf{f}_{l, \uparrow}^{\dag} \mathbf{f}_{l', \uparrow}^{\dag}
\right) (\mathbf{f}_{l, \downarrow}^{\dag})^{r_1} (\mathbf{f}_{l',
\downarrow}^{\dag})^{r_2} \mathbf{b}_{l, \zeta}^{\dag} \big{(} |
\mathbf{0}_l \rangle \otimes | \mathbf{0}_{l'} \rangle \big{)}
\nonumber \\
&=& \left[ (\mathbf{f}_{l, \downarrow}^{\dag})^{r_1} \mathbf{b}_{l,
\zeta}^{\dag} | \mathbf{0}_l \rangle \right] \otimes \left[
(\mathbf{f}_{l', \downarrow}^{\dag})^{r_2} | \mathbf{0}_{l'} \rangle
\right] \label{eq-comp-part-gs} \\
&& \mp \, i \left[ \mathbf{f}_{l, \uparrow}^{\dag} (-\mathbf{f}_{l,
\downarrow}^{\dag})^{r_1} \mathbf{b}_{l, \zeta}^{\dag} |
\mathbf{0}_l \rangle \right] \otimes \left[ \mathbf{f}_{l',
\uparrow}^{\dag} (\mathbf{f}_{l', \downarrow}^{\dag})^{r_2} |
\mathbf{0}_{l'} \rangle \right], \nonumber
\end{eqnarray}
where $\zeta = \{ 1,2 \}$ and $r_{1,2} = \{ 0,1 \}$. On the other
hand, the projection of the ground state $| \Psi_{\zeta, \pm}^{r_1,
r_2} \rangle$ onto the subspace with $\mathcal{K}_l^{\alpha} =
\mathcal{K}_{l'}^{\alpha} = 0$ is non-zero only if the overlap of $|
\Psi_{\zeta, \pm}^{r_1, r_2} \rangle$ is non-zero with either of the
two physical states $| \times_l \rangle \otimes | \uparrow_{l'}
\rangle$ or $| \times_l \rangle \otimes | \downarrow_{l'} \rangle$.
For $\zeta = 1$, we must choose $r_1 = 0$ and $r_2 = 1$, in which
case the first term in Eq.~(\ref{eq-comp-part-gs}) has a non-zero
overlap with $| \times_l \rangle \otimes | \downarrow_{l'} \rangle$.
For $\zeta = 2$, we must choose $r_1 = 1$ and $r_2 = 0$, in which
case the second term in Eq.~(\ref{eq-comp-part-gs}) has a non-zero
overlap with $| \times_l \rangle \otimes | \uparrow_{l'} \rangle$.
This means that the choice of exciting either $\mathbf{b}_{l,
1}^{\dag}$ or $\mathbf{b}_{l, 2}^{\dag}$ at the hole site $l$ before
the projection determines the local magnetization at the neighboring
site $l' = z(l)$ after the projection. We therefore conclude that
these two different choices correspond to different plaquette
quantum numbers $p = \{ 0,1 \}$.

Since the two-site system around the hole has only two physical
states that are distinguished by the plaquette quantum number $p$,
the remaining quantum numbers $h$ and $q$ are necessarily determined
by the spinons around the hole site. In the regime of slow hopping,
the definitions of these quantum numbers in Sec.~\ref{sec-stat-deg}
are straightforward to express in terms of the Majorana fermions
$b_l^{\alpha}$ and $c_l$, or equivalently, in terms of the spinon
operators $\mathbf{f}_{l, \uparrow}^{(\dag)}$ and $\mathbf{f}_{l,
\downarrow}^{(\dag)}$. Beyond the regime of slow hopping, the exact
expressions become more complicated, but the general principle
remains the same. For our purposes, it is enough to establish an
intuitive picture from the general principle by using the
interpretation in which the various hole types with different
quantum numbers $h$ and $q$ have different kinds of elementary
excitations bound to them. This interpretation has a simple
translation in the parton description: some of the spinons around
the hole site are bound to the holon at the hole site, and the
quantum numbers $h$ and $q$ are in turn determined by the structure
of these bound spinons.

\subsection{Mean-field treatment of the model} \label{sec-comp-mft}

In the mean-field treatment of Ref.~\onlinecite{You}, the
Hamiltonian of the model is first expressed in terms of the parton
operators and then subjected to an appropriate mean-field
decomposition. As a result of this treatment, the mean-field
Hamiltonian of the model with a small density $\rho \ll 1$ of mobile
holes takes the form [see Eqs.~(24) and (25) in
Ref.~\onlinecite{You}]
\begin{widetext}
\begin{eqnarray}
\tilde{H} &=& \sum_{l \in A} \, \sum_{\alpha = x,y,z} \Bigg{\{}
\bigg{[} J_{\alpha} u_{\alpha}^{\alpha} - \frac{t}{4}
\sum_{\zeta=1}^2 \left( w_{\alpha}^{\zeta} + \mathrm{c.c.} \right)
\bigg{]} \, i c_l c_{\alpha(l)} - \sum_{\alpha' = x,y,z} \bigg{[}
J_{\alpha} \delta_{\alpha, \alpha'} v_{\alpha} + \frac{t}{4}
\sum_{\zeta=1}^2 \left( w_{\alpha}^{\zeta} + \mathrm{c.c.} \right)
\bigg{]} \, i b_l^{\alpha'} b_{\alpha(l)}^{\alpha'}
\nonumber \\
&& + \, \frac{t}{4} \, \bigg{[} v_{\alpha} - \sum_{\alpha' = x,y,z}
u_{\alpha}^{\alpha'} \bigg{]} \, \sum_{\zeta=1}^2 \left[ i
\mathbf{b}_{l, \zeta}^{\dag} \mathbf{b}_{\alpha(l), \zeta} +
\textrm{H.c.} \right] \Bigg{\}} + \sum_l \sum_{\alpha = x,y,z}
\tilde{\beta}_l^{\alpha} \mathcal{K}_l^{\alpha} - \tilde{\mu} \sum_l
\sum_{\zeta=1}^2 \mathbf{b}_{l, \zeta}^{\dag} \mathbf{b}_{l, \zeta},
\label{eq-comp-mft-H}
\end{eqnarray}
\end{widetext}
where $u_{\alpha}^{\alpha'}$, $v_{\alpha}$, and $w_{\alpha}^{\zeta}$
are the respective expectation values of the generalized bond
fermion operators $\hat{u}_{\alpha}^{\alpha'} \equiv i b_l^{\alpha'}
b_{\alpha(l)}^{\alpha'}$, the generalized matter fermion operators
$\hat{v}_{\alpha} \equiv -i c_l c_{\alpha(l)}$, and the holon
coherence operators $\hat{w}_{\alpha}^{\zeta} \equiv i
\mathbf{b}_{l, \zeta}^{\dag} \mathbf{b}_{\alpha(l), \zeta}$. The
constraint $\sum_l \sum_{\zeta} \langle \mathbf{b}_{l, \zeta}^{\dag}
\mathbf{b}_{l, \zeta} \rangle = 2N \rho$ for the total number of
holes is enforced by the chemical potential $\tilde{\mu}$, while the
softened gauge constraint $\langle \mathcal{K}_l^{\alpha} \rangle =
0$ is enforced by the Lagrange multiplier $\tilde{\beta}_l^{\alpha}$
for all $l$ and $\alpha$. Importantly, the mean-field Hamiltonian in
Eq.~(\ref{eq-comp-mft-H}) is an extension of that in
Ref.~\onlinecite{You}. It is applicable to the gapped phase of the
model, where the coupling strengths $J_{\alpha}$ are different: $J_z
= 1$ and $J \equiv J_x = J_y \ll 1$.

The mean-field Hamiltonian in Eq.~(\ref{eq-comp-mft-H}) can be
solved by a self-consistent procedure in terms of the expectation
values $u_{\alpha}^{\alpha'}$, $v_{\alpha}$, and
$w_{\alpha}^{\zeta}$. In the absence of holes ($\rho = 0$), there is
a coupling of strength $J_{\alpha}$ between the bond fermion
expectation value $u_{\alpha}^{\alpha}$ and the matter fermion
expectation value $v_{\alpha}$ along each bond of $\alpha$ type.
Keeping only the lowest-order terms in $J \ll 1$, the
self-consistent solutions for these expectation values are
$u_{\alpha}^{\alpha} = v_z = 1$ and $v_x = v_y = J / 2$. Note that
the same expectation values are obtained from the exact solution of
the model in Sec.~\ref{sec-kit}. In the presence of holes ($\rho >
0$), the holons all condense into their lowest-energy state at zero
momentum, and hence the holon coherence expectation values are
$w_{\alpha}^{\zeta} \sim \rho$. This means that the original terms
$J_{\alpha} u_{\alpha}^{\alpha}$ and $J_{\alpha} \delta_{\alpha,
\alpha'} v_{\alpha}$ in the first two square brackets of
Eq.~(\ref{eq-comp-mft-H}) are in competition with new terms on the
order of $t \rho$. The expectation values $u_{\alpha}^{\alpha}$ and
$v_{\alpha}$ for $\rho > 0$ are then close to those for $\rho = 0$
as long as these new terms are negligible with respect to the
original terms. In particular, the bond fermion expectation values
$u_x^x$ and $u_y^y$ remain close to $1$ as long as $t \rho \ll
J_{x,y} v_{x,y} \sim J^2$, while the bond fermion expectation value
$u_z^z$ and the matter fermion expectation value $v_z$ remain close
to $1$ as long as $t \rho \ll J_z v_z, J_z u_z^z \sim 1$.

\subsection{Discussion of ground-state properties} \label{sec-comp-disc}

We are now ready to make a comparison between the mean-field ground
state obtained from Eq.~(\ref{eq-comp-mft-H}) and the exact ground
state discussed in Secs.~\ref{sec-gas-free} and \ref{sec-fast}.
Although there are general trends in the phase diagram of the model
that are common to both approaches, this comparison reveals several
interesting discrepancies between the exact description and the
mean-field treatment. In particular, there are two significant
discrepancies concerning the internal degrees of freedom and the
particle statistics of mobile holes.

The most important result of our exact study is that each hole has
three internal degrees of freedom and that it can be characterized
by three corresponding quantum numbers $h$, $q$, and $p$. The
quantum number $p$ describes a local magnetization around the hole,
while the quantum numbers $h$ and $q$ capture the possibility of an
elementary excitation (flux or fermion) being bound to it. The
parton description in Ref.~\onlinecite{You} incorporates the quantum
number $p$ via the introduction of two distinct holon species (see
Sec.~\ref{sec-comp-part}). However, the mean-field treatment is
unable to represent the quantum numbers $h$ and $q$: it ignores the
possibility of bound states between holes and elementary excitations
as it inherently neglects any correlations between these independent
degrees of freedom.

In the regime of slow hopping, it is straightforward to verify
explicitly that all holes in the mean-field ground state have
quantum numbers $h = 0$ and $q = 0$ as they have no elementary
excitations bound to them. Since $t \ll J^4$ and $\rho \ll 1$ in
this regime, the conditions $t \rho \ll J^2$ and $t \rho \ll 1$ are
both satisfied, and hence the mean-field expectation values
$u_{\alpha}^{\alpha}$ and $v_{\alpha}$ for $\rho > 0$ are close to
those at $\rho = 0$. On the other hand, these expectation values are
the same as those obtained from the exact solution of the undoped
model. Since the exact ground state of the undoped model is free of
elementary excitations by definition, the mean-field ground state of
the doped model has no elementary excitations either. Note that the
quantum numbers of the mean-field ground state are then consistent
with those of the exact ground state in the case of $J^8 \ll t \ll
J^4$ but not in the case of $t \ll J^8$ (see
Sec.~\ref{sec-gas-free}).

The particle statistics of the various hole types are further
important results of our exact study. Unsurprisingly, the particle
statistics depends on the quantum numbers $h$ and $q$ as the binding
of an elementary excitation can lead to a statistical transmutation.
Since only bare holes with $h = 0$ and $q = 0$ are captured by the
mean-field treatment, the relevant comparison is between the bare
holes of the exact description and the holons of the mean-field
treatment. We find a remarkable discrepancy in this respect: the
bare holes of our exact study are fermions, while the holons of the
parton description in Ref.~\onlinecite{You} are bosons. It would
then be interesting to resolve this discrepancy by considering a
fermionic analogue of the mean-field treatment in
Ref.~\onlinecite{You}. For example, an appropriate transformation
between spinful bosons and spinful fermions\cite{Balents} could be
used to relate the two species of bosonic holons and the fermionic
bare holes with $p = \{ 0,1 \}$. Alternatively, it is natural to ask
how an analysis going beyond the mean-field saddle point could
provide the correct statistics.

Beyond the regime of slow hopping, we can compare the evolution of
the mean-field ground state as a function of $t$, $J$, and $\rho$
with our picture of the exact ground state where holes are
surrounded by clouds of fluctuating excitations. In the mean-field
treatment, there are two important characteristic scales of $t
\rho$. First, the bond fermion expectation values $u_x^x$, $u_y^y$,
and $u_z^z$ are all close to $1$ only for $t \rho \ll J^2$, and flux
excitations then start appearing at $(t \rho)_P \sim J^2$. Second,
the matter fermion expectation values $v_z$ are close to $1$ only
for $t \rho \ll 1$, and fermion excitations then start appearing at
$(t \rho)_f \sim 1$. In the language of the exact description in
Sec.~\ref{sec-fast}, the critical value $(t \rho)_{P,f}$ corresponds
to the critical density at which the fluctuating fluxes (fermions)
around different holes merge. If we assume that our upper bounds on
the excitation cloud radii are good estimates so that $R_P \sim
\sqrt{t / J^4}$ and $R_f \sim \sqrt{t}$, the corresponding critical
values from Sec.~\ref{sec-fast} are $(t \rho)_P \sim J^4$ and $(t
\rho)_f \sim 1$. These results have a simple interpretation: each
kind of excitation starts appearing when the kinetic energy density
$t \rho$ reaches its excitation energy. However, by using this
interpretation, we obtain inconsistent values for the flux
excitation energy as it is $E_P \sim J^4$ in the exact description
and $E_P \sim J^2$ in the mean-field treatment. The reason for this
inconsistency is that flux excitations do not appear explicitly in
the mean-field treatment but instead are decoupled as independent
bond fermion excitations.

At the isotropic point of $J = 1$, for which the mean-field theory
in Ref.~\onlinecite{You} is devised, the regime of slow hopping is
unattainable for any hopping amplitude due to the existence of
gapless fermionic excitations. It is then not \emph{a priori}
possible to think of each hole as possessing well-defined internal
degrees of freedom.\cite{Trousselet} The innocuous choice of a
quasiparticle representing the hole can be non-obvious due to the
intricate many-body problem posed by the detailed hole dynamics, and
in the most extreme scenario, it can even become ill-defined as the
coupling between the hole and the gapless excitations renders the
quasiparticle description problematic in itself. It is a natural
extension of our present work to consider a single isolated hole in
the gapless phase of the model and discuss its potential
quasiparticle representation along with any internal degrees of
freedom possessed by it.

\section{Summary and outlook} \label{sec-out}

In this work, we presented a thorough and controlled microscopic
study of slow mobile holes hopping in the spatially anisotropic
(Abelian) gapped phase of the Kitaev honeycomb model. We found that
the mobile holes in the model have internal degrees of freedom as
they can bind the fractional excitations of the model and that the
resulting hole types with different fractional excitations bound to
them are fundamentally different in terms of their single-particle
and multi-particle properties. We now conclude the paper with two
suggestions for the future direction of this research.

The interest in doped topological states is in part due to their
identification as possible candidates for high-temperature
superconductors.\cite{Lee, Anderson-0} If Cooper pairs are formed by
extra electrons or missing electrons (holes) in such a doped
topological state, the condensation of these Cooper pairs can lead
to superconducting behavior. As discussed briefly in
Sec.~\ref{sec-stat-bound}, the holes in the Kitaev honeycomb model
form bound pairs if the Coulomb repulsion is strong enough to
counteract phase separation but not strong enough to counteract pair
formation. It is then natural to ask what kind of internal degrees
of freedom these hole pairs possess and what their manifestations
are in the superconducting behavior of hole pairs. Furthermore, the
question of superconductivity is of central importance in the
complementary mean-field works.\cite{You, MFT} Consequently, an
exact study of hole pairs could further clarify the relation between
the exact description and the mean-field treatments.

The binding of fractional excitations by mobile holes is interesting
in part because it provides a controlled way of introducing
fractional particles into the model and manipulating the resulting
quantum state by exploiting the anyonic statistics of these
fractional particles.\cite{Dusuel} Importantly, the Kitaev honeycomb
model has even more exotic fractional excitations in its spatially
isotropic (non-Abelian) gapped phase.\cite{Kitaev} It is then
natural to expect that these fractional excitations with non-Abelian
anyonic statistics can also be bound to mobile holes and that the
properties of the resulting fractional particles would be
interesting to explore.


\begin{acknowledgments}

We thank F. J. Burnell, I. Kimchi, C. R. Laumann, and S. H. Simon
for enlightening discussions. This work was supported by the EPSRC
under Grant No.~EP/I032487/1, and by the Helmholtz Virtual Institute
``New States of Matter and Their Excitations''.

\end{acknowledgments}




\begin{references}

\bibitem{Dagotto} E. Dagotto, Rev. Mod. Phys. \textbf{66}, 763 (1994).
\bibitem{Lee} P. A. Lee, N. Nagaosa, and X.-G. Wen, Rev. Mod. Phys.
\textbf{78}, 17 (2006).
\bibitem{Anderson-0} P. W. Anderson, \emph{The Theory of Superconductivity
in the High-$T_C$ Cuprates} (Princeton University Press, Princeton,
1997).
\bibitem{HTS} J. G. Bednorz and K. A. M\"{u}ller, Z. Phys. B: Condens.
Matter \textbf{64}, 189 (1986).
\bibitem{Misguich} G. Misguich and C. Lhuillier, arXiv:cond-mat/0310405.
\bibitem{Anderson-1} P. W. Anderson, Science \textbf{235}, 1196 (1987).
\bibitem{Anderson-2} P. W. Anderson, Mater. Res. Bull. \textbf{8}, 153
(1973).
\bibitem{Moessner} R. Moessner and K. S. Raman, arXiv:0809.3051.
\bibitem{Kivelson} S. A. Kivelson, D. S. Rokhsar, and J. P. Sethna,
Phys. Rev. B \textbf{35}, 8865(R) (1987).
\bibitem{QDM-1} D. S. Rokhsar and S. A. Kivelson, Phys. Rev. Lett.
\textbf{61}, 2376 (1988); R. Moessner and S. L. Sondhi, Phys. Rev.
Lett. \textbf{86}, 1881 (2001).
\bibitem{Rajaraman} R. Rajaraman, arXiv:cond-mat/0103366.
\bibitem{Wen-1} X. G. Wen and Q. Niu, Phys. Rev. B \textbf{41}, 9377
(1990).
\bibitem{Kitaev} A. Y. Kitaev, Ann. Phys. \textbf{321}, 2 (2006).
\bibitem{Willans} A. J. Willans, J. T. Chalker, and R. Moessner,
Phys. Rev. Lett. \textbf{104}, 237203 (2010); A. J. Willans, J. T.
Chalker, and R. Moessner, Phys. Rev. B \textbf{84}, 115146 (2011).
\bibitem{Trousselet} F. Trousselet, P. Horsch, A. M. Ole\'{s}, and
W.-L. You, Phys. Rev. B \textbf{90}, 024404 (2014).
\bibitem{Baskaran-1} G. Baskaran, Z. Zou, and P. W. Anderson, Solid
State Commun. \textbf{88}, 853 (1993).
\bibitem{GP} M. C. Gutzwiller, Phys. Rev. Lett. \textbf{10}, 159
(1963); T. A. Kaplan, P. Horsch, and P. Fulde, Phys. Rev. Lett.
\textbf{49}, 889 (1982).
\bibitem{Wen-2} X.-G. Wen, Phys. Rev. B \textbf{65}, 165113 (2002).
\bibitem{Baskaran-2} G. Baskaran and P. W. Anderson, Phys. Rev. B
\textbf{37}, 580(R) (1988).
\bibitem{Essin} A. M. Essin and M. Hermele, Phys. Rev. B
\textbf{87}, 104406 (2013).
\bibitem{Hermele} M. Hermele, T. Senthil, M. P. A. Fisher, P. A. Lee,
N. Nagaosa, and X.-G. Wen, Phys. Rev. B \textbf{70}, 214437 (2004).
\bibitem{MFT} F. J. Burnell and C. Nayak, Phys. Rev. B \textbf{84},
125125 (2011); T. Hyart, A. R. Wright, G. Khaliullin, and B.
Rosenow, Phys. Rev. B \textbf{85}, 140510(R) (2012); J.-W. Mei,
Phys. Rev. Lett. \textbf{108}, 227207 (2012); R. Schaffer, S.
Bhattacharjee, and Y. B. Kim, Phys. Rev. B \textbf{86}, 224417
(2012); S. Okamoto, Phys. Rev. B \textbf{87}, 064508 (2013).
\bibitem{You} Y.-Z. You, I. Kimchi, and A. Vishwanath, Phys. Rev.
B \textbf{86}, 085145 (2012).
\bibitem{QDM-2} S. Kivelson, Phys. Rev. B \textbf{39}, 259 (1989);
D. Poilblanc, Phys. Rev. Lett. \textbf{100}, 157206 (2008).
\bibitem{Yao} H. Yao, S.-C. Zhang, and S. A. Kivelson, Phys. Rev. Lett.
\textbf{102}, 217202 (2009).
\bibitem{Lieb} E. H. Lieb, Phys. Rev. Lett. \textbf{73}, 2158 (1994).
\bibitem{Dusuel} S. Dusuel, K. P. Schmidt, and J. Vidal, Phys. Rev. Lett.
\textbf{100}, 177204 (2008).
\bibitem{Balents} L. Balents, M. P. A. Fisher, and C. Nayak, Phys. Rev.
B \textbf{61}, 6307 (2000).

\end{references}
\end{document}